\documentclass[twocolumn,trackchanges]{aastex631}
\usepackage{hyperref}
\usepackage{courier}
\usepackage{multirow}
\usepackage{color}
\usepackage{makecell}

\hypersetup{linkcolor=red,citecolor=blue,filecolor=cyan,urlcolor=magenta}

\shorttitle{Distant Superclusters in the HSC Survey}
\shortauthors{Chen et al.}

\begin{document}

\title{A Systematic Search of Distant Superclusters with the Subaru Hyper Suprime-Cam Survey}

\correspondingauthor{Tsung-Chi Chen}
\email{r11244003@ntu.edu.tw}

\author[0009-0008-7522-4179]{Tsung-Chi Chen}
\affiliation{Institute of Astronomy and Astrophysics, Academia Sinica, Taipei 10617, Taiwan}
\affiliation{Department of Physics, National Taiwan University, Taipei 10617, Taiwan}

\author[0000-0001-7146-4687]{Yen-Ting Lin}
\affiliation{Institute of Astronomy and Astrophysics, Academia Sinica, Taipei 10617, Taiwan}
\affiliation{Graduate Institute of Astrophysics, National Taiwan University, Taipei 10617, Taiwan}
\affiliation{Institute of Physics, National Yang Ming Chiao Tung University, Hsinchu 30010, Taiwan}

\author[0000-0002-1249-279X]{Hsi-Yu Schive}
\affiliation{Department of Physics, National Taiwan University, Taipei 10617, Taiwan}
\affiliation{Graduate Institute of Astrophysics, National Taiwan University, Taipei 10617, Taiwan}
\affiliation{Center for Theoretical Physics, National Taiwan University, Taipei 10617, Taiwan}
\affiliation{Physics Division, National Center for Theoretical Sciences, Taipei 10617, Taiwan}

\author[0000-0003-3484-399X]{Masamune Oguri}
\affiliation{Center for Frontier Science, Chiba University, Chiba-shi, Chiba 263-8522, Japan}

\author[0000-0002-3839-0230]{Kai-Feng Chen}
\affiliation{Department of Physics, Massachusetts Institute of Technology, Cambridge, MA 02139, USA}

\author{Nobuhiro Okabe}
\affiliation{Physics Program, Graduate School of Advanced Science and Engineering, Hiroshima University, 1-3-1 Kagamiyama, Higashi-Hiroshima, Hiroshima 739-8526, Japan}
\affiliation{Hiroshima Astrophysical Science Center, Hiroshima University, 1-3-1 Kagamiyama, Higashi-Hiroshima, Hiroshima 739-8526, Japan}
\affiliation{Core Research for Energetic Universe, Hiroshima University, 1-3-1, Kagamiyama, Higashi-Hiroshima, Hiroshima 739-8526, Japan}

\author{Sadman Ali}
\affiliation{Subaru Telescope, National Astronomical Observatory of Japan, Hilo, HI 96720, USA}

\author[0000-0003-4758-4501]{Connor Bottrell}
\affiliation{International Centre for Radio Astronomy Research, University of Western Australia, 35 Stirling Highway, 6009 Perth, Australia}

\author[0000-0002-7998-9899]{Roohi Dalal}
\affiliation{Department of Astrophysical Sciences, Princeton University, Princeton, NJ 08544, USA}

\author{Yusei Koyama}
\affiliation{Subaru Telescope, National Astronomical Observatory of Japan, Hilo, HI 96720, USA}

\author[0000-0001-6419-8827]{Rogério Monteiro-Oliveira}
\affiliation{Institute of Astronomy and Astrophysics, Academia Sinica, Taipei 10617, Taiwan}

\author[0000-0003-4442-2750]{Rhythm Shimakawa}
\affiliation{Waseda Institute for Advanced Study (WIAS), Waseda University, 1-21-1, Nishi Waseda, Shinjuku, Tokyo 169-0051, Japan}
\affiliation{Center for Data Science, Waseda University, 1-6-1, Nishi-Waseda, Shinjuku, Tokyo 169-0051, Japan}

\author{Tomotsugu Goto}
\affiliation{Institute of Astronomy, National Tsing Hua University, Hsinchu 30013, Taiwan}

\author{Bau-Ching Hsieh}
\affiliation{Institute of Astronomy and Astrophysics, Academia Sinica, Taipei 10617, Taiwan}

\author{Tadayuki Kodama}
\affiliation{Astronomical Institute, Graduate School of Science, Tohoku University, Sendai, Miyagi 980-8578, Japan}

\author{Atsushi J. Nishizawa}
\affiliation{DX Center, Gifu Shotoku Gakuen University, Gifu, Gifu 501-6194, Japan}

\begin{abstract}
Superclusters, encompassing environments across a wide range of overdensities, can be regarded as unique laboratories for studying galaxy evolution. Although numerous supercluster catalogs have been published, none of them goes beyond redshift $z=0.7$. In this work, we adopt a {\it physically motivated} supercluster definition, requiring that superclusters should eventually collapse even in the presence of dark energy. Applying a friends-of-friends (FoF) algorithm to the CAMIRA cluster sample constructed using the Subaru Hyper Suprime-Cam survey data, we have conducted the first systematic search for superclusters at $z=0.5-1.0$ and identified 673 supercluster candidates over an area of 1027 deg$^2$. The FoF algorithm is calibrated by evolving $N$-body simulations to the far future to ensure high purity. We found that these high-$z$ superclusters are mainly composed of $2-4$ clusters, suggesting the limit of gravitationally bound structures in the younger Universe. In addition, we studied the properties of the clusters and brightest cluster galaxies (BCGs) residing in different large-scale environments. We found that clusters associated with superclusters are typically richer, but no apparent dependence of the BCG properties on large-scale structures is found. We also compared the abundance of observed superclusters with mock superclusters extracted from halo light cones, finding that photometric redshift uncertainty is a limiting factor in the performance of superclusters detection.  
\end{abstract}


\section{Introduction}\label{sec:intro}
In the standard model of cosmology, $\Lambda$CDM (cold dark matter with a cosmological constant), 
massive structures grow via gravitational interaction between structures of smaller scales (e.g., \citealt{white78,White91, Springel+05}). In this hierarchical formation scenario, superclusters, often loosely defined to be groups of clusters, are the most massive coherent structures, spanning length scales from tens to hundreds of megaparsecs. These structures largely preserve initial conditions and display nonspherical morphology \citep{Basilakos03, Einasto07b, Costa-Duarte11} as the time scale required for a galaxy cluster to traverse such a length scale exceeds the age of the Universe. This indicates that superclusters may be a potential test for the cosmological model \citep{Sheth&Diaferio11, Park12}. On the other hand, environments associated with superclusters often contain richer and more luminous galaxy groups and clusters compared to environments with lower global densities \citep{Einasto03, Luparello13, Chon13} and are accompanied by voids \citep{Einasto97}. This suggests that superclusters are unique laboratories for studying how environments influence galaxy formation and evolution at very large scales \citep{Einasto07c, Costa-Duarte13, Luparello13, Einasto14, Cohen17, Guglielmo18, Einasto20, Monteiro-Oliveira22}. However, the extent to which these high-density environments affect galaxy and galaxy cluster properties remains an open question. 

Two popular methods that have been adopted for supercluster finding are the luminosity field method \citep{Einasto07a, Luparello11, Liivamagi12} and the Friends-of-Friends (FoF) algorithm \citep{Basilakos03, Chon13, Sankhyayan23}. In the luminosity field method, a luminosity density map is constructed by smoothing the galaxy luminosity with a kernel of a certain length scale. Superclusters are then selected from the luminosity density field above a chosen density threshold $D_{T}$ and are associated with galaxy groups or clusters. As for the FoF approach, for a given linking length $l$, the FoF algorithm links clusters together if their separation is smaller than $l$ and concatenates any connections that share common members into FoF groups (of clusters). The linking length is conventionally parameterized by the mean cluster separation: $l = b\cdot(\bar{n})^{-\frac{1}{3}}$, where $\bar{n}$ is the global cluster number density at a given redshift and $b$ is a dimensionless quantity often called the linking parameter.  

Despite the long history of studies of superclusters, a universal definition of supercluster has not yet been reached by the community (see the introduction by \citealt{Sankhyayan23}). The definition of superclusters that a given catalog adopted is largely reflected by the density threshold or linking length/parameter chosen during the sample construction. Selecting a higher density threshold or shorter linking length fragmentizes structures, while a lower density threshold or larger linking length percolates networks. A density threshold $D_{T}=4.5$ and a linking length $l\approx 20\,$ comoving Mpc (hereafter cMpc) are adopted by \citet{Einasto07a} and \citet{Sankhyayan23}, respectively, to capture superclusters classified as unbound over-dense regions. According to the spherical collapse model (e.g., \citealt{Busha03}; \citealt{Dunner06}), such structures will not successfully decouple from the Hubble flow and will be separated into smaller structures, forming several isolated regions, the so-called ``island universes''. 

\citet{Chon15} proposed a physically-motivated definition that superclusters should survive the accelerating expanding Universe and collapse in the future. Building upon this idea, we show in this work that this definition provides a natural boundary for the length scale of the supercluster and in turn physically regulates the parameters for the supercluster cluster finding methods. Moreover, such a definition enables the construction of a more homogeneous catalog and permits more direct comparison among different works. \citet{Luparello11} and \citet{Chon13} have implemented this definition in their supercluster catalog construction by selecting the density threshold $D_T=5.5$ and the linking length parameter $b\approx 0.46$, respectively. In this work, we shall adopt the definition of superclusters proposed by \citet{Chon15}.

Many supercluster catalogs have been constructed over the past two decades (e.g., \citealt{Einasto07a, Luparello11, Liivamagi12, Chon13, Chow-Martinez14, Sankhyayan23}) due to the advent of large sky surveys, starting with the Sloan Digital Sky Survey (SDSS; \citealt{York00}) and the 2dF Galaxy Redshift Survey (2dFGRS; \citealt{Colless01}). Although the wide-area sky coverage enables the discoveries of large number of superclusters in the local Universe, the survey depth, in the case of SDSS, has restricted the redshift range to $z\lesssim0.4$. The discovery and confirmation of high redshift superclusters (e.g., $z\geq0.7$) is only possible through individual, targeted studies (e.g., \citealt{Rosati99, Lubin00, Nakata05, Swinbank07, Gilbank08, Kim16}). Therefore, there is not yet a systematic search of superclusters beyond redshift $z>0.7$ over a large area on the sky. 

The Hyper Suprime-Cam Subaru Strategic Program (HSC-SSP, hereafter the HSC survey) is a large optical imaging survey primarily designed to probe the dark matter distribution via weak lensing and to unveil the evolution of galaxies \citep{Aihara18a, Aihara18b}. Using the Hyper Suprime-Cam \citep{Miyazaki12, Miyazaki18, Komiyama18, Furusawa18} installed at the $8.2$m Subaru telescope in Hawai'i, the 330-night survey started in 2014 and finished the observations in 2022. It consists of three layers: wide, deep, and ultra-deep, characterized by different survey areas and depths in five broad-band filters ($grizy$; \citealt{Kawanomoto18}). The HSC survey data are processed through the \texttt{hscPipe} pipeline \citep{Bosch18, Huang18}, which originates from the Legacy Survey of Space and Time (LSST) pipeline \citep{Juric17, Ivezic19}. To conduct astrometric and photometric calibration, the HSC survey has utilized Pan-STARRS1 data \citep{Tonry12, Schlafly12, Magnier13}. In the latest public data release (PDR3; \citealt{Aihara22}), the HSC-wide layer has reached a limiting magnitude of $i_{\rm AB}\approx 26$ ($5\sigma$ for point sources within a $2''$ aperture) over $\approx 670$\,deg$^2$ observed in $grizy$ filters. 

With the extraordinary depth and area of the HSC survey, we have conducted a systematic search for superclusters at redshift $z=0.5-1.0$ over a sky area of $\approx 1027$ deg$^2$ based on the latest internal data release (see Section~\ref{subsec:camira}). This is the first systematic search of distant superclusters that extends to such a high redshift over a thousand deg$^2$. Our final supercluster catalog consists of 673 supercluster candidates with multiplicity between $2-5$. 

We structure this paper as follows. Section \ref{sec:data} describes the observational data and explains how we examine and evaluate the cluster photometric redshift (photo-$z$). We present our methodology, including the $N$-body simulations, linking length optimization and validation, in great detail in Section \ref{sec:method}. Our supercluster catalog is presented and compared with two known superclusters in Section \ref{sec:cat}.  In Section \ref{sec:BCG}, we show the results of an analysis of environmental effects on galaxies and clusters belonging to superclusters. Discussions and conclusions are given in Section \ref{sec:summary}. Throughout this paper, magnitudes are given in the AB magnitude system \citep{oke83}. We assume a flat $\Lambda$CDM model with cosmological parameters similar to \citet{Oguri18}: $\Omega_{M}=0.28$, $\Omega_{\Lambda}=0.72$, $h=0.7$, $n_{s}=0.96$, and $\sigma_8=0.82$. For the definition of halo mass, we adopt $M_{200m}=\frac{4\pi}{3}(200\overline{\rho_m})R_{200m}^3$, where $R_{200m}$ is the physical radius such that the enclosed spherical density is 200 times the mean matter density $\overline{\rho_m}$ at the cluster redshift. 

\section{Galaxy cluster and Photometric redshift data} \label{sec:data}
We construct our supercluster sample using the latest version of the CAMIRA (Cluster ﬁnding algorithm based on Multi-band Identiﬁcation of Red sequence gAlaxies) cluster catalog \citep{Oguri18}. We present an overview of the CAMIRA cluster catalog and describe our sample selection below. Furthermore, in order to enhance the purity of our cluster sample, we compare the CAMIRA cluster photo-$z$ with spectroscopic redshifts from the Early Data Release from Dark Energy Spectroscopic Instrument (DESI EDR; \citealt{DESI_EDR23}), as well as photo-$z$ of cluster member galaxies derived by the code {\it Direct Empirical Photometric method}  (DEmP; \citealt{Hsieh14}).

\subsection{CAMIRA cluster catalog\label{subsec:camira}}

The CAMIRA catalog is built by applying the CAMIRA algorithm \citep{Oguri14} to the internal release of HSC survey data ``S21A'' (with the latest bright star masks), with data obtained between April 2014 and January 2021.\footnote{The PDR3 is based on data obtained up to January 2020.} In short, the CAMIRA algorithm utilizes the calibrated stellar population synthesis (SPS) model of \citet{BC03} to estimate the likelihood of a galaxy being a red-sequence galaxy and derived a ``number parameter'' based on it for each galaxy. Two-dimensional richness maps are then constructed by convolving a stellar mass filter $F_M$ and a spatial filter $F_R$ with the number parameter of each galaxy at a given redshift slice. For each  peak in the richness maps, the CAMIRA cluster photo-$z$ $z_{CAM}$ is obtained by maximizing the likelihood (see Eqn.~15 of \citealt{Oguri14}) and iteratively identifying the Brightest Cluster Galaxy (hereafter BCG) position until convergence. \citet{Oguri18} have shown that the redshifts are highly accurate, with a bias of $|\delta_{z}| \approx 0.005(1+z)$ and a scatter of $\sigma_{z} \approx 0.01(1+z)$ throughout the redshift range of $z=0.1-1.0$. For full details of the algorithm and performance of CAMIRA, please refer to \citet{Oguri14} and \citet{Oguri18}. 

In the CAMIRA cluster catalog, the richness $N_{mem}$ and the cluster photo-$z$ $z_{CAM}$ are provided. In addition, the (red) members of each CAMIRA cluster are provided in CAMIRA cluster member catalog with stellar mass $M_{*}$ and membership probability $w$ (see Eqn.~20 of \citealt{Oguri14}). The stellar mass $M_{*}$ is converted from the {\tt cmodel} magnitude \citep{Bosch18} using the mass-to-light ratio given by the calibrated SPS model. The membership probability $w$ scales with the multiplication between the number parameter, the stellar mass filter $F_M$, and the spatial filter $F_R$ (Eqn.~5, 8, and 9 in \citealt{Oguri14}). The stellar mass filter downweights galaxies with very high ($\ge  10^{13}\,M_{\odot}$) and low stellar mass   ($\lessapprox 10^{10.2}\,M_{\odot}$) for the reduction of the projection effect and for more accurate richness estimation. The spatial filter reduces the weight for galaxies located at the outskirts of  clusters. The richness quantifies the $w$-weighted count of red member galaxies located within a circular aperture of size $1$\,$h^{-1}\,$ physical Mpc (hereafter pMpc) and having stellar mass $M_{*} \gtrsim 10^{10.2} M_{\odot}$. In the latest version of the CAMIRA sample, a lower limit in richness of $N_{mem}=10$ is imposed. The cluster catalog also provides spectroscopic redshifts of BCGs $z_{BCG}$ if available. These spectroscopic redshift are from several spectroscopic surveys overlapping with the HSC survey footprint, including zCOSMOS DR$3$ \citep{2009ApJS..184..218L}, UDSz (\citealt{2013MNRAS.433..194B}; \citealt{2013MNRAS.428.1088M}), 3D-HST (\citealt{2014ApJS..214...24S}; \citealt{2016ApJS..225...27M}), FMOS-COSMOS \citep{2015ApJS..220...12S}, VVDS \citep{2013A&A...559A..14L}, VIPERS DR$2$ \citep{2018A&A...609A..84S}, SDSS DR$15$ \citep{2019ApJS..240...23A}, SDSS QSO DR$14$ \citep{paris18}, GAMA DR$3$ \citep{2018MNRAS.474.3875B}, WiggleZ DR$1$ \citep{2010MNRAS.401.1429D}, DEEP$2$ DR$4$ \citep{newman13}, DEEP$3$ \citep{cooper11}, PRIMUS DR$1$ (\citealt{2011ApJ...741....8C}; \citealt{2013ApJ...767..118C}), $2$dFGRS \citep{2003astro.ph..6581C}, $6$dFGRS (\citealt{2004MNRAS.355..747J}, \citealt{2009MNRAS.399..683J}), C$3$R$2$ DR$2$ (\citealt{2017ApJ...841..111M}; \citealt{2019ApJ...877...81M}), DEIMOS $10k$ sample \citep{Hasinger18}, LEGA-C DR$2$ \citep{2018ApJS..239...27S} and VANDELS DR$4$ \citep{2021A&A...647A.150G}. 
If $z_{BCG}$ of a cluster is available and satisfies
\begin{equation}
    |z_{BCG} - z_{CAM}| < \sigma_z(1+z_{CAM})
\end{equation}
(where $\sigma_z \approx 0.01$ is the scatter), we adopt $z_{BCG}$ as the cluster redshift. In Appendix \ref{subsec:appendix_BCG}, we provide an  analysis showing that the redshift of BCG could be representative of cluster redshift.

\begin{figure}[htb]
    \centering
    \begin{minipage}{0.475 \textwidth}
        \begin{center}
            \includegraphics[width = 0.99\hsize]{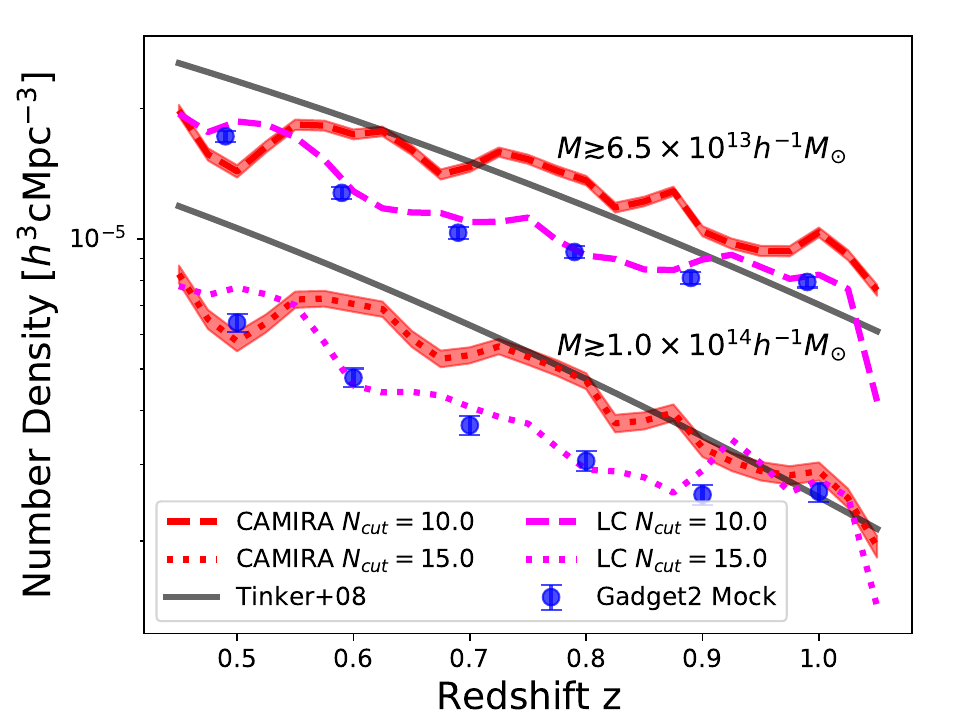}  
        \end{center}
    \end{minipage}
    \caption{The  comoving number density as a function of redshift of CAMIRA S21A cluster sample (red curves) and mock catalogs made from our simulation and halo light cone from \citet{Takahashi17} (blue points and magenta curves; described later in Section \ref{subsec:mock}). The curves in dashed style and the upper blue points correspond to a richness cut $N_{cut}=10.0$. The curves in the dotted style and the lower blue points correspond to a richness cut $N_{cut}=15.0$. The black solid curves are the halo number density obtained by integrating the halo mass function of \citet{Tinker08} from $6.5\times10^{13}\,h^{-1}M_{\odot}$ and $1.0\times10^{14}\,h^{-1}M_{\odot}$, respectively. The error bands and error bars are calculated assuming Poisson errors. We do not show the errors of the mock catalog made from \citet{Takahashi17} for clarity of the Figure.}
    \label{fig:n_cl}
\end{figure}

Although the CAMIRA cluster catalog contains cluster candidates with photo-$z$ up to $z\approx 1.4$, we focus on the redshift range between $0.5$ and $1.0$ in this work, as the richness of $z>1$ CAMIRA clusters is not corrected for incompleteness \citep{Oguri14, Oguri18} and therefore would be underestimated. Figure \ref{fig:n_cl} shows the comoving number density of the CAMIRA clusters as a function of redshift. Compared to the halo number density predicted by the halo mass function of \citet{Tinker08}, the richness cuts of $N_{mem}\ge 10.0$ and $N_{mem}\ge 15.0$ correspond to  halo masses $M_{200m}\approx 6.5\times 10^{13}\,h^{-1}\,M_{\odot}$ and $\approx 1.0\times 10^{14}\,h^{-1}\,M_{\odot}$, respectively.

As the CAMIRA algorithm utilizes red-sequence galaxies to search for galaxy clusters, blue clusters may possibly be missed in our sample. However, according to \citet{Misato22}, who presented a sample of 43 blue clusters (median blue galaxy fraction $f_b \approx 0.54$) identified at redshift $z=0.84$, the weak-lensing masses of these blue clusters are $\approx 4$ times lower than those of CAMIRA clusters with richness $N_{mem}=10-19$ (see Figure 7 of \citealt{Misato22}). 
Therefore, we believe that there should be no concern for incompleteness in our supercluster sample construction due to the use of a red-sequence-based cluster sample.

\subsection{DESI EDR\label{subsec:DESI}}

We next examine the reliability of the photo-$z$ of our cluster sample by comparing with spectroscopic redshifts of potential cluster member galaxies and quasars from the DESI EDR, which have not yet been included in the HSC data release. We select extragalactic objects from their HEALPix-based redshift summary catalog that meet the criteria below: (1) $\texttt{ZCAT\_PRIMARY}=\texttt{True}$, (2) $\texttt{ZWARN}=0$, and (3)$\texttt{SPECTYPE}=\texttt{GALAXY}$ or $\texttt{QSO}$. 

The cluster photo-$z$ validation is done as follows. Considering that cluster photo-$z$ $z_{CAM}$ might be offset from the true value, we first search for DESI galaxy concentrations by calculating the $1\sigma$-clipped median $z_0$ of all DESI galaxies within a ``long'' cylinder of radius $1h^{-1}\,$pMpc and line-of-sight distance of $\pm \Delta z(1+z_{CAM})$. We then place a ``shorter'' cylinder with radius $1h^{-1}\,$pMpc and line-of-sight distance of $\Delta v = \pm 3000\,$km/s at the location of $1\sigma$-clipped median $z_0$ derived above. If we find at least $N_{min}$ DESI galaxies within the shorter cylinder of a given cluster, we then designate the cluster as confirmed and derive a new redshift $z_{cl, DESI}$ for it using the median spectroscopic redshift of the DESI galaxies that falls inside the shorter cylinder. We select $\Delta z=0.02$, $N_{min}=4$, and detail the choice of these parameters in Appendix \ref{subsec:appendix_DESI}. This way, photo-$z$ of 883 clusters are validated.

\subsection{DEmP\label{subsec:photo-z}}

To compensate for the absence of spectroscopic redshifts in our cluster sample, we further compare the CAMIRA cluster photo-$z$ with the member galaxy photo-$z$ derived by DEmP.
As described in \citet{Hsieh14}, DEmP implements regional polynomial fitting and uniformly-weighted training set to mitigate the issues regarding proper fitting function and biased training set, respectively. However, in the HSC survey, only the regional polynomial fitting is implemented, as the uniformly weighted training set does not improve the overall performance due to the large training set \citep{Tanaka18}. According to \citet{Nishizawa20}, the photo-$z$, stellar mass, and star formation rate (SFR) for each galaxy are derived from the linear fitting of $40$ nearest objects in a nine-dimensional parameter space.  

We first obtain the DEmP photo-$z$ for each CAMIRA member galaxy by cross-matching the DEmP catalog and CAMIRA cluster member galaxy catalog with a separation less than $0.08''$ and select the nearest primary source \footnote{Using object IDs to retrieve the photometric redshift of cluster member galaxies is a more straightforward approach. Unfortunately, information on object IDs is not provided in the CAMIRA cluster member galaxy catalog when the work is done.}. The \texttt{photoz\_mode} is adopted as the point estimator of the redshift. For each CAMIRA cluster in our sample, we construct its photo-$z$ cumulative distribution function (CDF) after removing member galaxies with \texttt{photoz\_mode\_risk} $\geq0.1$ and membership probability $w < w_{thres}$. We consider an interval centered at the CAMIRA photo-$z$, $[z_l, z_u]$, where $z_u=z_{CAM} + \Delta z(1+z_{CAM})$, $z_l=z_{CAM} - \Delta z(1+z_{CAM})$, and $\Delta z=0.02$. The cluster photo-$z$ is considered to be consistent with the photo-$z$ of the member galaxies if the following criteria hold: (1) $CDF(z_u) - CDF(z_l) > 0.5$, and (2) Number of member galaxies within the interval $[z_l, z_u]$ greater than or equal to $N_{min}$. The parameters $w_{thres}$ and $N_{min}$ are empirically determined; we adopt $w_{thres}=0.5$ and $N_{min}=10$ here. The justification on the choices of these hyperparameters is described in Appendix \ref{subsec:appendix_DEmP}. We also note that, due to the extraordinary depth of the HSC wide layer, the completeness is $>95\%$ in the redshift range $z=0.3-1.1$ for galaxies with stellar mass $\gtrapprox 3\times 10^{9}M_{\odot}$ \citep{Lin17}. Therefore, we conduct the confirmation process above without considering the effect of completeness.

\begin{figure}[htb]
    \centering
    \begin{minipage}{0.475\textwidth}
        \begin{center}
            \includegraphics[width=0.99\textwidth]{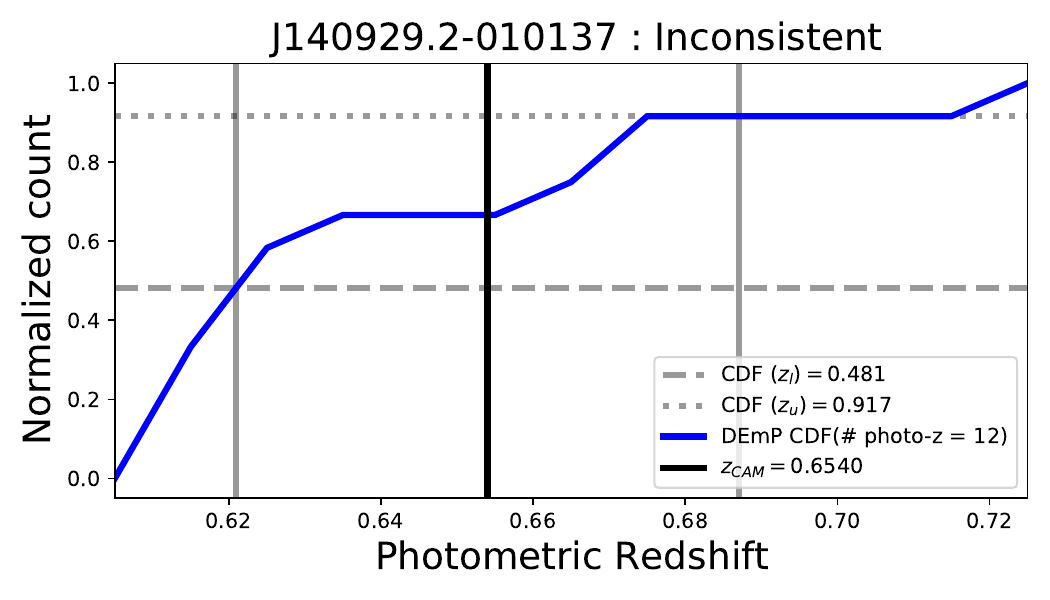}
        \end{center}
    \end{minipage}
    \begin{minipage}{0.475\textwidth}
        \begin{center}
            \includegraphics[width=0.99\hsize]{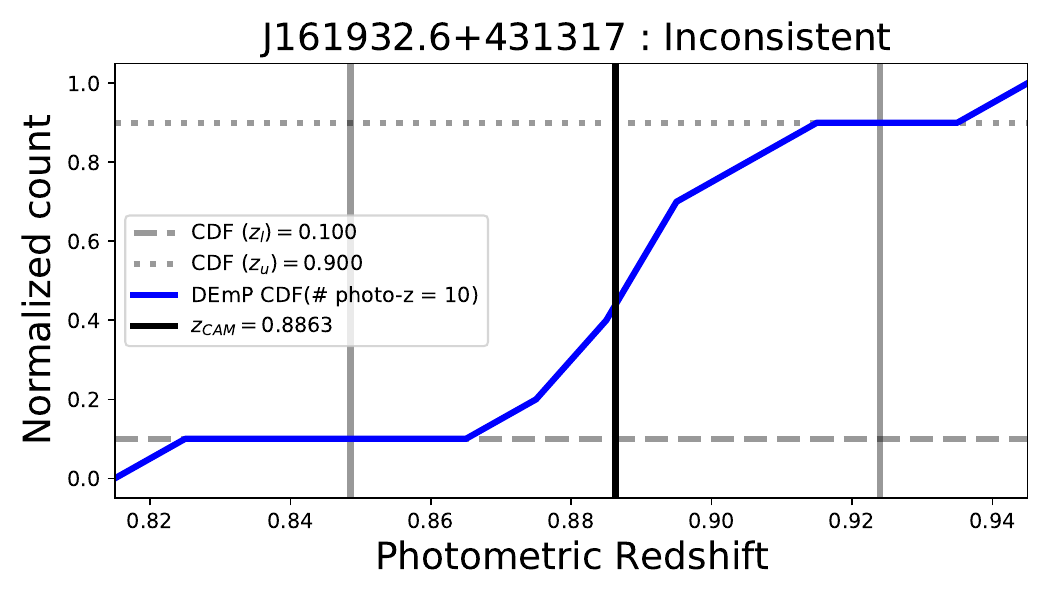}
        \end{center}
    \end{minipage}
    \begin{minipage}{0.475\textwidth}
        \begin{center}
            \includegraphics[width=0.99\hsize]{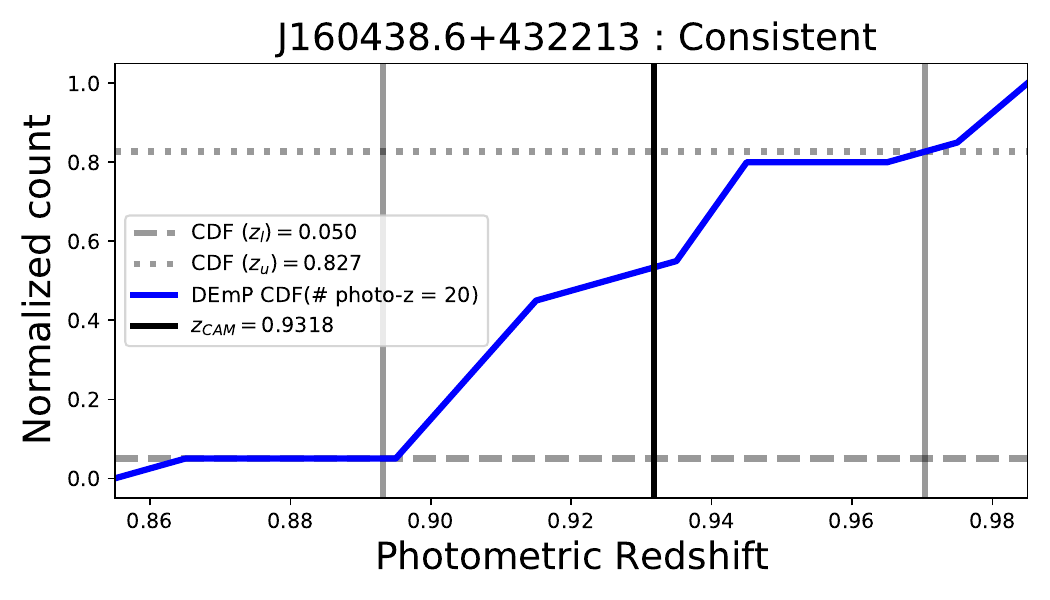}
        \end{center}
    \end{minipage}
    \caption{Demonstration of comparing CAMIRA cluster photometric redshift with DEmP photometric redshift of member galaxies. The blue curve shows the normalized cumulative distribution function (CDF) of member galaxies'  photo-$z$ estimated by DEmP. The black vertical curve indicates the cluster photo-$z$. The gray dashed line and the gray dotted line indicate the value of CDF at the lower and upper bound of the redshift interval $[z_l, z_u]$ (two vertical gray lines) considered, respectively. According to criteria described in Section \ref{subsec:photo-z}, the cluster photo-$z$ is inconsistent with member galaxies' photo-$z$ in the top and middle panels, while the bottom panel demonstrates a consistent case.}
    \label{fig:compare_DEmP}
\end{figure}

Figure \ref{fig:compare_DEmP} shows example CDFs of some CAMIRA clusters to illustrate our procedure. The CDF is shown by the blue curve, and the solid black vertical line indicates the cluster photo-$z$. The two gray horizontal lines indicate the fraction of member galaxies within the interval $[z_l, z_u]$ (represented by the vertical gray lines). Following our criteria described above, the bottom panel shows consistency between the photo-$z$ of the cluster and its member galaxies, while the other two do not satisfy the conditions. 

\subsection{Summary of Photo-z examinations} \label{subsec:photo-z_summary}

\begin{figure}[htb]
    \centering
    \begin{minipage}{0.475\textwidth}
        \begin{center}
            \includegraphics[width=0.99 \textwidth]{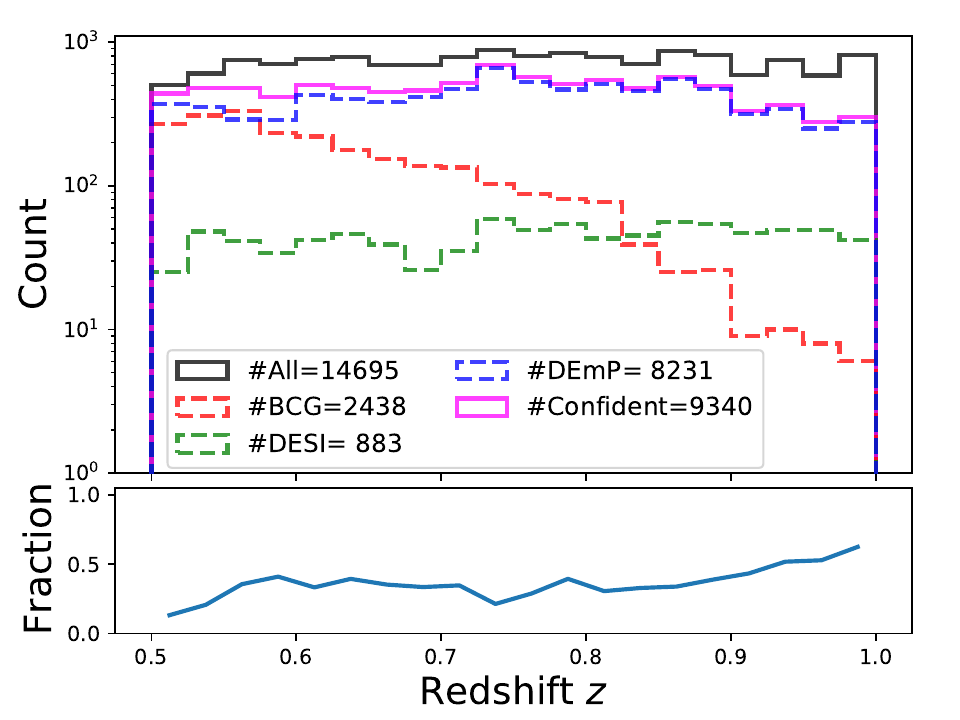}
        \end{center}
    \end{minipage}
    \caption{Upper panel: Redshift distributions of different subsets: the entire cluster sample (black histogram), consistent BCG spectroscopic redshift (red histogram),  DESI confirmation (green histogram), member galaxy DEmP photometric redshift check (blue histogram) and clusters with confident redshift (magenta histogram). The total cluster count for each subset is shown in the legend. Lower panel: Fraction of cluster photometric redshift without any confirmation as a function of redshift.}
    \label{fig:z_distribution}
\end{figure}

In the previous three subsections, we have used three different approaches to examine the photometric redshift accuracy of CAMIRA clusters. The upper panel of Figure \ref{fig:z_distribution} summarizes the results of these three examination process: confirmation with DESI,  BCG spectroscopic redshift consistency examination, and member galaxy DEmP photo-$z$ check. The fraction of clusters validated by DESI EDR, BCG spec-$z$, and DEmP photo-$z$ are $6\% \,(883/14695)$, $17\% \,(2438/14695)$, and $56\% \,(8231/14695)$, respectively. The low fraction of DESI and BCG spec-z examination result from the limited area of available spectroscopic surveys. Although based on the same photometric catalog, the DEmP examination validates around half of the CAMIRA clusters in our sample. This is expected as photometric redshift of galaxies typically has a larger scatter ($\approx 0.025(1+z_{phot})$, see \citealt{Nishizawa20}) than that of clusters.  In addition, among the clusters validated by BCG spec-$z$, approximately $63\%$ of clusters are also validated by DEmP photo-$z$, which is close to our completeness estimation described in Appendix \ref{subsec:appendix_DEmP}.

We refer to the subset of clusters that satisfies one or more of the confirmation tests as having {\bf confident} redshifts. These clusters constitute $64\%$ of the clusters at redshift $z=0.5-1.0$. When a cluster is validated by multiple confirmation tests described above, we use the redshift of the cluster following the order: $z_{cl, DESI}$ (Section \ref{subsec:DESI}), $z_{BCG}$ (Section \ref{subsec:camira}), and $z_{CAM}$ (Section \ref{subsec:photo-z}). Following this redshift priority,
14,695 clusters at $z=0.5-1.0$ from the HSC wide layer are selected as our sample. The lower panel of Figure \ref{fig:z_distribution}  shows the fraction of clusters that do not have a confident redshift as a function of the CAMIRA photo-$z$. This curve will later be utilized to perturb the redshift of the halo light cone to mimic the CAMIRA photo-$z$ uncertainty in Section \ref{subsec:SC_n}.

\section{Construction of CAMIRA supercluster catalog} \label{sec:method}

\begin{figure*}[htb]
    \centering
    \begin{minipage}{0.49\textwidth}
        \begin{center}
            \includegraphics[width=0.99\textwidth]{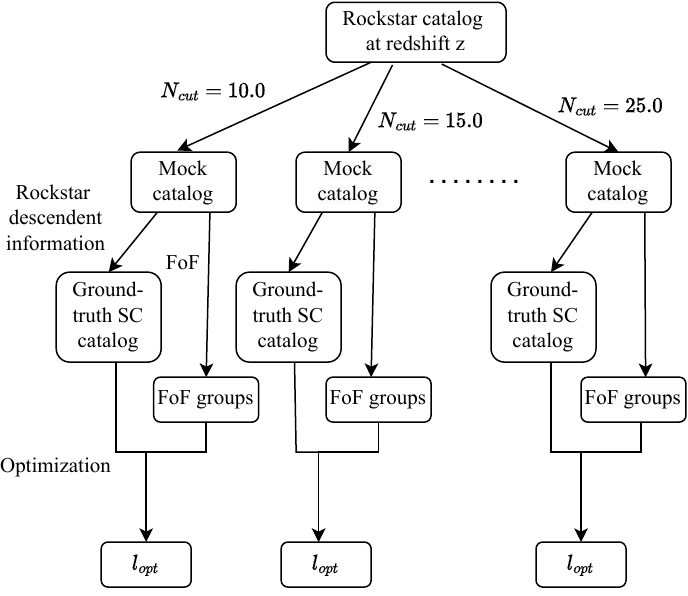} 
        \end{center}
    \end{minipage}\hfill
    \begin{minipage}{0.49\textwidth}
        \begin{center}
            \includegraphics[width=0.99\textwidth]{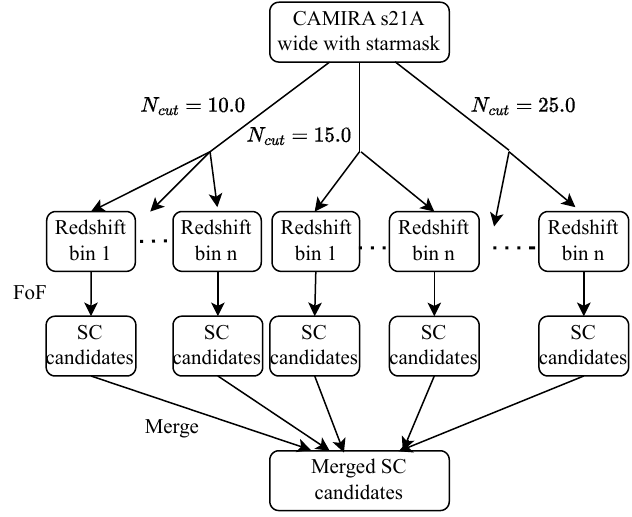} 
        \end{center}
    \end{minipage}   
    \caption{Left: The workflow of linking length calibration. We start by applying the richness cut $N_{cut}$ to the Rockstar catalogs to construct mock catalogs (Section \ref{subsec:mock}). The optimized linking length $l_{opt}$ (Section \ref{subsec:optimization}) is obtained by iteratively comparing the results of the FoF algorithm with ground-truth superclusters (referred to as SCs in the Figure).  Right: The workflow of the supercluster finding algorithm. First, we apply different richness cuts and divide the CAMIRA clusters into different redshift bins. We then apply the calibrated linking length to each redshift bin and obtain supercluster candidates. The last step involves merging supercluster candidates from different redshift bins and the richness cuts into the final supercluster catalog.} 
    \label{fig:flow}
\end{figure*}

Our supercluster finding algorithm is designed on the basis of the classic FoF algorithm \citep{huchra82}. We modify the \texttt{FoFGroups} class provided by \texttt{Halotools}  \citep[][{\tt v0.7}]{Hearin17} to meet our needs. In order to connect groups of clusters that will eventually experience gravitational collapse, we employ numerical simulations to calibrate the linking length. The left panel of  Figure \ref{fig:flow} shows the workflow of the linking length calibration. Our strategy here is to first evolve our $N$-body simulations to the far future (Section \ref{subsec:simulation}), and we construct mock catalogs from simulation snapshots (Section \ref{subsec:mock}). The optimized linking length is calibrated for each mock catalog by utilizing the descendent information from the halo finder (Section \ref{subsec:optimization}). Moreover, we also estimate the theoretical performance for our supercluster finding method (Section \ref{subsec:validation}). 

\subsection{N-body simulation\label{subsec:simulation}}

By definition, our supercluster candidates must undergo gravitational collapse {\it eventually}.  We therefore rely on $N$-body simulations that are run well beyond $z=0$ (or $a=1$, where $a$ is the cosmic expansion factor) to find the optimal linking lengths for our FoF algorithm. We first generate the unigrid initial condition at redshift $z=100.0$ by MUSIC (MUlti-Scale-Initial-Conditions; \citealt{Hahn11}) with the transfer function provided by CAMB (Code for Anisotropies in the Microwave Background; \citealt{Lewis00}). The initial condition is then evolved by GADGET-2 (Galaxies with dark matter and gas interact; \citealt{Springel05}) up to scaling factor $a=15.0$ \footnote{The optimized linking length is only reduced by approximately $0.1\,h^{-1}\,$cMpc if we set the final snapshot at $a=10.0$. Hence, the optimization process converges with respect to the end of simulation time.} when the age of the Universe is approximately 57.4 Gyr. Dark matter halos are extracted from each snapshot by the halo finder ROCKSTAR (Robust Overdensity Calculation using K-Space Topologically Adaptive Refinement; \citealt{Behroozi13}). We have verified that the halo mass function output from ROCKSTAR is consistent with the analytical fitting formula of \citet{Tinker08}.

After an extensive convergence test (Appendix \ref{sec:convergence}), we set the fiducial simulations to have particle number $N=1024^3$, comoving box size $L=700\,h^{-1}\,$cMpc and gravitational softening length $\epsilon=15\,h^{-1}\,$ckpc. The corresponding dark matter mass resolution is then $2.5\times10^{10}\,h^{-1}\,M_{\odot}$. As described in Section \ref{subsec:mock} below, the lowest halo mass we considered, $M_{200m}=2.0\times 10^{13}\,h^{-1}\,M_{\odot}$, is resolved by $\approx 800$ dark matter particles. We dump snapshots at redshift $z_i=0.5, 0.6, 0.7, 0.8, 0.9, 1.0$ and scaling factor $a_i=1, 2, 5, 10, 15$. Snapshots at redshift $z>0.0$ are used to construct mock catalogs, while snapshots at  $a\geq 1$ are used to identify superclusters. As the simulation box volume is comparable to the survey volume (i.e., the $z=0.5-1$ over the HSC survey footprint), we run six simulation boxes, each with a different realization of the initial condition, to account for cosmic variance.

\subsection{Construction of mock catalogs \label{subsec:mock}}

As our linking length relies on calibration by simulations, it is essential to tailor the ROCKSTAR catalogs to mock catalogs that faithfully reflect the richness distribution of the CAMIRA clusters.  Following \citet{Murata19}, the richness--mass relation $P(lnN|M,z)$ of CAMIRA clusters is modeled by a log-normal distribution with scatter $\sigma_{lnN|M,z}$, which depends on halo mass and redshift. For a given cosmology, the eight parameters in the model of \citet{Murata19} are constrained by jointly fitting the measurements of the cluster abundance and the stacked lensing profile. Here, we adopt their parameter estimations derived from the WMAP9 cosmology, which does not cause any differences in our results even though it is not the same as our assumed cosmology (Table 2 in \citealt{Murata19}). 

Inspired by Lin et al.~(in prep), we construct the mock catalog by assigning a richness to each halo in a Monte Carlo fashion. In practice, for a given ROCKSTAR halo catalog at redshift $z_i$, we consider halos with  mass $M_{200m}\geq 2.0\times 10^{13}\,h^{-1}\,M_{\odot}$ and designate the richness $N_{mem}$ to each halo by drawing from the richness-halo mass relation $P(lnN|M=M_{200m},z=z_i)$ \citep{Murata19}. The lower limit of the halo mass considered is selected such that less massive halos have a negligible probability of receiving richness $N_{mem}>10.0$, the lower limit of the CAMIRA catalog. After assigning richness to each halo, we apply the richness cut $N_{mem} \geq 10.0$ to construct mock catalogs. By this approach, we create ten Monte Carlo realizations for each of the six $N$-body simulations; therefore we have 60 different realizations in total. 

In order to better interpret our observational results, we also generate mock catalogs from the halo light cones produced by \citet{Takahashi17}, who presented a suite of 108 all-sky ray-tracing simulations. The light cones were constructed on the basis of 14 nested-arranged $N$-body simulations of different box sizes. As the number of particles was fixed to $N=2048^3$, the mass resolutions and softening lengths decrease as the size of the box increases (see Table 1 of \citealt{Takahashi17}). The initial conditions were generated by the second-order Lagrangian perturbation theory (2LPT; \citealt{Crocce06, Nishimichi09}) and CAMB, and were then evolved by GADGET-2. Dark matter halos with minimum particle number of 50 were captured by ROCKSTAR. Following \citet[][see Appendix C therein]{Shirasaki15}, \citet{Takahashi17} computed the projected angular position of halos in the sky using the ray-tracing simulation. Moreover, the radial distance and peculiar velocity of each halo were converted to the corresponding redshift. 

In practice, after retrieving the halo light cone data, we make a cutout between $0.0^{\circ}\leq R.A. \leq 150^{\circ}$ and $-2^{\circ} \leq Dec. \leq 5^{\circ}$, assign richness to each halo following the similar procedure described above, and apply a richness cut $N_{cut}=10.0$. We note that the mass resolution of the \citet{Takahashi17} light cone in the redshift range we consider is lower than our fiducial simulations, i.e., a halo of mass $M_{200m}=2.0\times10^{13}\,h^{-1}\,M_{\odot}$ is resolved by only $100$ particles. Nevertheless, we stress here that the mock catalogs generated from \citet{Takahashi17} are used to explain the observational results, whereas the mock catalogs generated from our own simulations are used for calibration. Hence, the lower resolution light cone-based mock catalog is sufficient as long as the matter distribution follows the theoretical expectation (see Figure 19, 20, 21 of \citealt{Takahashi17}).

\begin{figure}[htb]
    \begin{minipage}{0.475\textwidth}
    \centering
        \includegraphics[width=0.99\textwidth]{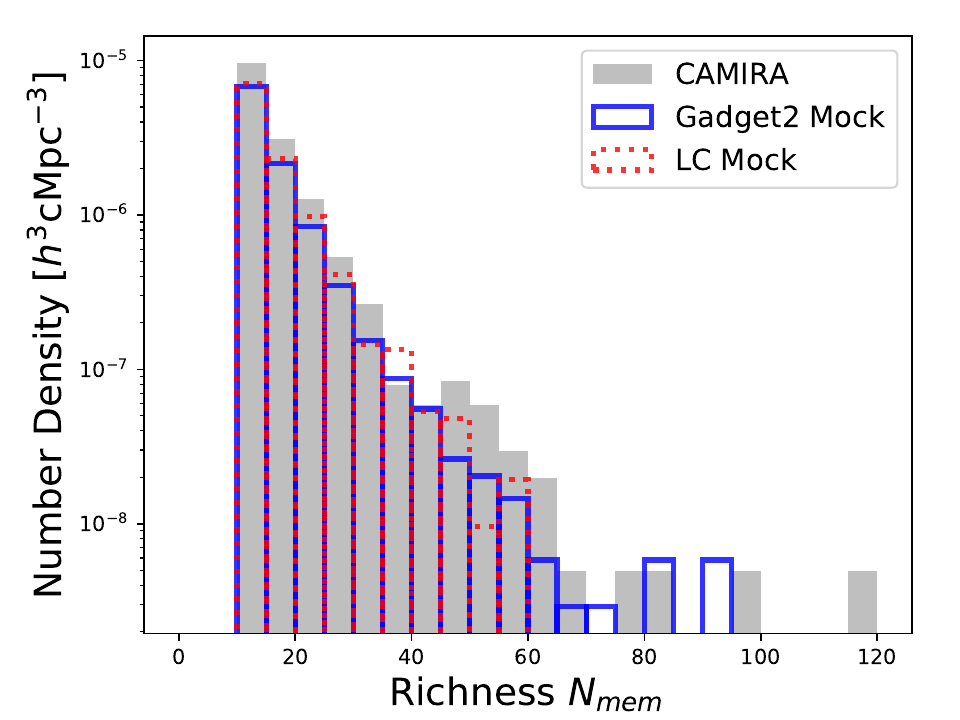}
    \end{minipage}
    \caption{The richness distributions of the CAMIRA clusters (gray), a mock catalog  constructed from our simulations (blue), and a mock catalog extracted from the  \cite{Takahashi17} light cone (red) at redshift between 0.65 and 0.75. The consistency of richness distribution between observational data and mock data validates our approach.}
    \label{fig:N_mem}
\end{figure}

Figure~\ref{fig:n_cl} also compares the comoving halo number density of the mock catalogs generated from our GADGET-2 simulation and the mock catalog obtained from the \citet{Takahashi17} light cone with the CAMIRA cluster comoving number density as a function of redshift. Figure~\ref{fig:N_mem} shows the richness distribution of the CAMIRA clusters, one of the mock catalogs generated from GADGET-2, and one of the mock catalogs generated from the \citet{Takahashi17} light cone at $z=0.65-0.75$. The general agreement of the comoving number density and richness distribution validates our approach to construct mock catalog. We note that the CAMIRA cluster number density appears to be offset from the mock halo number density by a maximum $\approx 50\%$ between redshift $z=0.7-0.8$ (Figure~\ref{fig:n_cl}) for a reason yet to be understood. However, this poses negligible concern to our goal here, since we would see in Section \ref{subsec:optimization} next that our optimization strategy for linking length ensures the purity of the supercluster candidates selected. 

\subsection{Optimization of the linking length\label{subsec:optimization}}

As mentioned in Section \ref{sec:intro}, an appropriate choice of the linking length $l$ is critical for clustering analysis, since it directly determines the properties of the outputs of the FoF algorithm. \citet{Chon13} constructed a supercluster catalog from the ROSAT-ESO Flux Limited X-ray (REFLEX $\rm II$; \citealt{Chon12}) cluster sample adopting an overdensity parameter $f=10$, which is equivalent to linking parameter $b\approx0.46$. Their linking length, which is derived from the spherical collapse model, increases monotonically from $l\approx 11\,$cMpc at $z=0$ to $l\approx 300\,$cMpc at $z=0.4$. \citet{Sankhyayan23} recently used a modified FoF algorithm to build a supercluster catalog from the cluster sample by \citet{WH15}. They selected a linking length of $l\approx20\,$cMpc such that the total number of superclusters found was maximized.

In this work, we empirically determine the linking length from a systematic comparison of the FoF groups obtained by applying different linking lengths to the ground-truth superclusters in our $N$-body simulations. The ground-truth superclusters are the groups of halos in mock catalogs that will eventually collapse in the future (by $a=15.0$). In practice, for a mock catalog at a given redshift $z_i$, we trace the descendants following the merger trees of the halos. If a group of halos merges into one halo before or at $a=15.0$, that is, sharing the same \texttt{parent ID} or \texttt{ID}, then it is identified as a bona fide supercluster. We collectively called the set containing all ground-truth superclusters at a given redshift $S_{TSC}$. On the other hand, we study the impact of the linking length by looking at the sets of FoF groups obtained with each choice of linking lengths. For a given linking length $l$, we collectively call the set containing all FoF groups $S_{FoF}(l)$. We require that each FoF group in $S_{FoF}(l)$ and each ground-truth supercluster in $S_{TSC}$ have at least two member halos. The set $S_{TSC}$ is linking length independent, while set $S_{FoF}(l)$ depends on the chosen linking length.

For a given linking length $l$, we evaluate the performance of the FoF algorithm by two metrics\footnote{Although the performance metrics here are weighted by halo mass, we also have tried optimizing the linking length based on unweighted metrics and found that our results are  unbiased. The details are presented in Appendix \ref{sec:appendix_uw_metric}.}, namely purity $P$ and completeness $C$:
\begin{equation}
    P(l\,;z_i\,,N_{cut}) = \frac{\sum M_{200m}(S_{FoF}(l)\land S_{TSC})}{\sum M_{200m}(S_{FoF}(l))}
    \label{purity}
\end{equation}
\begin{equation}
    C(l\,;z_i\,,N_{cut}) = \frac{\sum M_{200m}(S_{FoF}(l)\land S_{TSC})}{\sum M_{200m}(S_{TSC})}
    \label{completeness}
\end{equation}
where $z_i$ is the redshift of the snapshot,  $N_{cut}$ (e.g., 10.0) is the richness cut. The denominators of purity and completeness represent the total FoF group halo mass and the total mass of the ground-truth supercluster member halos, respectively. For the numerator of purity and completeness, we describe the operator $\land$ in detail using the examples below. Consider a ground-truth supercluster composed of four member halos, say TSC=[CL1, CL2, CL3, CL4], the following 4 cases are possible:
\begin{enumerate}
    \item A FoF group with exactly the same composition as the TSC is captured by the FoF algorithm (i.e., FoF1=[CL1, CL2, CL3, CL4]). In this case, we include the total halo mass of FoF1 in the numerator.
    \item A FoF group that contains part of the TSC is found by the FoF algorithm, say FoF1=[CL1, CL2, CL3]. In this case, we add the total halo mass of FoF1 to the numerator.
    \item The TSC is identified as two FoF groups, say FoF1=[CL1, CL2] and FoF2=[CL3, CL4]. In this case, we only consider the more massive FoF group of the two in the numerator, so that the fragmentation situation degrades both purity and completeness.
    \item A FoF group could contain additional halo members other than halo members of TSC, say FoF1=[CL1, CL2, CL3, CL5]. In this case, FoF1 will not be added to the numerator so that the completeness does not increase monotonically.
\end{enumerate}

Following the rules described above, each FoF group in the ``intersection'' set, $S_{FoF}(l)\land S_{TSC}$, is contained in a ground-truth supercluster from $S_{TSC}$. This way, the purity $P$ quantifies the fraction of the total halo mass of $S_{FoF}(l)$ in ground-truth superclusters for a given $l$. The completeness $C$ quantifies the fraction of the total halo mass of $S_{TSC}$ recovered for a given $l$.

\begin{figure}[htb] 
    \centering
    \begin{minipage}{0.475 \textwidth}
        \begin{center}
            \includegraphics[width=0.99\hsize]{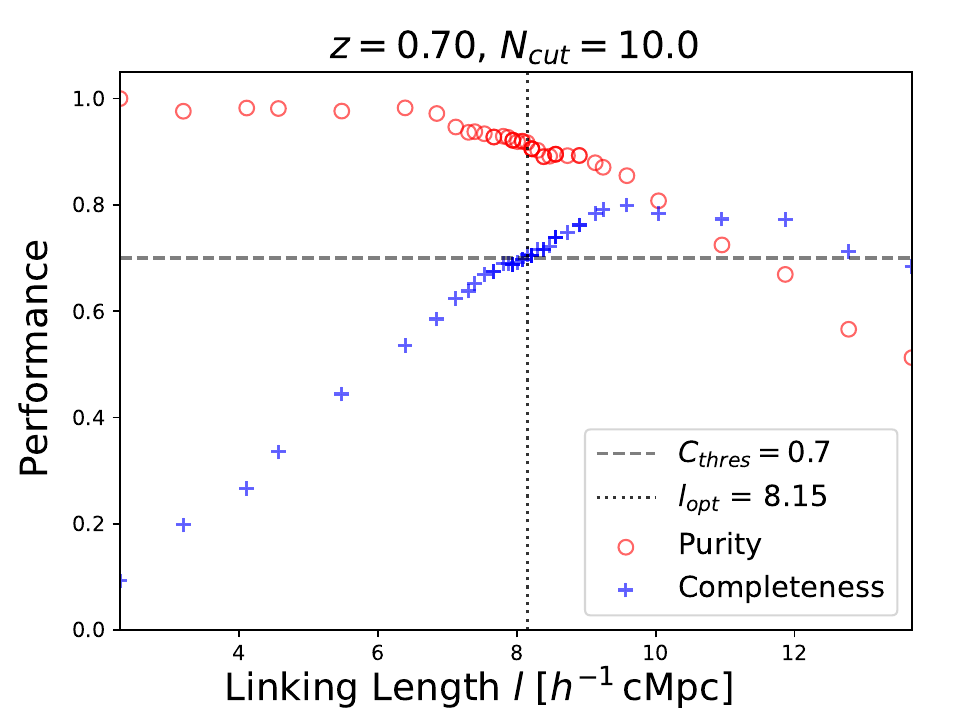}
        \end{center}
    \end{minipage}
    \begin{minipage}{0.475 \textwidth}
        \begin{center}
            \includegraphics[width=0.99\hsize]{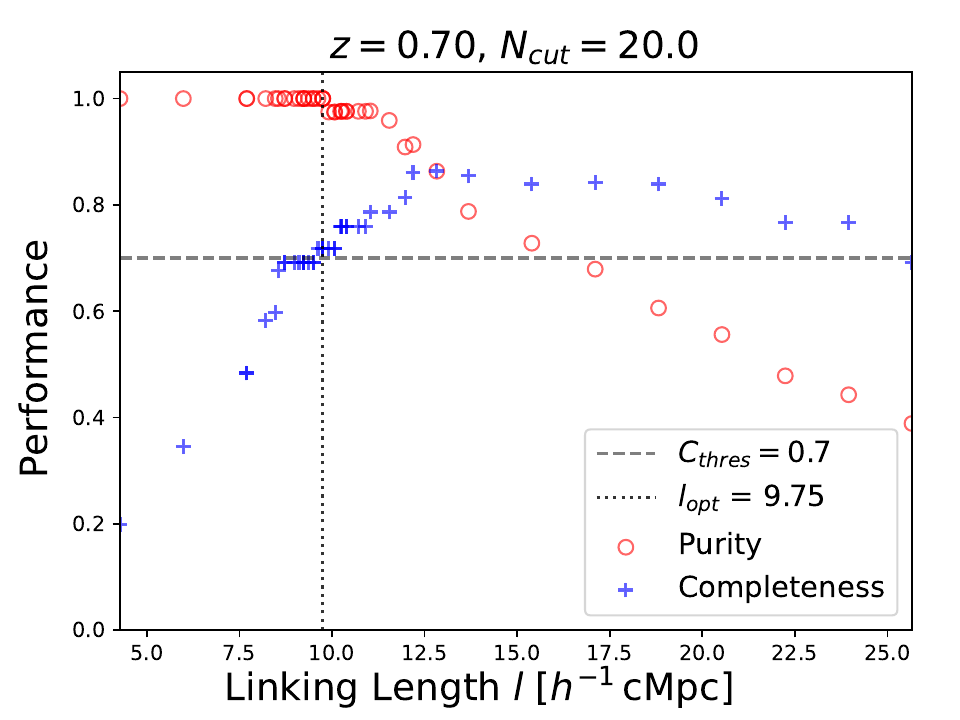}
        \end{center}
    \end{minipage}
    \caption{The optimization of linking length at redshift $z_i=0.7$ with richness cut $N_{cut}=10.0$ (upper panel) and $N_{cut}=20.0$ (lower panel). The y-axis indicates purity (red circle) and completeness (blue cross) defined by Eqn.~(\ref{purity}) and Eqn.~(\ref{completeness}) respectively. The horizontal dashed line indicates the completeness threshold $C_{thres}=0.7$, and the vertical dotted line shows the optimization result $l_{opt}$.}
    \label{fig:optimization}
\end{figure}

For a mock catalog at a given redshift $z_i$, the optimized linking length is found after $5$ iterations. In the first iteration $k=1$, we adopt a wide range of initial guesses of the linking parameters $b_{k=1,i}$ and calculate the performances for each linking parameter. We select the parameter that results in $C\ge C_{thres}=0.7$ while maximizing the purity as the optimized linking parameter $b_{1, opt}$. This reflects our philosophy of constructing a supercluster catalog that is suitable for follow-up observations and conducting studies of environmental effect on both clusters and their member galaxies; therefore, we prioritize purity over completeness here. For the following $k=2,3,4,5\,$-th iterations, we test the linking parameters in the neighbor of the optimized linking parameters from the previous iteration $b_{k-1, opt}$ and also refine the separations between the new guesses of the linking parameters. The parameters mentioned above are decided empirically. In practice, during the iterations, we parameterize the linking length by linking parameters, and record the optimized result in terms of linking length $l_{opt}$. 

Furthermore, following a very similar procedure, we also construct mock catalogs and obtain the optimized linking lengths for the richness cuts $N_{cut}=15.0, \,20.0,\,25.0$. This is motivated by the possibility that a massive halo could attract and merge with another massive halo with greater separation due to stronger gravitational force. Hence, we expect a  longer linking length for halos of greater richness. Effectively, we assume the optimized linking length is dependent on the richness cut. 

Figure \ref{fig:optimization} demonstrates the optimization process for richness cut $N_{cut}=10.0$ and $N_{cut}=20.0$ at redshift $z_i=0.70$. The optimized linking lengths $l_{opt}$ for these two cases, indicated by the black dotted vertical lines, are $8.15\,h^{-1}\,$cMpc and $9.75\,h^{-1}\,$cMpc, respectively. As expected, we observe a trade-off between completeness and purity.  When the linking length increases, the purity decreases because of the percolation of large structures that could not survive in an accelerating expanding universe, but the completeness increases because more structures are captured. The completeness curve do not increase monotonically as linking length increases because a FoF group extending beyond its ground-truth supercluster counterpart is not added into the numerator of equations (\ref{purity}) and (\ref{completeness}). 

\begin{figure}[htb]
    \centering
    \begin{center}
        \begin{minipage}{0.475\textwidth}
            \includegraphics[width=0.99\hsize]{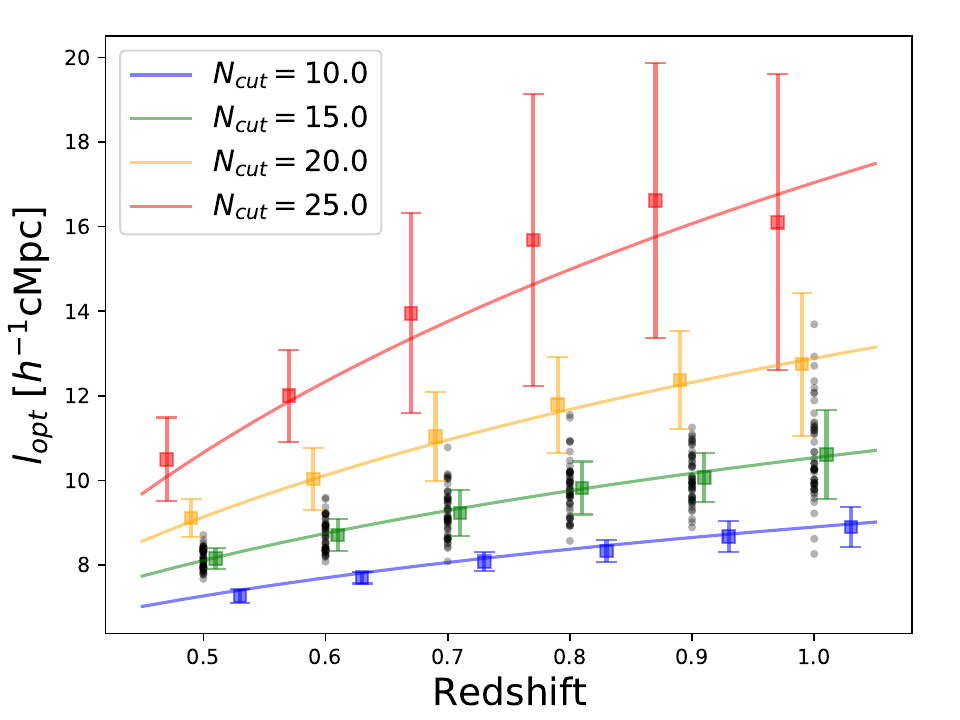}
        \end{minipage}
    \end{center}
    \caption{Optimized linking length $l_{opt}$ as a function of redshift for different richness cuts $N_{cut}$. The colors corresponding to different $N_{cut}$ are indicated by legend at the upper left corner. The square points and error bars are mean and $1\sigma$ standard deviation computed over all 60 realizations. We  offset the square points and error bars along x-axis slightly for clarity. The curves indicate the best fit to the data points (equation \ref{eq:lzfit}). The background gray points show the optimized results with $N_{cut}=15$ distributed in $l_{opt}-z$ space. For the sake of simplicity, we do not show optimized results for other richness cut.}
     \label{fig:LD}
\end{figure}

The optimization process is repeated for all snapshots between redshift $z=0.5-1.0$, all richness cuts $N_{cut}=10.0,\,15.0,\,20.0,\,25.0$, and all different realizations. In the end, for each richness cut, we obtain optimization results distributed in the $l_{opt}-z$ space. In order to describe the optimized linking length as a smooth function of redshift, we fit the distribution by a power-law:
\begin{equation}
    l_{opt}(z;\,N_{cut})= az^b + c
\label{eq:lzfit}    
\end{equation} 
where $a$, $b$ and $c$ are free parameters, $z$ is redshift and $N_{cut}$ is the richness cut. Figure \ref{fig:LD} shows the fitting result using data from all 60 realizations. 

\subsection{Validation of supercluster finding method\label{subsec:validation}}

We assess the theoretical performance of our calibrated supercluster finding method by cross-validation. We first derive the optimized linking length as a function of redshift for a given richness cut $l_{opt}(z; N_{cut})$ using realizations from  five (out of six) arbitrarily chosen simulation boxes. The FoF algorithm using the calibrated linking lengths is then applied to the sixth simulation. We construct the final supercluster catalog using two different approaches: (1) We only apply the FoF algorithm with linking length $l_{opt}(z;  N_{cut}=10.0)$ to the mock catalog of the richness cut $N_{cut}=10.0$. This supercluster catalog is referred to as ``without merging''. (2) We apply the FoF algorithm with linking length $l_{opt}(z;  N_{cut})$ to the mock catalog with richness cuts $N_{cut}$, where $N_{cut}=10.0,\,15.0\,,20.0\,,25.0$. For each snapshot, these four supercluster catalogs based on different richness cuts are merged by concatenating any FoF groups that share common members. This supercluster catalog is referred to as ``with merging''. The underlying motivation for the second approach is inspired by the intuition that more massive halos have deeper gravitational potential and attract more distant halos (in other words, more massive halos have larger correlation lengths). The performance of these two different approaches at each snapshot is evaluated by comparing to the ground-truth superclusters and calculating the purity and completeness using equations (\ref{purity}) and (\ref{completeness}).

\begin{figure}[htb]
  \centering
  \begin{center}
      \begin{minipage}{0.475 \textwidth}
            \includegraphics[width=0.99\hsize]{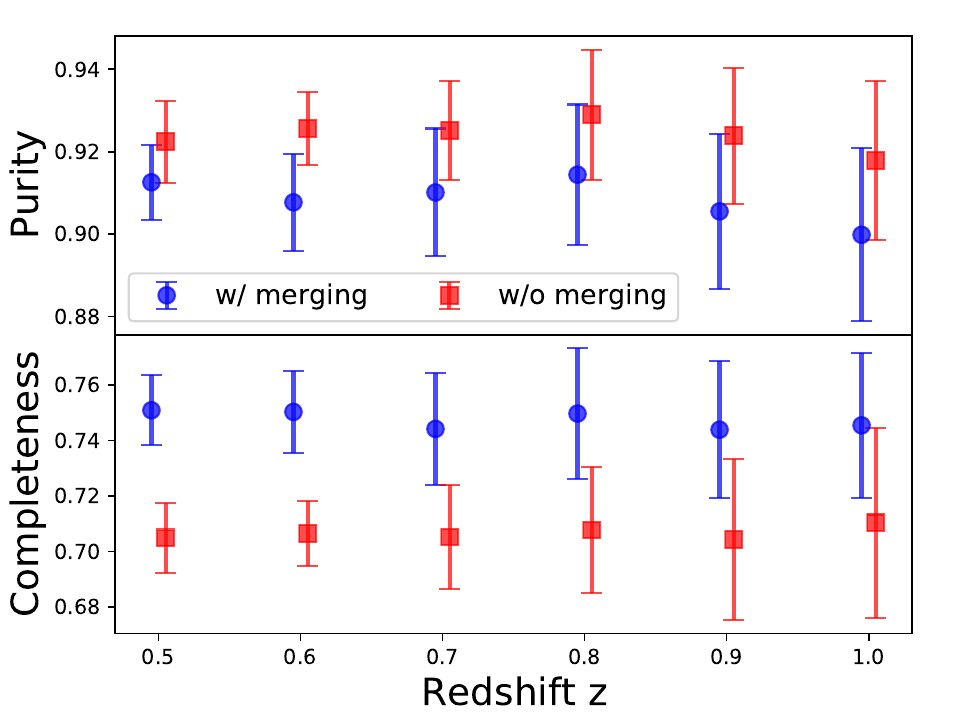}
      \end{minipage}
  \end{center}
  \caption{Performance as a function of redshift derived from the cross-validation procedure described in Section \ref{subsec:validation}. Points and error bars indicates mean and $1\sigma$ standard deviation calculated over all 60 realizations. The blue and  red points indicate supercluster catalog constructed ``without merging'' and ``with merging'' respectively. Using the ``with merging'' approach, the completeness throughout $z=0.5-1.0$ is enhanced by $5\%$ compared to the ``without merging'' approach.}
  \centering
  \label{fig:performanance}
\end{figure}

Figure \ref{fig:performanance} demonstrates the performance as a function of redshift calculated over 10 Monte Carlo realizations in each simulation box following the two different approaches described above. We see that, although merging supercluster catalogs of different richness cut sacrifices $\approx 1\%$ purity throughout all snapshots, its  completeness is increased by $\approx 5\%$ throughout all snapshots. This shows that considering the $N_{cut}$ dependence of linking length enhances the ability to recover the ground-truth superclusters.  

\section{CAMIRA supercluster catalog} \label{sec:cat} 

\begin{figure*}
    \centering
    \begin{center}
        \begin{minipage}{0.95\textwidth}
            \includegraphics[width = 0.95\hsize]{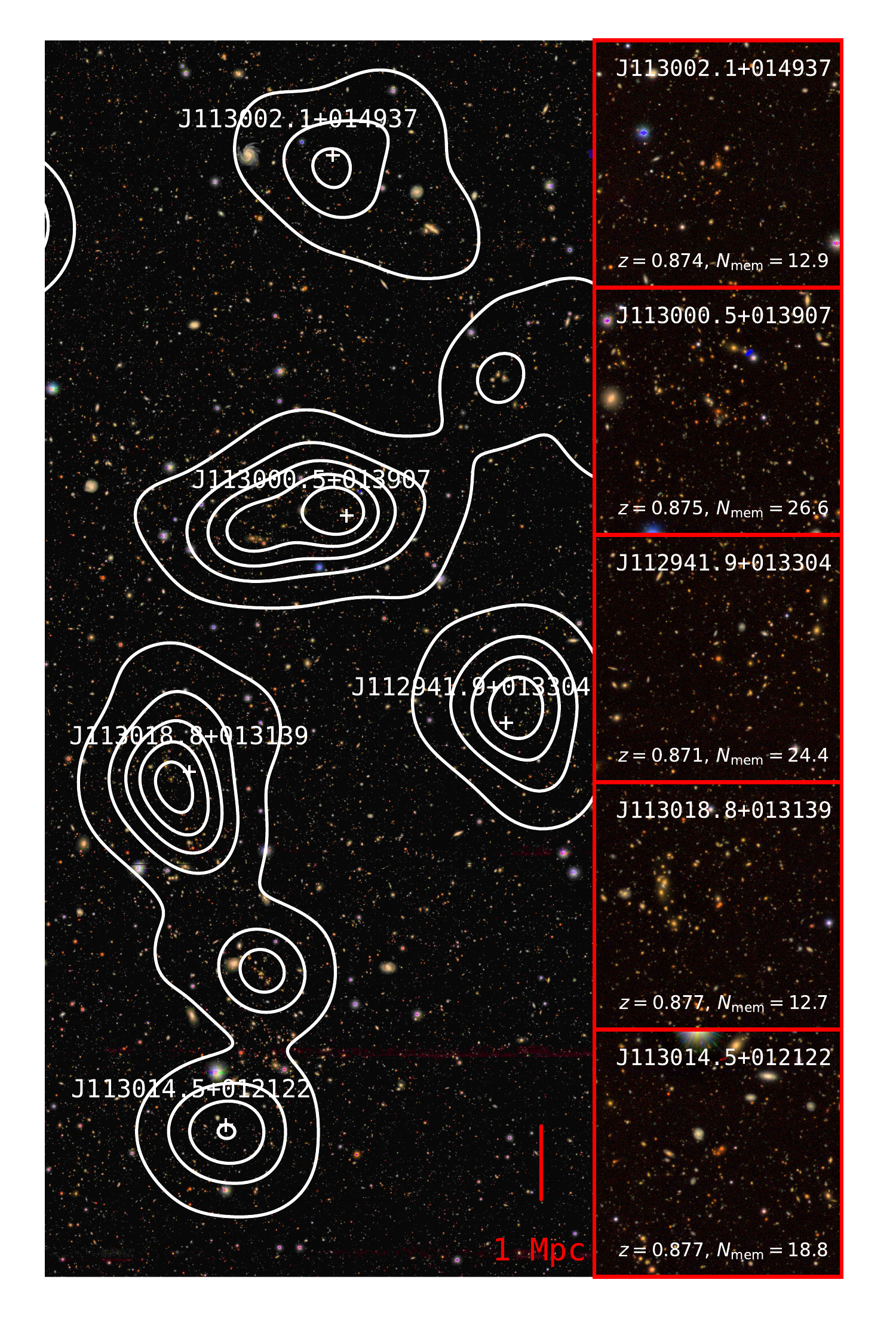}
        \end{minipage}
    \end{center}
    \vspace{-1.25cm}
    \caption{Left panel: HSC $riz$ color composite image of one of the  supercluster candidates with the highest multiplicity (5), at redshift $z\approx 0.87$, in our  catalog. The field of view is $16' \times 36'$. White contours represent galaxy number density, stepping three galaxies/pixel from one galaxy/pixel.  Right panel: zoom-in HSC $riz$ composite image of each member cluster. The field of view is $3' \times 3'$.}
    \label{fig:SC_image}
\end{figure*}

The right panel of Figure \ref{fig:flow} explains the workflow of our supercluster finding method applied to the observational data. Since the redshift distribution of the cluster samples is continuous, we divide them into overlapping redshift bins of width $\Delta z =0.05$, with an overlap of 0.025, which is to prevent potential superclusters from being separated by the edges of the redshift bins. The FoF algorithm is applied to clusters in each redshift bin with the linking length value at the central redshift of the bin. With all supercluster candidates obtained from different redshift bins and richness cuts, we concatenate them as described in Section \ref{subsec:validation}, forming our final supercluster catalog.

\begin{table*}[htb]
    \centering
    \begin{tabular}{ccccrcrrcr}  
    \hline \hline
    Supercluster ID  & CAMIRA ID & Multiplicity & R.A. & Dec. & $z_{CAM}$ & $z_{BCG}$ & $z_{cl, DESI}$ & $N_{mem}$ & Conf-z\\ 
    (1) & (2) & (3) & (4) & (5) & (6) & (7) & (8) & (9) & (10)\\
    \hline \hline
        1 &  J011124.2+014921 & 2 & 17.850688 & 1.822534 & 0.8077 & 0.81178 & $-$1.00000 & 35.5 & 1 \\
        1 &  J011132.0+015238 & 2 & 17.883536 & 1.877400 & 0.8076 & $-$1.00000 & $-$1.00000 & 32.4 & 1 \\
        2 &  J013636.9+004213 & 3 & 24.153902 & 0.703794 & 0.7910 & 0.79443 & $-$1.00000 & 38.4 & 1 \\
        2 &  J013708.9+002551 & 3 & 24.287286 & 0.431092 & 0.7918 & 0.79053 & $-$1.00000 & 20.5 & 1 \\
        2 &  J013728.5+003623 & 3 & 24.368642 & 0.606420 & 0.7920 & 0.79216 & $-$1.00000 & 36.0 & 1 \\
        3 &  J020832.7-043315 & 2 & 32.136452 & $-$4.554213 & 0.7618 & $-$1.00000 & $-$1.00000 & 38.7 & 1 \\
        3 &  J020846.1-042624 & 2 & 32.192284 & $-$4.440218 & 0.7580 & $-$1.00000 & $-$1.00000 & 42.0 & 1 \\
        4 &  J023527.9-010129 & 2 & 38.866387 & $-$1.024991 & 0.7580 & $-$1.00000 & $-$1.00000 & 27.1 & 1 \\
        4 &  J023554.8-010029 & 2 & 38.978506 & $-$1.008286 & 0.7532 & $-$1.00000 & $-$1.00000 & 27.5 & 1 \\
        5 &  J084911.7+024148 & 2 & 132.298617 & 2.696919 & 0.8292 & $-$1.00000 & $-$1.00000 & 33.6 & 1 \\
        5 &  J084941.2+025543 & 2 & 132.421754 & 2.928781 & 0.8304 & $-$1.00000 & $-$1.00000 & 26.6 & 1 \\
        6 &  J090755.5+025422 & 2 & 136.981335 & 2.906202 & 0.7989 & 0.79685 & $-$1.00000 & 36.7 & 1 \\
        6 &  J090823.8+025132 & 2 & 137.099013 & 2.859133 & 0.8008 & $-$1.00000 & $-$1.00000 & 30.9 & 1 \\
        7 &  J095611.7+015934 & 2 & 149.048780 & 1.992959 & 0.9954 & $-$1.00000 & 0.96491 & 49.1 & 1 \\
        7 &  J095624.3+015558 & 2 & 149.101237 & 1.932960 & 0.9700 & $-$1.00000 & 0.96481 & 26.5 & 1 \\
        8 &  J104319.9+041319 & 3 & 160.833029 & 4.222015 & 0.5700 & 0.57200 & $-$1.00000 & 27.8 & 1 \\
        8 &  J104355.6+041313 & 3 & 160.981508 & 4.220447 & 0.5720 & $-$1.00000 & $-$1.00000 & 33.6 & 1 \\
        8 &  J104401.7+040340 & 3 & 161.007286 & 4.061376 & 0.5700 & 0.57345 & $-$1.00000 & 24.3 & 1 \\
        9 &  J104915.4+043307 & 2 & 162.314297 & 4.552191 & 0.7810 & $-$1.00000 & $-$1.00000 & 25.1 & 1 \\
        9 &  J104926.6+043822 & 2 & 162.360939 & 4.639518 & 0.7802 & $-$1.00000 & $-$1.00000 & 39.0 & 1 \\
        10 &  J114545.0+015648 & 2 & 176.437664 & 1.946910 & 0.6060 & 0.63232 & $-$1.00000 & 28.1 & 1 \\
        10 &  J114559.6+020239 & 2 & 176.498319 & 2.044430 & 0.6040 & 0.63247 & $-$1.00000 & 26.2 & 1 \\
        11 &  J115658.8+012525 & 3 & 179.245111 & 1.423803 & 0.5820 & 0.57220 & 0.56916 & 44.6 & 1 \\
        11 &  J115724.7+013829 & 3 & 179.352939 & 1.641495 & 0.5736 & $-$1.00000 & 0.56814 & 12.7 & 1 \\
        11 &  J115745.9+012736 & 3 & 179.441162 & 1.460048 & 0.5539 & 0.56218 & 0.56827 & 27.7 & 1 \\
        12 &  J124250.2+035922 & 2 & 190.709164 & 3.989536 & 0.6458 & $-$1.00000 & $-$1.00000 & 28.1 & 1 \\
        12 &  J124312.6+035945 & 2 & 190.802338 & 3.996053 & 0.6503 & $-$1.00000 & $-$1.00000 & 25.9 & 1 \\
        13 &  J134341.1+032930 & 3 & 205.921269 & 3.491779 & 0.7510 & $-$1.00000 & $-$1.00000 & 29.4 & 1 \\
        13 &  J134343.1+033630 & 3 & 205.929784 & 3.608402 & 0.7560 & $-$1.00000 & $-$1.00000 & 26.5 & 1 \\
        13 &  J134405.3+032517 & 3 & 206.021919 & 3.421467 & 0.7559 & $-$1.00000 & $-$1.00000 & 16.7 & 1 \\
        14 &  J141026.0-011432 & 2 & 212.608325 & $-$1.242480 & 0.6122 & 0.63922 & 0.64088 & 31.1 & 1 \\
        14 &  J141050.4-010930 & 2 & 212.710020 & $-$1.158503 & 0.6300 & 0.63465 & 0.63915 & 34.6 & 1 \\
        15 &  J142044.0+030004 & 4 & 215.183491 & 3.001134 & 0.8723 & $-$1.00000 & $-$1.00000 & 28.2 & 1 \\
        15 &  J142105.4+025611 & 4 & 215.272457 & 2.936596 & 0.8693 & $-$1.00000 & $-$1.00000 & 13.8 & 1 \\
        15 &  J142114.8+030736 & 4 & 215.311775 & 3.126804 & 0.8789 & $-$1.00000 & $-$1.00000 & 26.6 & 1 \\
        15 &  J142141.9+030351 & 4 & 215.424449 & 3.064213 & 0.8820 & $-$1.00000 & $-$1.00000 & 12.2 & 1 \\
        16 &  J145943.8+014807 & 2 & 224.932406 & 1.802151 & 0.5133 & 0.52117 & $-$1.00000 & 25.4 & 1 \\
        16 &  J150032.1+015321 & 2 & 225.133663 & 1.889420 & 0.5250 & 0.52039 & $-$1.00000 & 26.8 & 1 \\
        17 &  J152916.9+434057 & 4 & 232.320532 & 43.682659 & 0.6420 & $-$1.00000 & $-$1.00000 & 41.7 & 1 \\
        17 &  J152922.6+435005 & 4 & 232.344015 & 43.834781 & 0.6400 & $-$1.00000 & $-$1.00000 & 12.4 & $-$1 \\
        17 &  J152942.6+433459 & 4 & 232.427692 & 43.583191 & 0.6280 & 0.64033 & $-$1.00000 & 28.4 & 1 \\
        17 &  J152957.7+432446 & 4 & 232.490454 & 43.413040 & 0.6348 & 0.63663 & $-$1.00000 & 45.1 & 1 \\
        18 &  J221359.5+015127 & 2 & 333.497897 & 1.857536 & 0.6860 & 0.68319 & $-$1.00000 & 57.5 & 1 \\
        18 &  J221452.3+014439 & 2 & 333.717980 & 1.744227 & 0.6961 & 0.68270 & $-$1.00000 & 46.0 & 1 \\
    \hline \hline
    \end{tabular}
    \caption{The first 18 supercluster candidates in the CAMIRA supercluster catalog. The columns are: (1) Supercluster ID (2) Cluster ID (Cluster ID in the released catalog will have a prefix ``HSCCL'', indicating the object type) (3) Multiplicity of superclusters (4) Right ascension [degree, J2000] (5) Declination [degree, J2000] (6) Photometric redshift of the CAMIRA cluster (7) Spectroscopic redshift of the BCG if available (8) Cluster redshift derived using DESI EDR data, see Section \ref{subsec:DESI} for detail (9) Richness (10) The confident redshift flag. It is $1$ if the photometric redshift of the cluster meets one of the criteria described in Section \ref{sec:data}, otherwise it is $-1$. The complete catalog is available online.}
    \label{tab:catalog}
\end{table*}

The final catalog consists of 673 supercluster candidates between redshift 0.5 and 1.0. It is presented in Table \ref{tab:catalog}. Figure \ref{fig:SC_image} shows the HSC $riz$ composite image of one of the supercluster candidates with the highest multiplicity in the final catalog. We believe that this is the first supercluster catalog that extends beyond redshift $z=0.7$. However, since approximately half of the clusters do not have confident redshifts, we must carefully examine the impact of photo-$z$ uncertainty and conservatively call them ``supercluster candidates'' (Section \ref{subsec:SC_n}). We also examine the multiplicity function (Section \ref{subsec:multiplicity}) and total mass of our supercluster candidates (Section \ref{subsec:M200}). Finally, we compare the supercluster candidates that we detect with two known high-$z$ superclusters (Section \ref{subsec:other_SC}). 

\subsection{Effect of photometric redshift uncertainty\label{subsec:SC_n}}
\begin{figure}[htb]
    \centering
    \begin{minipage}{0.475 \textwidth}
        \begin{center}
            \includegraphics[width = 0.99\hsize]{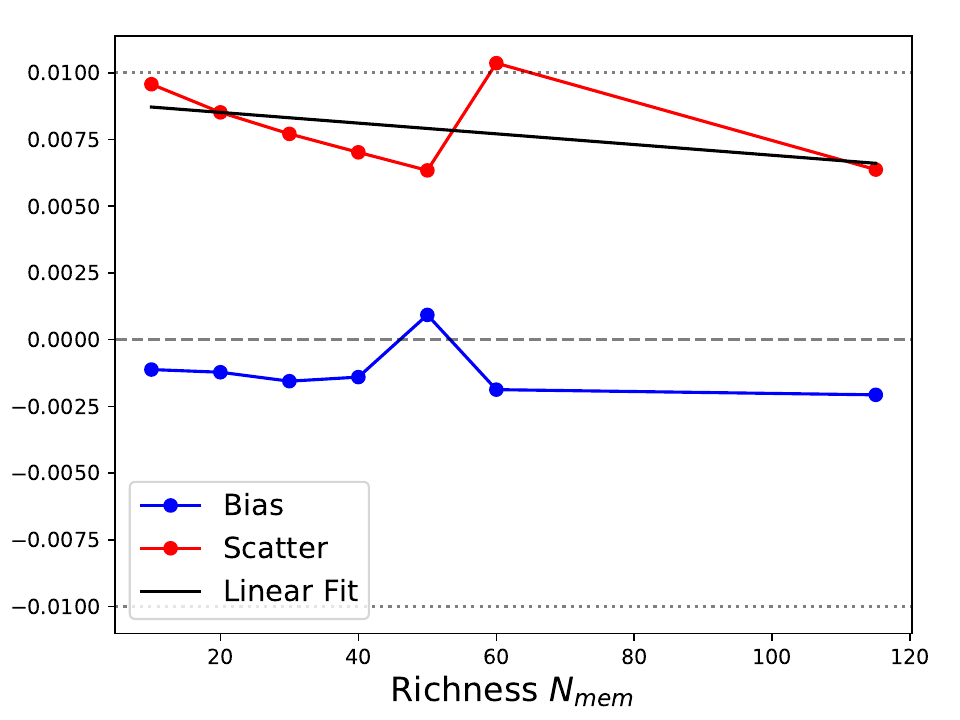}  
        \end{center}
    \end{minipage}
    \caption{The cluster photo-$z$ bias (blue curve) and scatter (red curve) as a function of richness. The black curve represents the best linear fit for the scatter as a function of richness.}
    \label{fig:photo-z_scatter}
\end{figure}

We first re-estimate the quality of the CAMIRA photo-$z$ as a function of richness $N_{mem}$, since we intuitively expect the scatter of photo-$z$ to decline when count of galaxy members increases. For this exercise, we select CAMIRA clusters with BCG spectroscopic-$z$ (spec-$z$) available in photometric redshift $z_{CAM}=0.1-1.0$ and assume BCG spec-$z$ as the ground truth redshift. Then, the bias $\delta_z$ and scatter $\sigma_z$ of the residual $\Delta_z$ are defined by:
\begin{equation} \label{eqn:photoz_bias}
    \delta_z = \mbox{median}(\Delta_z)
\end{equation}
\begin{equation} \label{eqn:photoz_scatter}
    \sigma_z = 1.48\times \mbox{MAD}(\Delta_z)
\end{equation}
where $\Delta_z = (z_{CAM}-z_{BCG})/(1+z_{BCG})$ and MAD stands for median absolute deviation. Furthermore, we apply $4\sigma$ clipping to $\Delta_z$ before computing $\delta_z$ and $\sigma_z$. Figure \ref{fig:photo-z_scatter} shows $\delta_z$ and $\sigma_z$ as a function of the richness $N_{mem}$. The black curve indicates a linear fit of $\sigma_z$. The trend of $\sigma_z$ follows our expectation, while $\delta_z$ is negligibly small.

\begin{figure*}[hbt]
  \centering
  \begin{center}
      \begin{minipage}{0.95 \textwidth}
          \includegraphics[width=0.975\hsize]{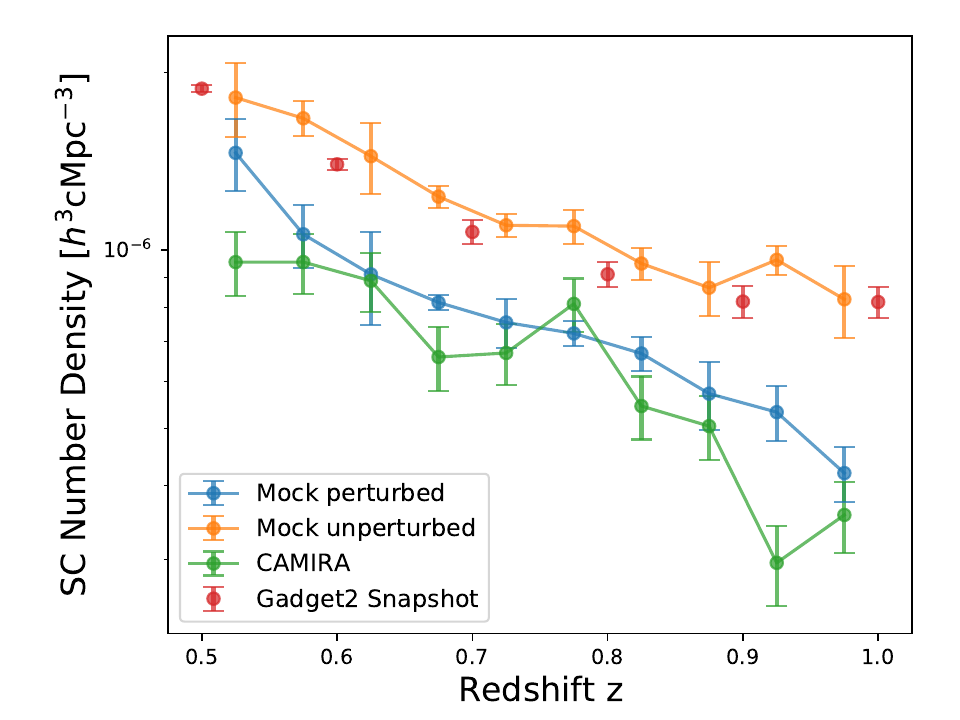}
      \end{minipage}
  \end{center}
  \vspace{-0.5cm}
  \caption{The number density of the supercluster candidates as a function of redshift. The green curve is calculated from the CAMIRA supercluster catalog. The red points are estimated from the cross-validation of Gadget snapshots. The blue and orange curves are calculated from the halo light cones with and without redshift perturbation, respectively. Except for the green curve whose error bars are assumed to be poisson noise, the error bars of other curves/points are 1$\sigma$ standard deviation calculated over six realizations. The consistency between the supercluster candidates number density of the CAMIRA superclusters and the perturbed mock superclusters suggests that the photo-$z$ uncertainty presents a limiting factor for the supercluster finding.}
  \centering
  \label{fig:SC_n}
\end{figure*}

We use the mocks generated from \citet{Takahashi17} to interpret the observational results. In addition to assigning richness to each halo as described in Section \ref{subsec:mock}, for a halo of richness $N_{mem}$ at redshift $z$, we perturb its redshift by adding a gaussian noise of scatter $\sigma_{z,lin}(N_{mem})(1+z)$ to the original halo redshift to mimic the cluster photo-$z$ uncertainty. Here, $\sigma_{z,lin}$ refers to the linear function that describes the scatter as a function of richness (the black line in Figure \ref{fig:photo-z_scatter}). We note that the perturbed fraction of halos as a function of redshift is determined according to the observation described in the lower panel of Figure \ref{fig:z_distribution}. After applying the supercluster finding method to the unperturbed light cones and perturbed light cones, two versions of mock supercluster catalogs are constructed. In the following analysis, we examine the supercluster catalogs constructed from the six different $N$-body simulations.

Figure \ref{fig:SC_n} shows the comoving number density of supercluster candidates as a function of redshift. At the first glance, we detect only half of the expected abundance of supercluster candidates from CAMIRA clusters compared to the abundance in the unperturbed light cones at all redshifts. However, the abundance of supercluster candidates in the perturbed light cones agree well with the observed values. This result is predictable, as the length scale corresponding to the photo-$z$ scatter of $\Delta z =0.01(1+z)$ is approximately $40\,h^{-1}\,$cMpc at redshift range $z=0.5-1.0$, which is $2-4$ times greater than our linking length (Figure \ref{fig:LD}). Hence, this result highlights the importance of having 
clusters confirmed by spectroscopic redshifts when searching for superclusters. 

\begin{figure}[htb]
  \centering
  \begin{center}
      \begin{minipage}{0.475\textwidth}
            \includegraphics[width=0.99\hsize]{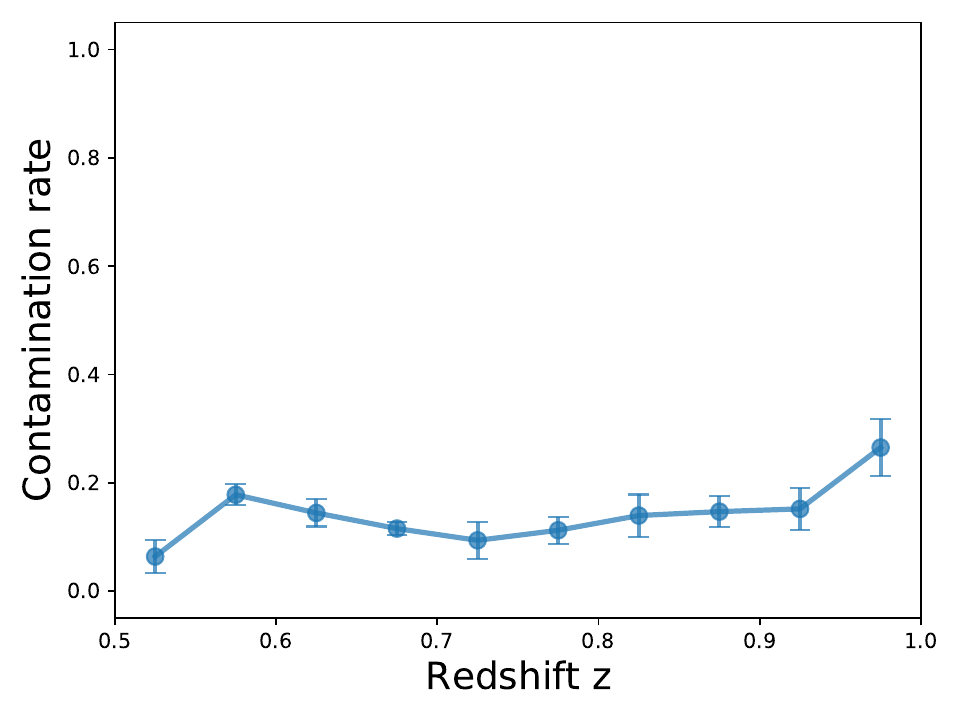}
      \end{minipage}
  \end{center}
  \vspace{-0.5cm}
  \caption{The contamination rate estimation as a function of redshift, which is derived by comparing superclusters extracted from the unperturbed light cones with those extracted from the perturbed light cones. The points and error bars represent the mean and standard deviation of contamination rate over six $N$-body simulations.}
  \centering
  \label{fig:contamination}
\end{figure}

Taking into account the photo-$z$ uncertainty, we next estimate the contamination rate of the supercluster catalog. To do so, we compare the supercluster catalogs obtained from the unperturbed light cones with those from the perturbed light cones. Assuming the supercluster catalogs extracted from the unperturbed light cones as standard, we define the contamination rate as the ratio between the number of perturbed superclusters not belonging to any unperturbed superclusters and the number of perturbed superclusters. Figure \ref{fig:contamination} shows the contamination rate as a function of redshift. The contamination rate increases from around $0.05$ at $z\approx0.5$ to $0.25$ at $z\approx1.0$ due to the increasing perturbed fraction in Figure \ref{fig:z_distribution}. 

We regard a supercluster candidate as {\bf confirmed} if all of its member clusters possess a confident redshift (see Figure \ref{fig:z_distribution}). 436 superclusters are confirmed among the 673 supercluster candidates, constituting $65\%$ of the catalog presented here. However, we note that this does not imply that the remaining supercluster candidates are spurious. This is because the three criteria described in Section \ref{sec:data} can not falsify the photo-$z$ of a cluster. A cluster photo-$z$ may not satisfy the criteria because of the incompleteness of the spectroscopic survey, the limited coverage of DESI EDR or the greater photo-$z$ scatter of DEmP.

\subsection{Multiplicity Function\label{subsec:multiplicity}}

\begin{figure*}[htb]
    \centering
    \begin{center}
        \begin{minipage}{0.95 \textwidth}
            \includegraphics[width=0.5\hsize]{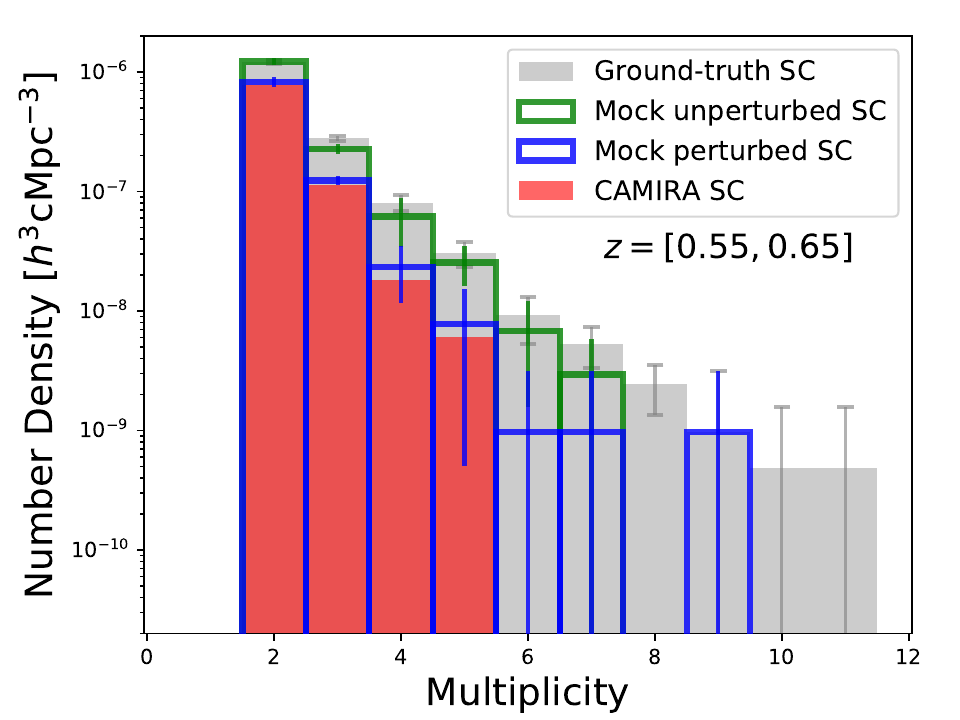}
            \includegraphics[width=0.5\hsize]{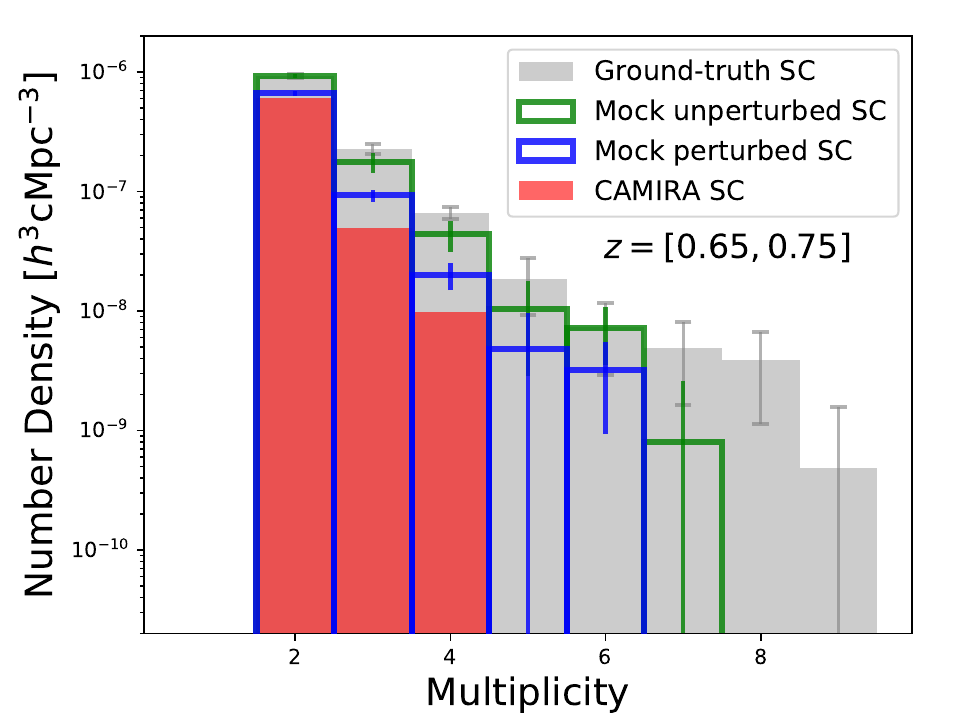}
            \includegraphics[width=0.5\hsize]{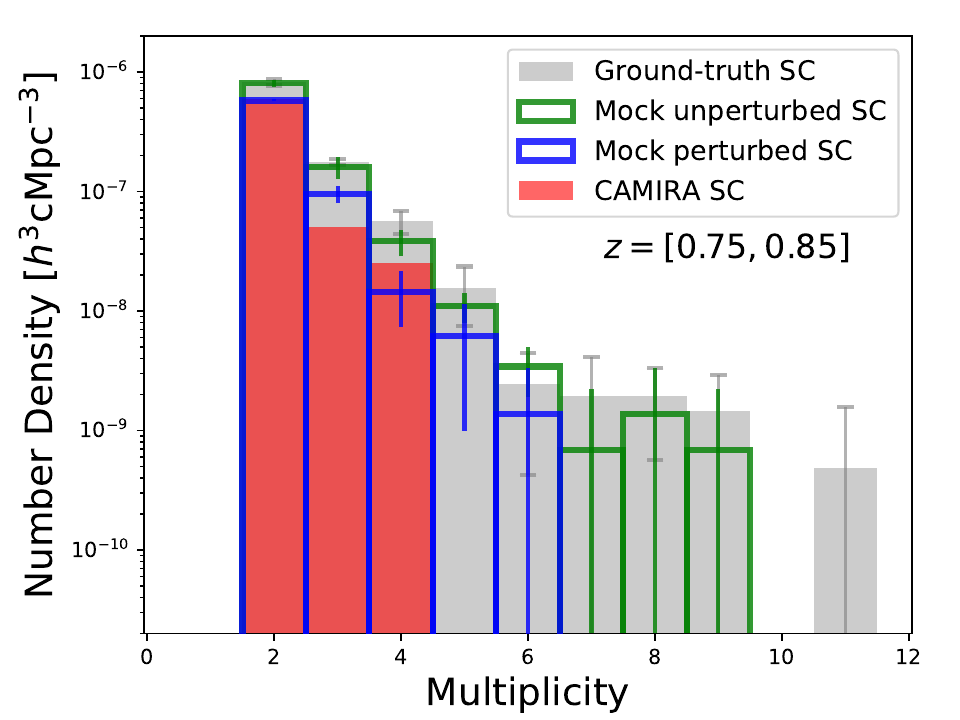}
            \includegraphics[width=0.5\hsize]{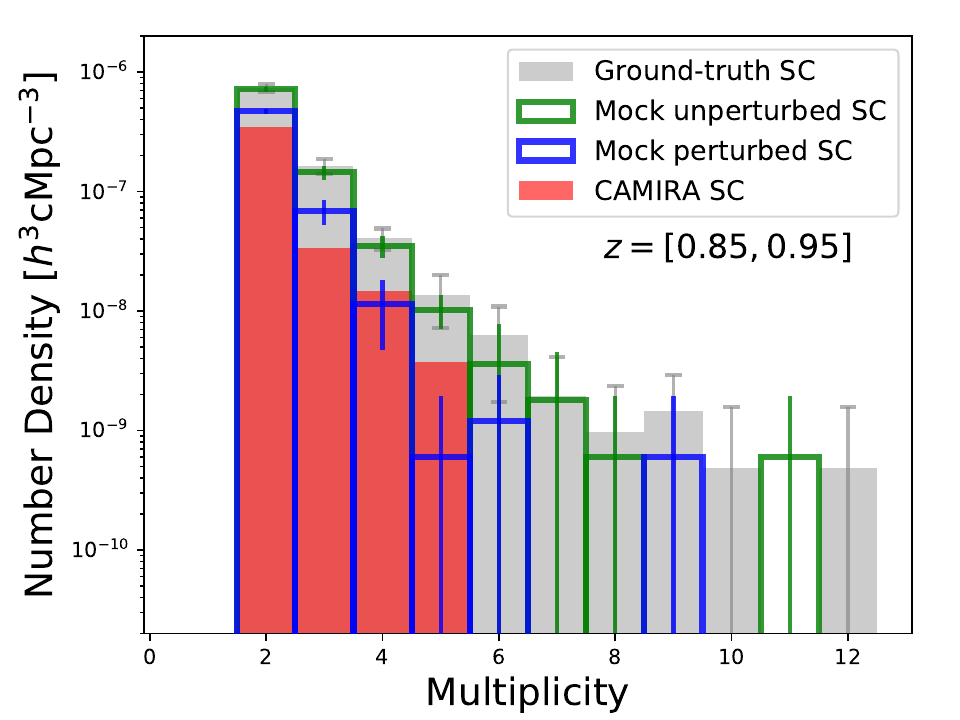}
        \end{minipage}
    \end{center}
    \caption{The multiplicity function at redshift $z=[0.55, 0.65]$ (top left panel), $z=[0.65, 0.75]$ (top right panel), $z=[0.75, 0.85]$ (bottom left panel), and $z=[0.85, 0.95]$ (bottom right panel). The gray histograms are multiplicity functions of ground-truth superclusters. The green histograms and blue histograms are multiplicity functions of superclusters extracted from unperturbed light cones and perturbed light cones, respectively. The error bars here indicate the $1\sigma$ standard deviation in six realizations. The red histogram shows the multiplicity function of CAMIRA supercluster candidates. The consistency between multiplicity function of CAMIRA supercluster candidates and mock perturbed superclusters again suggests the impact of photo-$z$ uncertainty.}
    \label{fig:multiplicity}
\end{figure*}

\begin{figure*}[hbt]
    \centering
    \begin{center}
        \begin{minipage}{0.95\textwidth}
            \includegraphics[width=0.5\hsize]{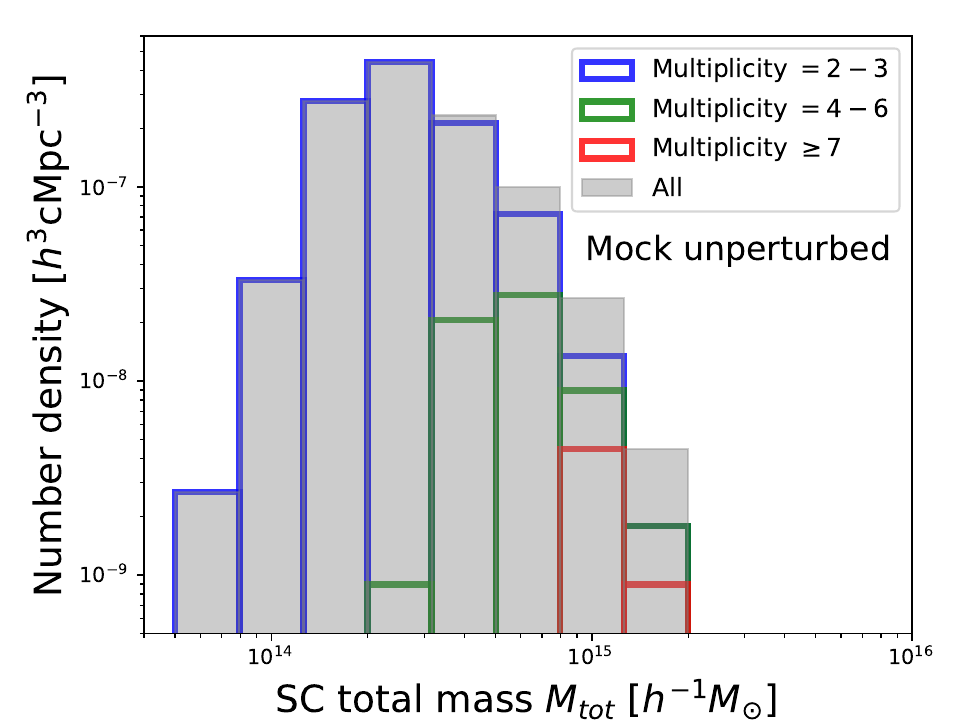}
            \includegraphics[width=0.5\hsize]{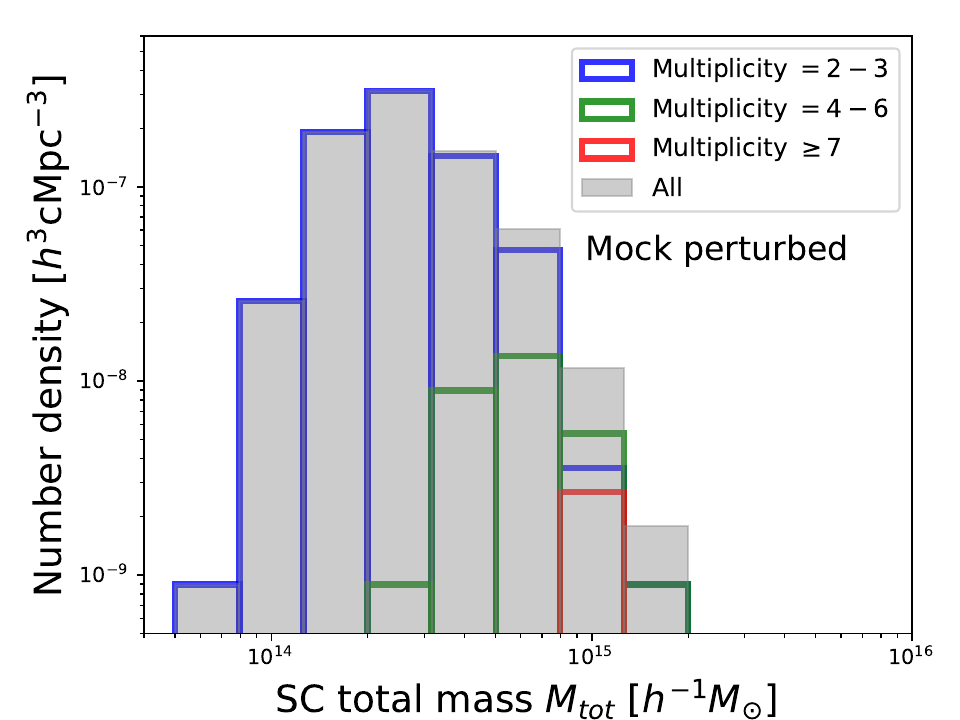}
        \end{minipage}
    \end{center}
    \caption{Supercluster total mass $M_{tot}$ distribution of unperturbed mock catalog (left) and perturbed mock catalog (right). Histograms of different colors correspond to different multiplicity bins indicated by the legend in the upper right corner.}
    \centering
    \label{fig:SC_mass}
\end{figure*}

The multiplicity is defined as the number of member clusters in a supercluster, and the multiplicity function describes the multiplicity distribution. We show the multiplicity function of ground-truth superclusters, supercluster candidates from unperturbed/perturbed light cones, and CAMIRA supercluster candidates in Figure \ref{fig:multiplicity}. The multiplicity function of ground-truth superclusters declines rapidly as the multiplicity increases. Superclusters having four members are ten times rarer than cluster pairs, and superclusters having six members are 100 times more scarce. This demonstrates that the supercluster population is dominated by cluster pairs and triplets, suggesting the limit of gravitationally bounded structures. The multiplicity functions extracted from unperturbed light cones well recover the ground-truth multiplicity functions by our method, at least below multiplicity of eight. We attribute the absence of high-multiplicity superclusters in unperturbed light cones to the following reasons: (1) Our optimization strategy prioritizes purity over completeness as described in Section \ref{subsec:optimization}, which prefers a shorter linking length due to the purity-completeness trade-off. (2) Although we weigh each supercluster by its mass in the performance metrics, the contribution of these high-multiplicity superclusters is still low, since they are extremely rare. (3) The existence of these high-multiplicity superclusters is extremely rare and stochastic, as revealed by the large error bars. (4) Some high-multiplicity superclusters are split or identified as low-multiplicity superclusters as we show in Appendix \ref{sec:appendix_fragmentation}. Capturing these scarce high-multiplicity superclusters will be as a major goal for improvement of our method in the future. 

In the current state, the presence of photo-$z$ uncertainty poses a greater obstacle. Although the observational multiplicity function is consistent with the multiplicity function of the perturbed light cones, we find that the ability of the supercluster finding method to recover superclusters decreases as the multiplicity increases compared to the ground-truth supercluster multiplicity function. This again indicates the importance of having spectroscopic redshift for supercluster searching at high redshift.

\subsection{Supercluster Total Mass Distribution\label{subsec:M200}}
\begin{table*}[htb]
    \centering
    \begin{tabular}{|c|cc|cc|}
    \hline
    &\multicolumn{2}{c|}{Unperturbed}&\multicolumn{2}{c|}{Perturbed} \\
    \hline
        \textbf{Mutiplicity Bins} & \textbf{Mean $M_{tot, med}$} & \textbf{Std $M_{tot, med}$} & \textbf{Mean $M_{tot, med}$} & \textbf{Std $M_{tot, med}$} \\ \hline \hline
        $[2, 3]$ & 2.44 & 0.02 & 2.46  & 0.03 \\ \hline
        $[4, 6]$ & 5.92 & 0.34 & 6.02  & 0.52 \\ \hline
        $\geq7$  & 10.82& 2.04 & 11.55 & 2.55 \\ \hline
        All      & 2.52 & 0.03 & 2.51  & 0.03 \\ \hline
    \end{tabular}
    \caption{The mean and standard deviation of median supercluster total mass over six realizations for different multiplicity bins. The masses are all given in units of $10^{14}\,h^{-1}\,M_{\odot}$}
    \label{tab:SC_mass}
\end{table*}

Considering the large scatter of the halo mass--richness scaling relation, we estimate the supercluster total mass in our supercluster catalog by characterizing the supercluster total mass in the unperturbed and perturbed light cones. The total mass of the supercluster $M_{tot}$ is calculated by summing the spherical overdensity mass $M_{200m}$ of the member clusters (thus ignoring contributions from smaller halos). Figure \ref{fig:SC_mass} shows the supercluster total mass distribution of a light cone with redshift perturbation and without redshift perturbation throughout redshift range $z=0.5-1.0$. Although the photo-$z$ uncertainty has a significant impact on the abundance of supercluster candidates detected, the overall supercluster total mass distributions between the two cases are very similar. As the superclusters of multiplicity two dominate the SC population, the total mass of the supercluster peaked around $2.5\times 10^{14}\,h^{-1}\,M_{\odot}$. The most massive superclusters are those with the highest multiplicity, reaching $M_{tot}\approx1\times10^{15}\,h^{-1}\,M_{\odot}$. In table \ref{tab:SC_mass}, we summarize the mean of the median supercluster total mass in six different $N$-body realizations for different multiplicity bins. However, we emphasize that the values here represent the lower limit of the supercluster total mass since we do not take the mass contribution of less massive halos into account. 

In principle, we could calculate the total dark matter particles mass encompassed in the volume of superclusters in the simulations. Nevertheless, defining the volume (or boundary) of a supercluster is not a trivial task, since superclusters remain nonvirialized. A rough incompleteness correction factor $C_{tot}$ could be estimated from the halo mass function:
\begin{equation}
    C_{tot}(z)=\frac{\int^{M_{max}}_{10^{10}} M\phi(M;z) \,dM}{\int^{M_{max}}_{M_{cl, min}} M\phi(M;z) \,dM}
\end{equation}
where we assume $M_{cl,min}\approx6.5\times10^{13}\,h^{-1}\,M_{\odot}$ corresponding to richness cut of $10.0$, $M_{max}$ is arbitrarily set to be $10^{17}\,h^{-1}\,M_\odot$ (although the exact value does not affect the results),
and $\phi(M;z)$ is the halo mass function \citep{Tinker08} at a given redshift $z$. The resulting correction factor $C_{tot}$ is $\approx$ 11 at $z=0.5$ and increases to $\approx$ 34 at $z=1.0$. This suggests that the total underlying mass taking into account the less massive halos could be $\gtrapprox 2.5\times10^{15}\,h^{-1}\,M_{\odot}$.

\subsection{Comparisons with known superclusters\label{subsec:other_SC}}

\subsubsection{King Ghidorah Supercluster\label{subsubsec:KG_SC}}
\begin{figure*}[htb]
    \centering
    \begin{center}
        \begin{minipage}{0.95\textwidth}
            \includegraphics[width = 0.99\hsize]{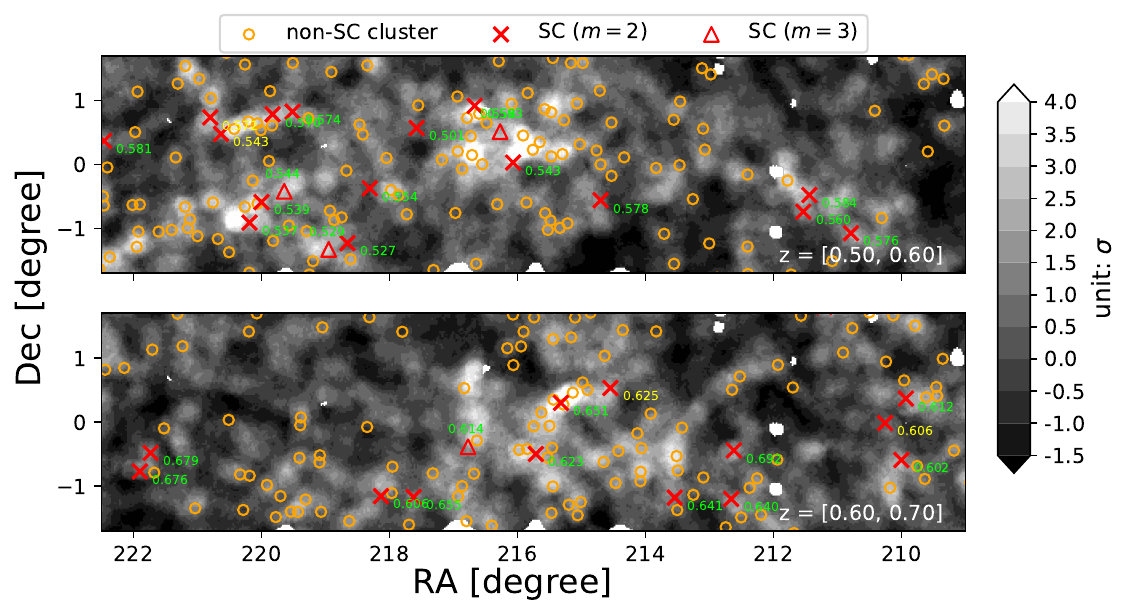}
        \end{minipage}
    \end{center}
    \centering
    \vspace{-0.5cm}
    \caption{CAMIRA supercluster candidates and isolated clusters in King Ghidorah Supercluster field overlaid by HSC PDR3 density map  in two different redshift slices. The orange open circles represent the positions of CAMIRA clusters that are not associated with any supercluster candidates. Red crosses and red triangles indicate the positions of supercluster candidates with multiplicity two ($m=2$) and three ($m=3$), respectively (here the positions are the mean positions of member clusters). The number labeled under each supercluster candidate is the mean redshift of member clusters. If all members of a given supercluster have confident redshift (see Section \ref{subsec:photo-z_summary}), then the redshift number is colored in green; otherwise redshift number is colored in yellow. Under the definition that superclusters will eventually experience gravitational collapse, our results suggest that KGSC might be fragmented into numerous supercluster candidates. }
    \centering
    \label{fig:KG_SC}
\end{figure*}

King Ghidorah Supercluster (KGSC), which is associated with two elongated filaments, extending toward northeast and southeast at redshift $z\approx 0.55$, was recently discovered by \citet{Shimakawa23}. The discovery was made by searching for density excess above $3\sigma$ in the redshift slice $[0.5,0.6)$ of the HSC PDR3 density map  \citep{Shimakawa21}. The main body of KGSC, exhibiting three density peaks in both their weak lensing (WL) map and stellar mass map, is found to be connected with more than 15 CAMIRA clusters. The southeast filament, consisting of two WL mass peaks, is also associated with more than 9 CAMIRA clusters, while the northeast filament is associated with less rich clusters and no WL mass peaks. \citet{Shimakawa23} also estimated the total mass of the main body and the southeast filament via the correlation between the total mass of the supercluster and the overdensity areas, suggesting total masses of $\approx 1.1\times10^{16}M_{\odot}$ and $5.0\times 10^{15}M_{\odot}$, respectively.

Figure \ref{fig:KG_SC} depicts the positions of the supercluster candidates of different multiplicities along with the CAMIRA clusters that are not associated with any supercluster candidates in the KGSC field. Under the definition that superclusters will eventually experience gravitational collapse, KGSC is fragmented into numerous supercluster candidates. Three supercluster candidates are associated with the density peak in the northern main body, while the rest of the CAMIRA clusters may eventually evolve into island universes. Although we find four supercluster candidates connected with the northeast filaments, two of them  could not be confirmed. Interestingly, there appear to be at least three supercluster candidates associated with the southeast filaments, including one supercluster candidate with three confident redshift cluster members. This indicates an extremely dense environment here.

\subsubsection{CL1604 \label{sububsec:CL1604}}

Supercluster CL1604 in the HECTOMAP field is one of the few massive structures confirmed at redshift close to unity \citep{Lubin00, Gal08}. Located at redshift around $z\approx0.9$, three galaxy clusters with halo mass $M_{vir}>10^{14}\,h^{-1}\,M_{\odot}$ and five galaxy groups with halo mass $M_{vir}>10^{13}\,h^{-1}\,M_{\odot}$ comprise the main body of CL1604 \citep{Lemaux12, Hayashi19}. A recent study by \citet{Hayashi19} further found structures extended toward north and south using the HSC wide field images and Subaru/FOCAS and  Gemini/GMOS spectroscopic confirmation, revealing coherent structures of length scale $\approx 50$ cMpc. Moreover, through spectral index analysis, both the main body and the extend structures appear to have 3 density peaks along the line-of-sight direction that formed at the same epoch \citep{Lemaux12, Hayashi19}. 

\begin{figure*}[htb]
    \centering
    \begin{center}
        \begin{minipage}{0.95\textwidth}
            \includegraphics[width = 0.99\hsize]{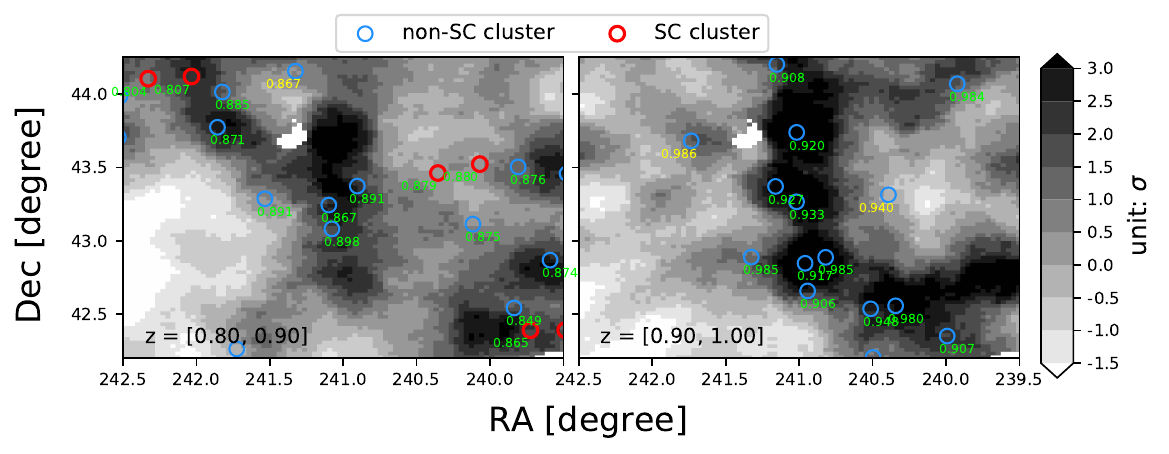}
        \end{minipage}
    \end{center}
    \centering
    \vspace{-0.5cm}
    \caption{CAMIRA supercluster candidates and isolated clusters in CL1604 field overlaid by HSC PDR3 density map  in two different redshift slices. The blue open circles represent the positions of CAMIRA clusters that are not associated with any supercluster candidates. The red open circles are CAMIRA clusters associated with supercluster candidates. The number labeled under each cluster is the redshift of the cluster. If we are confident (see Section \ref{subsec:photo-z_summary}) in the redshift of the clusters, then the redshift number is colored in green; otherwise redshift number is colored in yellow. We note that we invert the color map here compared to Figure \ref{fig:KG_SC} for display purposes.}
    \centering
    \label{fig:CL1604}
\end{figure*}

Figure \ref{fig:CL1604} shows the CAMIRA clusters that are associated with supercluster candidates and the isolated clusters with HSC PDR3 density map  \citep{Shimakawa21} overlaid. Similarly to the case of the KGSC, we find that only parts of the structures are identified as superclusters under the constraint of future collapse. In this case, there is one cluster pair at the west of the main body and another in the northeast extending structure, suggesting the rest of clusters associated with Cl1604 will evolve into island universes. 

\section{Clusters and Brightest Cluster Galaxies in superclusters} \label{sec:BCG}

As a demonstration of the use of this supercluster catalog, in this section, we investigate the environmental effects of superclusters on their member clusters and BCGs. Our cluster sample is divided into two redshift bins: $z=[0.5, 0.7]$ and $z=[0.7, 1.0]$. We refer to clusters that reside inside any supercluster candidates as supercluster clusters (SC-Cls), while we define our control sample, isolated clusters (Iso-Cls), as clusters that are not associated with any supercluster candidates. We further require that the clusters in Iso-Cls meet the following two criteria to avoid any potential associations with supercluster candidates: (1) no other clusters are located within a projected distance of $15\,h^{-1}\,$cMpc for lower redshift bin and $20\,h^{-1}\,$cMpc for higher redshift bin,  (2) no other clusters are within the line-of-sight distance $\pm0.03(1+z_{cl})$ for both redshift bins. This set of samples is referred to as \texttt{All}. 

\begin{table}[!ht]
    \centering
    \begin{tabular}{|c|cc|cc|}
    \hline
    &\multicolumn{2}{c|}{All}&\multicolumn{2}{c|}{Conf-z} \\
    \hline
        \textbf{Redshift Bins} & \textbf{SC-cl} & \textbf{Iso-Cl} & \textbf{SC-cl} &  \textbf{Iso-Cl}  \\ \hline \hline
        $[0.5, 0.7]$ & 611 & 529 & 440 & 303 \\ \hline
        $[0.7, 1.0]$ & 840 & 599 & 506 & 294 \\ \hline
    \end{tabular}
    \caption{Summary of cluster number count of each sample.}
    \label{tab:cl_sample}
\end{table}

Taking into account the possible impact of the photo-$z$ uncertainty as demonstrated in Section \ref{subsec:SC_n}, we construct another set of SC-Cls and Iso-Cls using only clusters with confident redshift described in Section \ref{subsec:photo-z_summary} and Figure \ref{fig:z_distribution}. For this second sample, we regard a supercluster candidate to be confirmed if all of its member clusters have a confident redshift. For SC-Cl, we consider clusters from confirmed superclusters. For the control sample, we include clusters with confident redshift that are not associated with any confirmed superclusters. We also require that clusters in Iso-Cls meet the two criteria mentioned above. This second sample is referred to as \texttt{Conf-z}. Table \ref{tab:cl_sample} summarizes the number of our SC-Cls and Iso-Cls in each redshift bin. For our BCG analysis, we determine the BCG for each CAMIRA cluster following \citet{2021MNRAS.507.4016D}, who selected the most massive (red) galaxy in a cluster with membership probability $w>0.1$.

\subsection{Cluster richness and BCG stellar mass\label{subsec:mstar-Nmem}}

\begin{figure*}[htb]
    \centering
    \begin{center}
        \begin{minipage}{0.95\textwidth}
            \includegraphics[width=0.525\hsize]{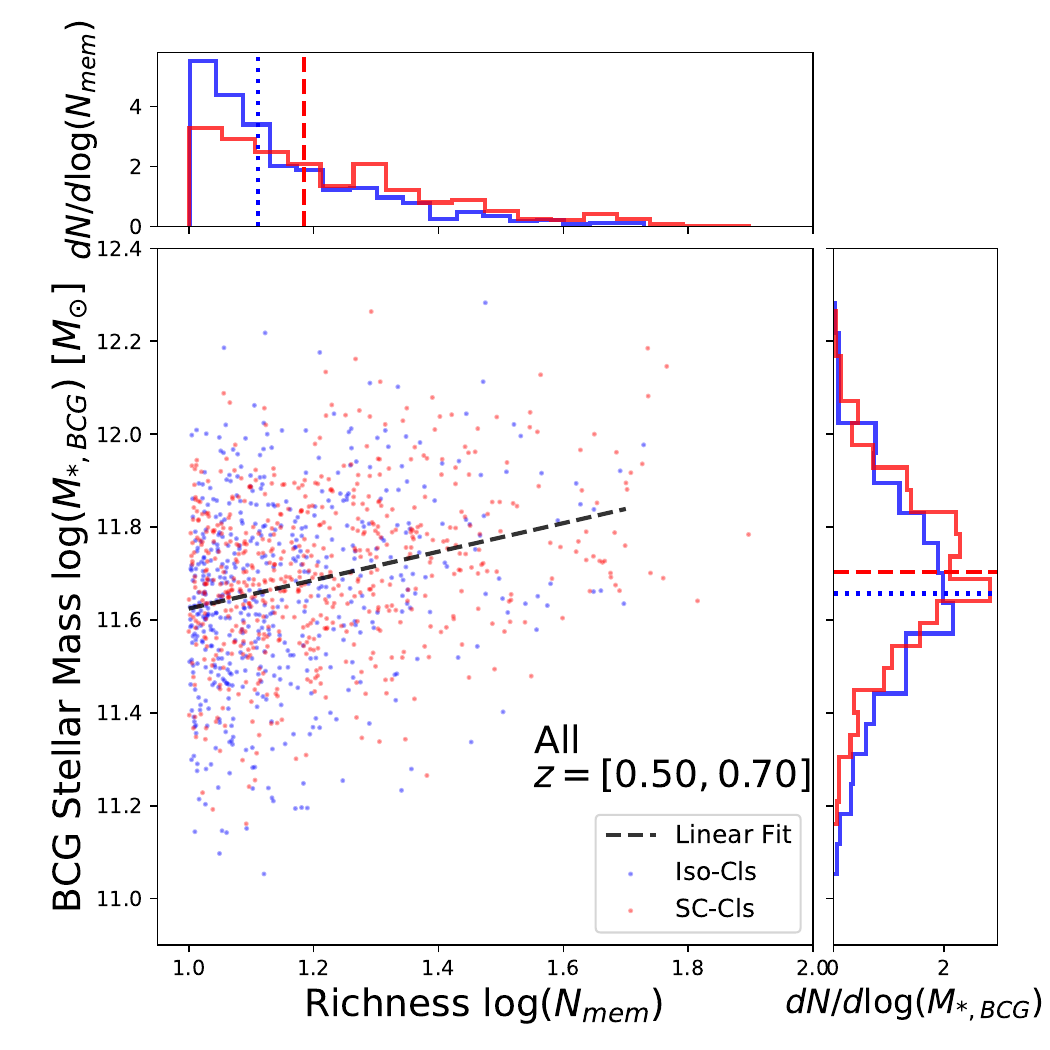}
            \includegraphics[width=0.525\hsize]{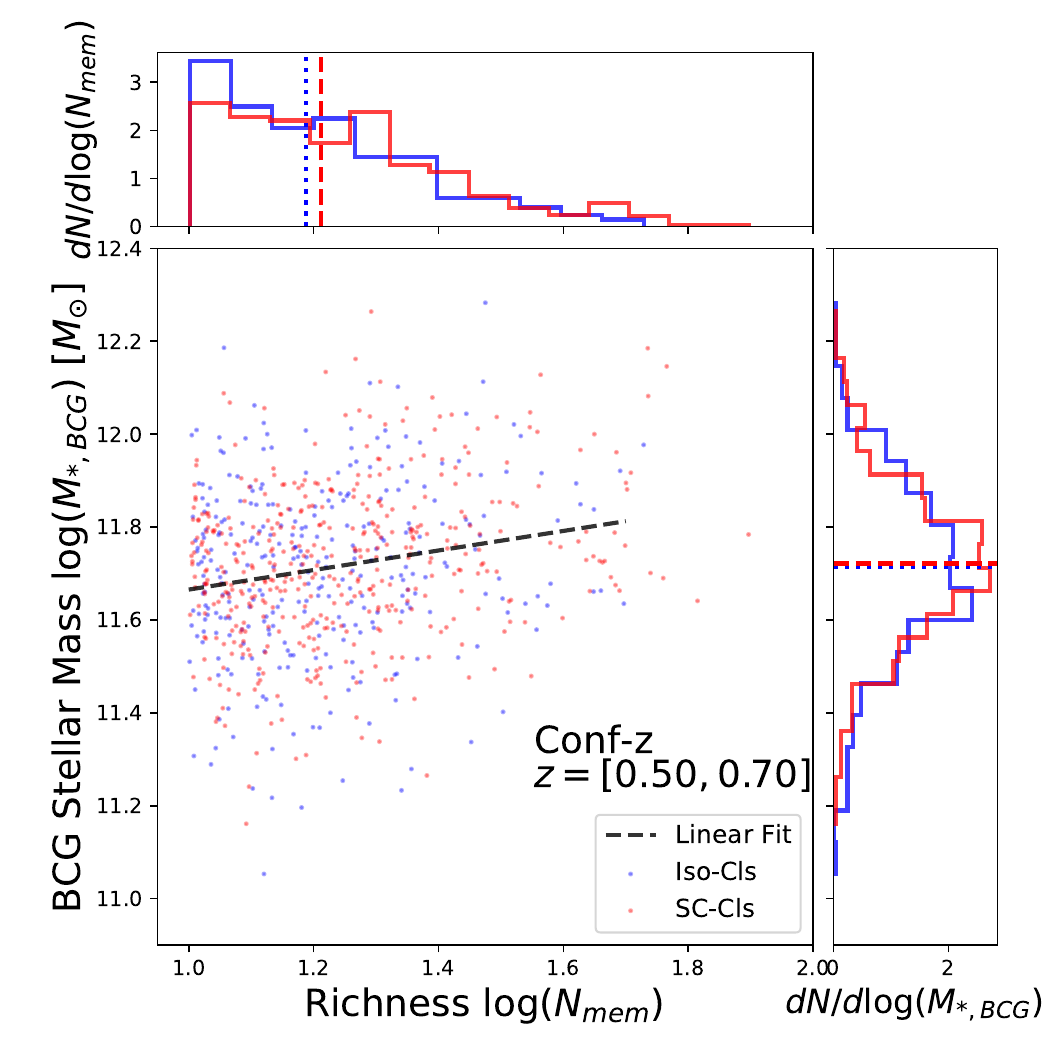}
            \includegraphics[width=0.525\hsize]{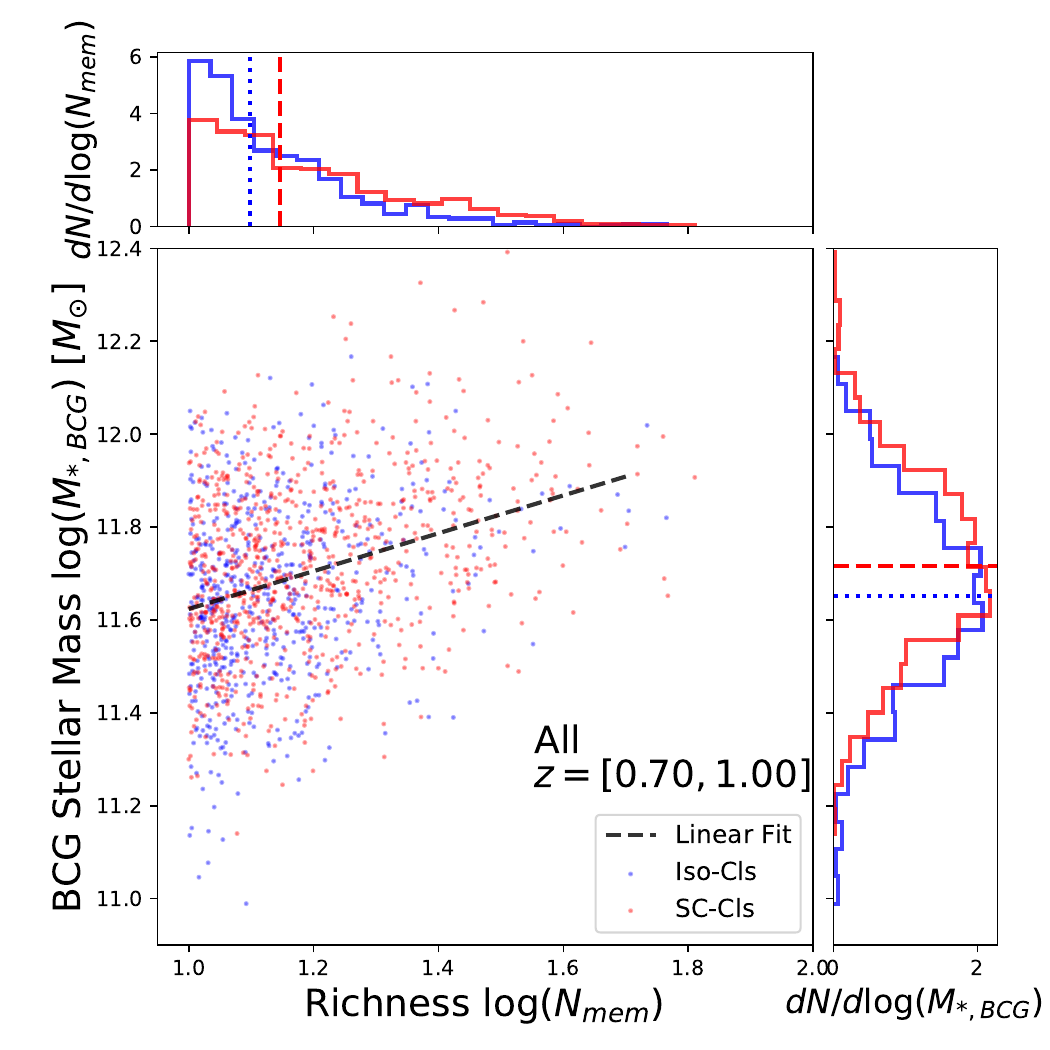}
            \includegraphics[width=0.525\hsize]{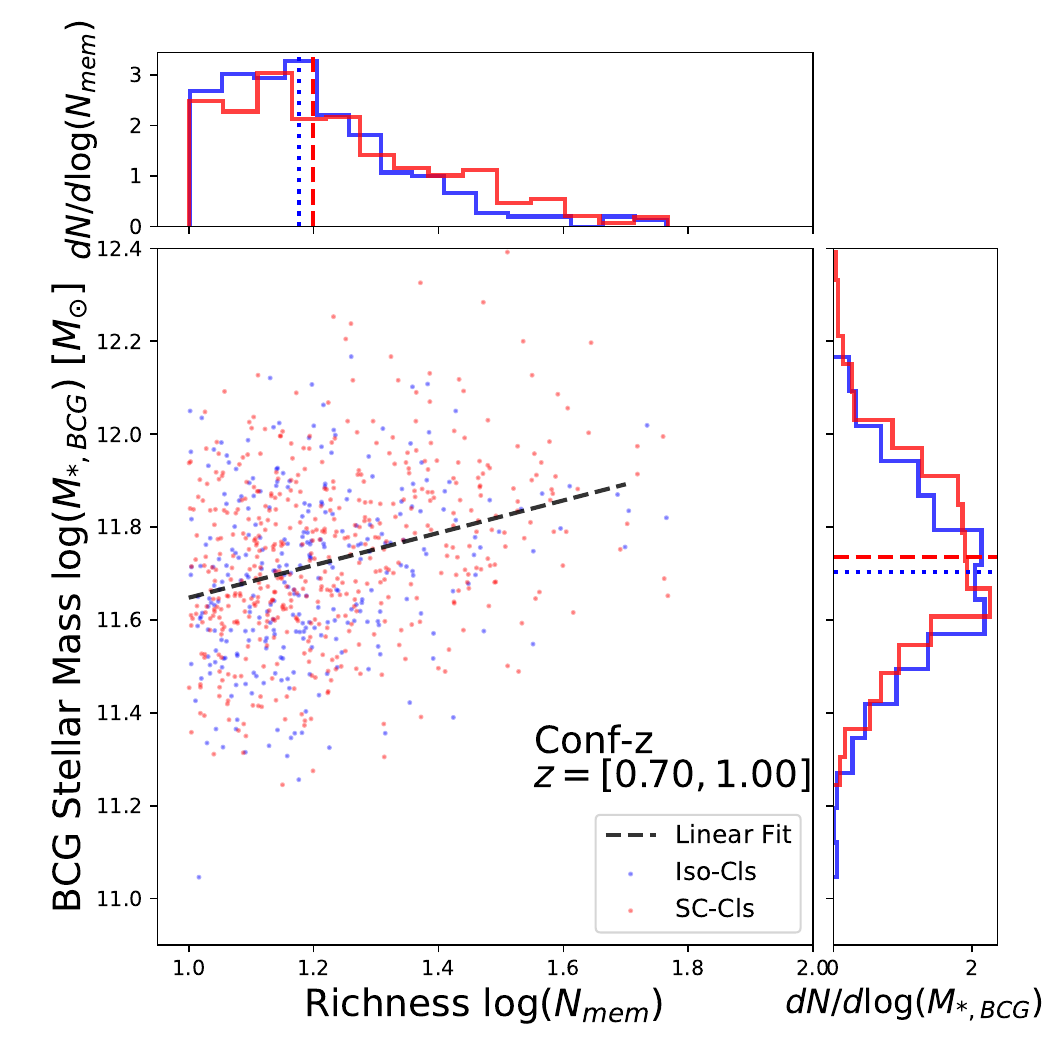}
        \end{minipage}
    \end{center}
    \caption{The stellar mass of BCGs as a function of cluster richness. The blue and red points represent isolated clusters and supercluster clusters, respectively. The upper and right side panels show the distributions of richness and stellar mass. The corresponding redshift bin and sample set are indicated in each panel. The black dashed line indicates the power-law fit of all data points in a given redshift bin and sample set. The richness distributions and stellar mass distributions in the side panels all suggest that clusters and BCGs in the supercluster environment tend to be more massive (see Figures~\ref{fig:mstar_residual} and \ref{fig:residual_richness_corr} for more details).}
    \label{fig:mstar-Nmem}
\end{figure*}

Figure \ref{fig:mstar-Nmem} shows the comparison of BCG stellar mass as a function of richness between SC-Cls and Iso-Cls for the two sample sets in two redshift bins. The BCG stellar mass is taken from the CAMIRA catalog. Side panels on the top and right show the normalized distributions of richness and stellar mass, respectively. The richness distribution of SC-Cls is evidently more top-heavy than that of Iso-Cls in all four cases, suggesting that clusters residing in superclusters are generally more massive. Moreover, the BCG stellar mass distributions also imply similar results: BCGs in SC-Cls are slightly more massive than BCGs in Iso-Cls, although the difference between the median stellar mass of the two distributions is only around $0.05\,$dex. It is  well-known that the BCG stellar mass is positively correlated with its host cluster mass \citep[e.g.,][]{lin04}. The trend of more massive BCGs in supercluster environments could arise from the presence of the richer clusters in such environments, as found in the literature \citep[e.g.,][]{Einasto03, Chon13}. Therefore, to investigate the effect of large-scale environments, we must decouple the environmental effect of different spatial scales. 

\begin{figure*}[htb]
    \centering
    \begin{center}
        \begin{minipage}{0.95\textwidth}
            \includegraphics[width=0.525\hsize]{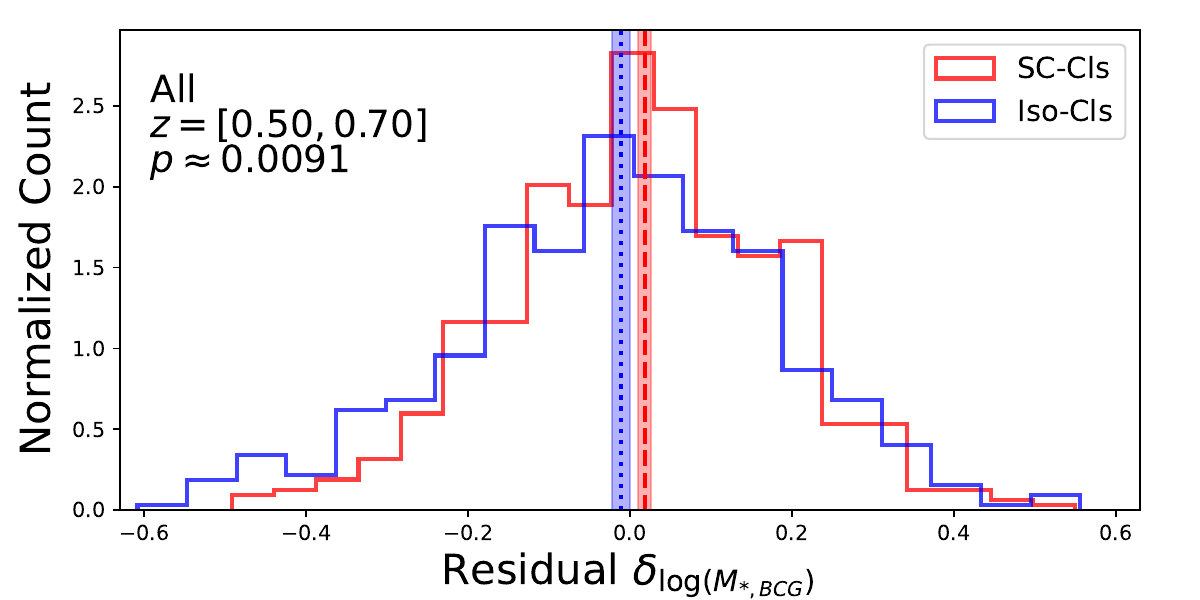}
            \includegraphics[width=0.525\hsize]{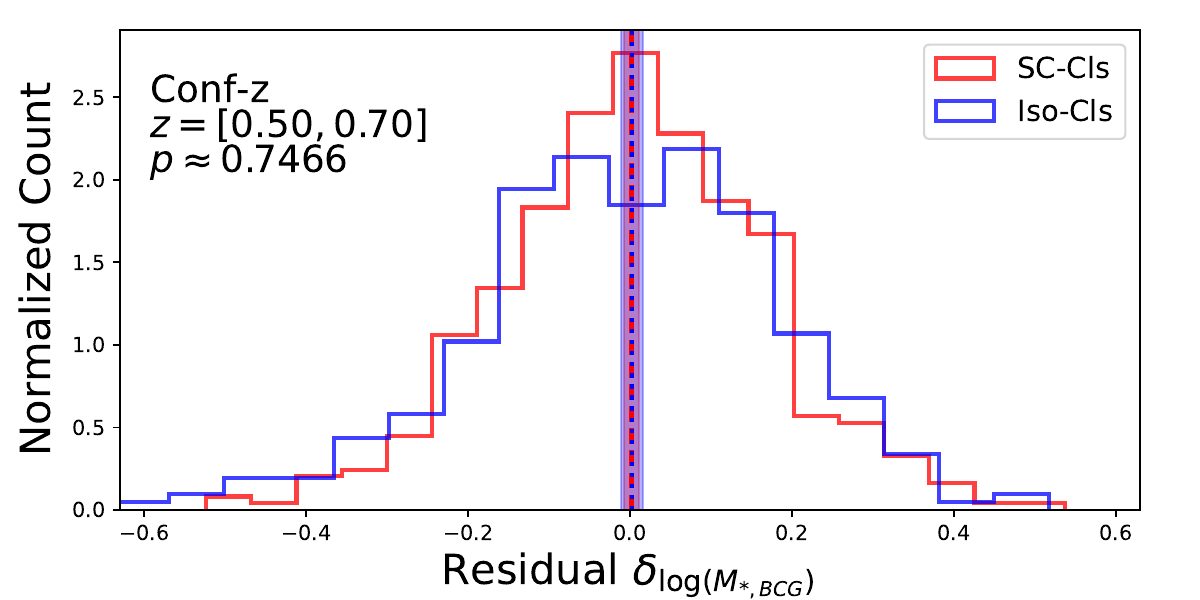}
            \includegraphics[width=0.525\hsize]{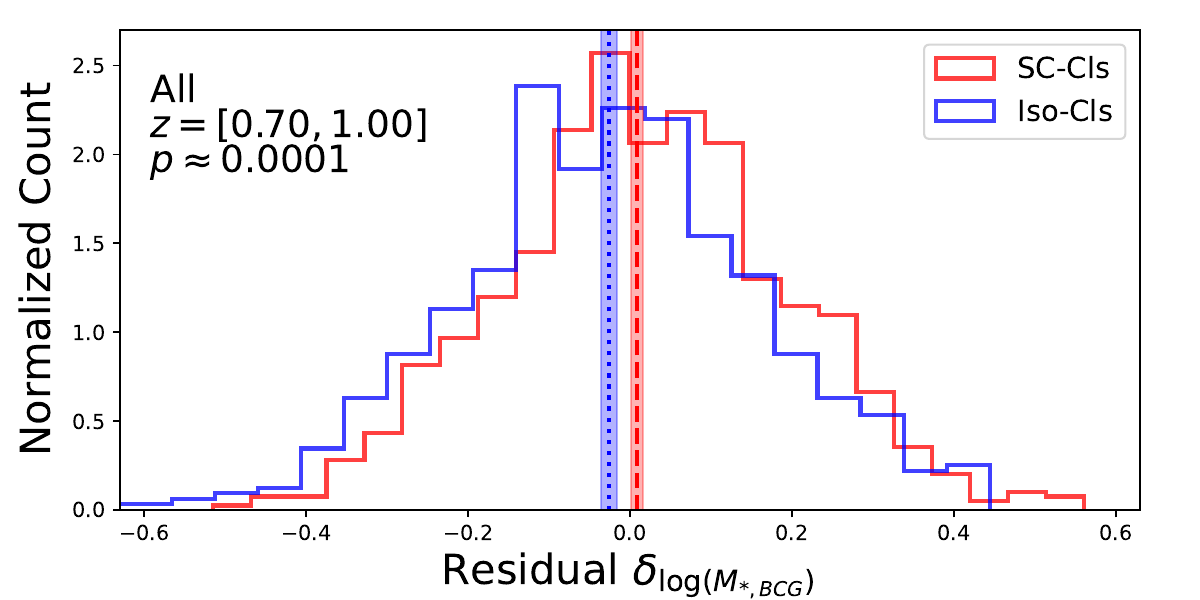}
            \includegraphics[width=0.525\hsize]{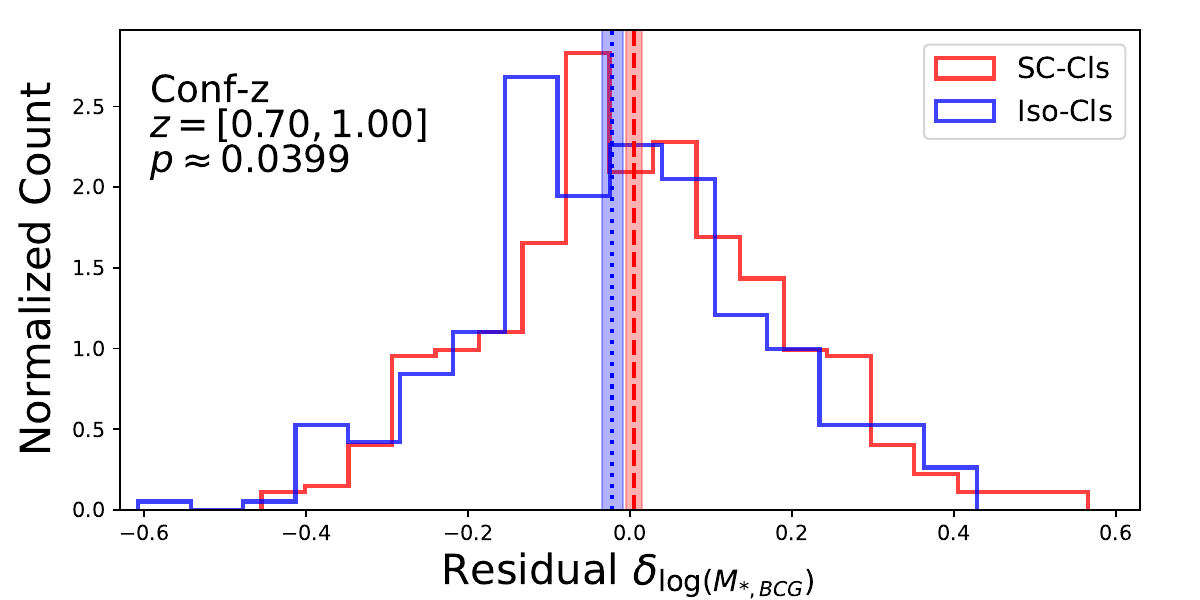}
        \end{minipage}
    \end{center}
    \caption{The normalized distributions of stellar mass residual $\delta_{\log(M_{*,BCG})}$ for supercluster clusters (SC-Cls; red) and isolated clusters (Iso-Cls; blue). The vertical dashed  and the dotted lines show the median values of the residual distributions of SC-Cls and Iso-Cls, respectively. The width of the vertical band indicates the median uncertainty calculated by $\sigma\sqrt{\pi/(2N)}$ where $N$ and $\sigma$ are the sample size and the standard deviation. The corresponding redshift bin, sample set, and Kolmogorov-Smirnov test $p$ value of each panel are shown at the upper left corner. Given the offset of median in the supercluster environments and the median uncertainty, all panels suggest BCG stellar mass has a marginal dependence on the supercluster environments.}
    \label{fig:mstar_residual}
\end{figure*}

Quantitatively, we follow the approach described in \citet{Luparello15}, who compared Brightest Group Galaxies (BGGs) living inside supercluster environments with those outside of such environments. For each redshift bin and sample set, we fit a power-law  to the combined SC-Cls and Iso-Cls samples, as shown by the black dashed line in each panel of Figure \ref{fig:mstar-Nmem} and denoted as $L$. The residual of BCG stellar mass $\delta_{\log(M_{*,BCG})}$ for each cluster is then computed as 
\begin{equation}
    \delta_{\log(M_{*,BCG})} = \log(M_{*,BCG}) - L[\log(N_{mem})],
\end{equation}
where $M_{*,BCG}$ and $N_{mem}$ is stellar mass and host cluster richness of a given BCG. We then examine the distributions of the stellar mass residual $\delta_{\log(M_{*,BCG})}$ for SC-Cls and Iso-Cls. Figure \ref{fig:mstar_residual} shows the BCG stellar mass residual distributions and the median of the distributions for the two redshift bins and two sample sets. As the residual distributions of SC-Cls are slightly offset from that of Iso-Cls toward the massive end, all panels seem to suggest that the supercluster environments marginally 
make the BCGs more massive. We further calculate the median uncertainty $\sigma_{med}$ by:
\begin{equation}
    \sigma_{med}= \sigma \sqrt{\pi/(2N)}
\end{equation} 
where $\sigma$ and $N$ are standard deviation and sample size. Considering the uncertainty in the median values (shown as the bands around the vertical lines in Figure \ref{fig:mstar_residual} and tabulated in Table \ref{tab:residual_mass_median}), the differences in the median are only significant for the \texttt{All} sample.

We also perform a two-sided Kolmogorov-Smirnov (KS test) and show the $p$-value in each panel of Figure \ref{fig:mstar_residual}. Again, only \texttt{All} sample in both redshift bins show a weak indication that the null hypothesis that two distributions are identical may be ruled out. However, we should cautiously interpret the greater deviation between two distributions in the left panels, as the projection effect (e.g., a foreground galaxy may be mis-identified by CAMIRA as the BCG) and contamination rate could affect the \texttt{All} sample.  

\begin{table*}[htb]
    \centering
    \begin{tabular}{|c|c|cc|}
    \hline
    \multicolumn{2}{|c|}{} &\multicolumn{2}{c|}{\textbf{Residual $\delta_{\log(M_*, BCG)}$}}\\
    \hline
        \textbf{z bins} & \textbf{Sample} & \textbf{SC-Cl Median } & \textbf{Iso-Cl Median}   \\ (1) & (2) & (3) & (4) \\ \hline \hline
        \multirow{2}{4em}{$[0.5, 0.7]$}  & All   & $0.0180\pm 0.0084$ & $-0.0111\pm 0.0107$  \\
                                         & Conf-z& $0.0018\pm 0.0097$ & $0.0023\pm 0.0134$  \\ \hline
        \multirow{2}{4em}{$[0.7, 1.0]$}  & All   & $0.0086\pm 0.0076$ & $-0.0260\pm 0.0095$  \\ 
                                         & Conf-z& $0.0045\pm 0.0098$ & $-0.0218\pm 0.0129$   \\ \hline
    \end{tabular}
    \caption{Statistics of the BCG stellar mass residual $\delta_{\log(M_{*,BCG})}$ distributions for the two redshift bins and two samples in Figure~\ref{fig:mstar_residual}. Residual medians and median uncertainties of SC-Cls are listed in column (3). Residual medians and median uncertainties of Iso-Cls are listed in column (4).}
    \label{tab:residual_mass_median}
\end{table*}

\begin{figure*}
    \centering
    \begin{center}
        \begin{minipage}{0.95\textwidth}
            \includegraphics[width = 0.5 \hsize]{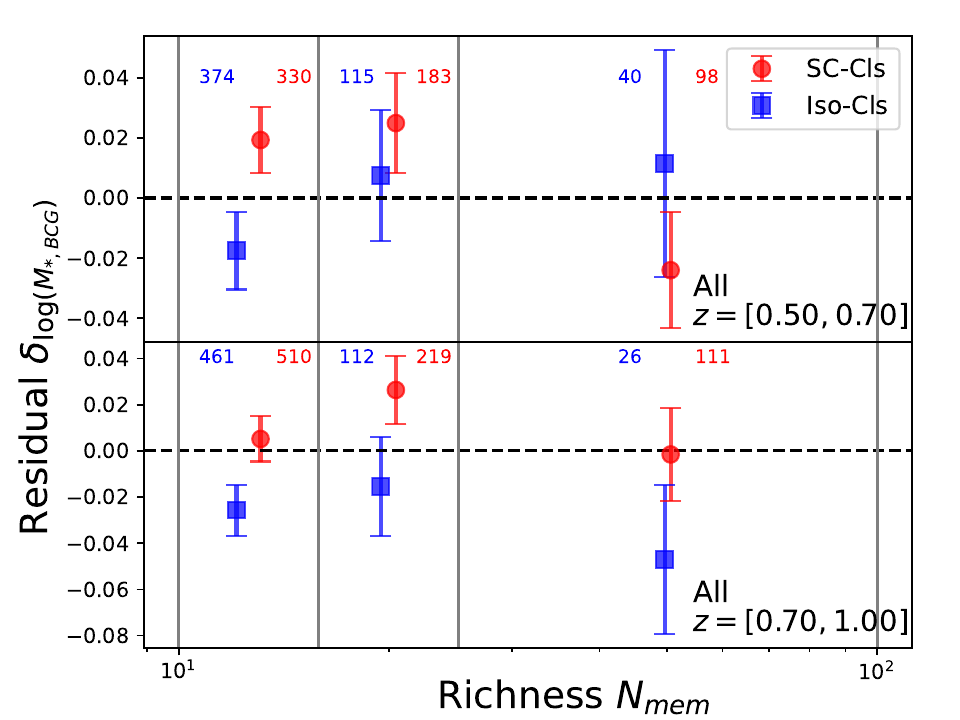}
            \includegraphics[width = 0.5 \hsize]{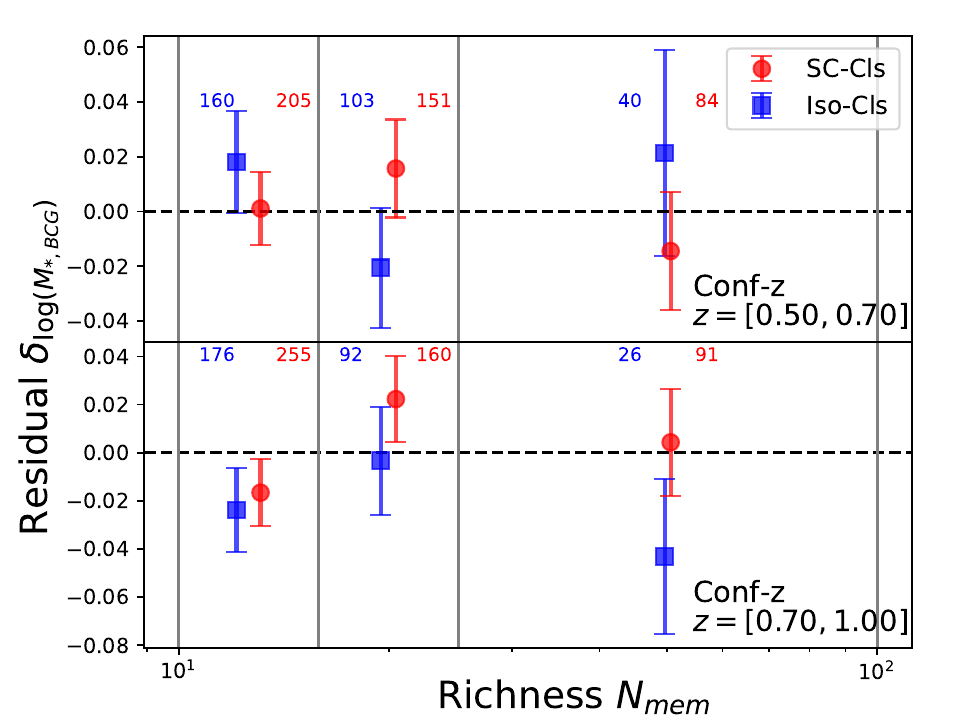}
        \end{minipage}
    \end{center}
    \caption{Stellar mass residual $\delta_{\log(M_{*,BCG})}$ as function of richness for each redshift bin and each sample. The vertical gray lines show the edges of the richness bin. The red circles  (blue squares) and error bars are the medians and median uncertainties of $\delta_{\log(M_{*,BCG})}$ calculated within each richness bin for supercluster clusters (isolated clusters). The number in each richness bins shows the sample size. }
    \label{fig:residual_richness_corr}
\end{figure*}

As the residual distributions of $M_{*,BCG}$ reflect the general deviation of $M_{*,BCG}$ from the average behavior of clusters covering a wide range of richness, we further investigate the residual $\delta_{\log(M_{*,BCG})}$ as a function of richness. We recalculate the median and median uncertainties of $\delta_{\log(M_{*,BCG})}$ in three different richness bins.\footnote{The bin widths are chosen such that the number of clusters in the highest $N_{mem}$ bin is still statistically meaningful (especially for the {\tt Conf-z} sample).} In Figure \ref{fig:residual_richness_corr}, although the medians of SC-Cls exceed the medians of Iso-Cls in every richness bin, the  uncertainties undermine the difference. 

We have further examined the distributions in finer redshift bins ($\Delta z=0.1$), and found that only in one out of five bins, the $p$-value of the {\tt All} is less than 0.01. However, we note that the uncertainty in stellar mass measurements would generally be (much) larger than 0.1\,dex; this, coupled with the fact that the differences are only at 0.02\,dex level (c.f.~the large scatter of BCG mass at fixed richness as shown in Figure \ref{fig:mstar-Nmem}) imply that the BCG stellar mass has at most a marginal dependence on the supercluster environment. 

\subsection{BCG color and SFR distributions\label{subsec:color_SFR}}
\begin{table*}[htb]
    \centering
    \begin{tabular}{|c|c|cc|cc|}
    \hline
    \multicolumn{2}{|c|}{} &\multicolumn{2}{c|}{\textbf{Rest frame color $M_g-M_r$}}&\multicolumn{2}{c|}{\textbf{$\log$(SFR)}} \\
    \hline
        \textbf{z bins} & \textbf{Sample} & \textbf{SC-Cl Median } & \textbf{Iso-Cl Median} & \textbf{SC-Cl Median} &\textbf{Iso-Cl Median}  \\ (1) & (2) & (3) & (4) & (5) & (6)\\ \hline \hline
        \multirow{2}{4em}{$[0.5, 0.7]$}  & All   & $0.6996 \pm 0.0009$ & $0.6994 \pm 0.0010$ & $-0.6550 \pm 0.0271$ & $-0.7160 \pm 0.0300$ \\
                                         & Conf-z& $0.6996 \pm 0.0010$ & $0.6996 \pm 0.0013$ & $-0.6465 \pm 0.0312$ & $-0.7070 \pm 0.0405$ \\ \hline
        \multirow{2}{4em}{$[0.7, 1.0]$}  & All   & $0.7137 \pm 0.0014$ & $0.7104 \pm 0.0017$ & $-1.3815 \pm 0.0333$ & $-1.4360 \pm 0.0362$ \\ 
                                         & Conf-z& $0.7137 \pm 0.0018$ & $0.7104 \pm 0.0025$ & $-1.3760 \pm 0.0449$ & $-1.4100 \pm 0.0525$ \\ \hline
    \end{tabular}
    \caption{Statistics of the rest-frame color and star formation rate distributions for the two redshift bins and two samples in Figures \ref{fig:color_hist} and  \ref{fig:sfr_hist}. Columns (3) and (5) (columns 4 \& 6) are the median values for the distributions of $M_g-M_r$ and SFR for SC-Cls (Iso-Cls), respectively.}
    \label{tab:color_sfr_stats}
\end{table*}

\begin{figure*}[htb]
    \centering
    \begin{center}
        \begin{minipage}{0.95\textwidth}
            \includegraphics[width = 0.5 \hsize]{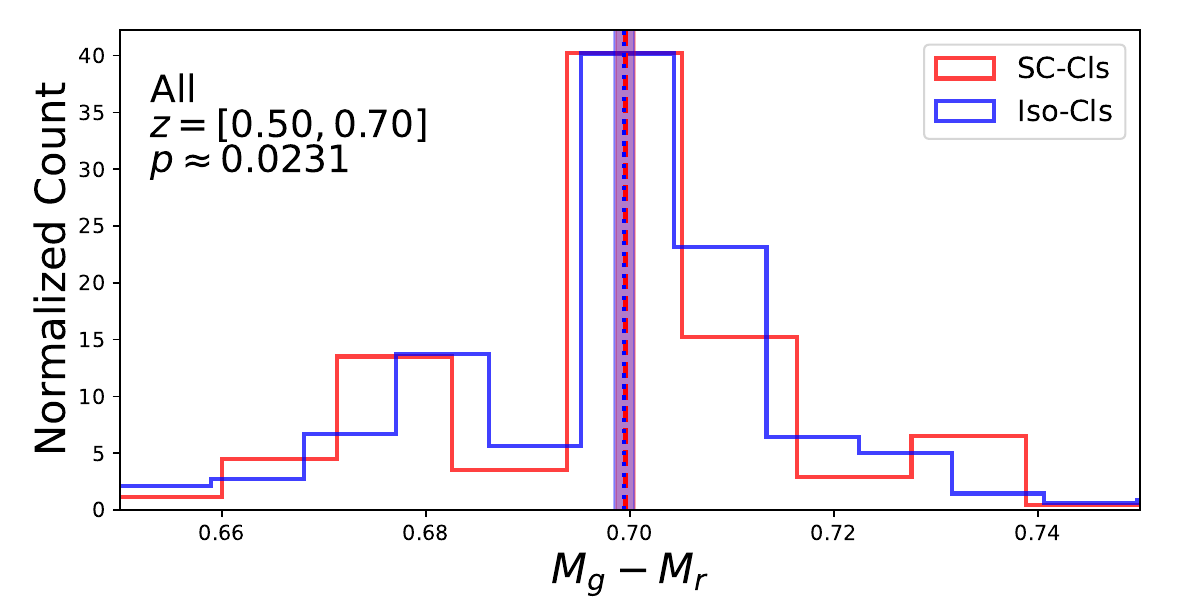}
            \includegraphics[width = 0.5 \hsize]{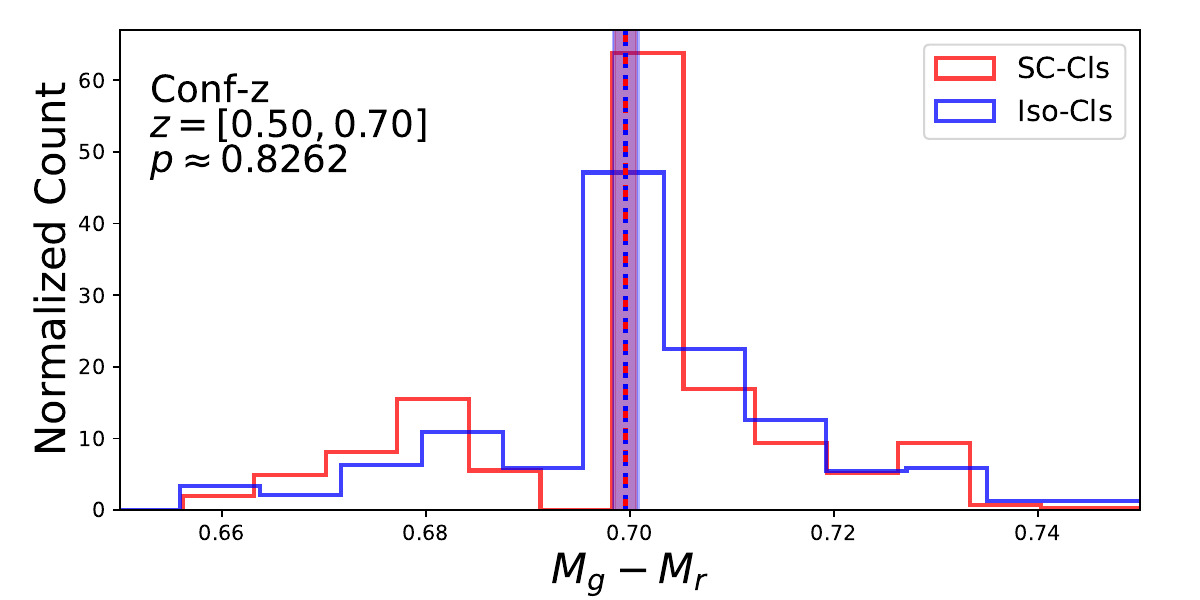}
            \includegraphics[width = 0.5 \hsize]{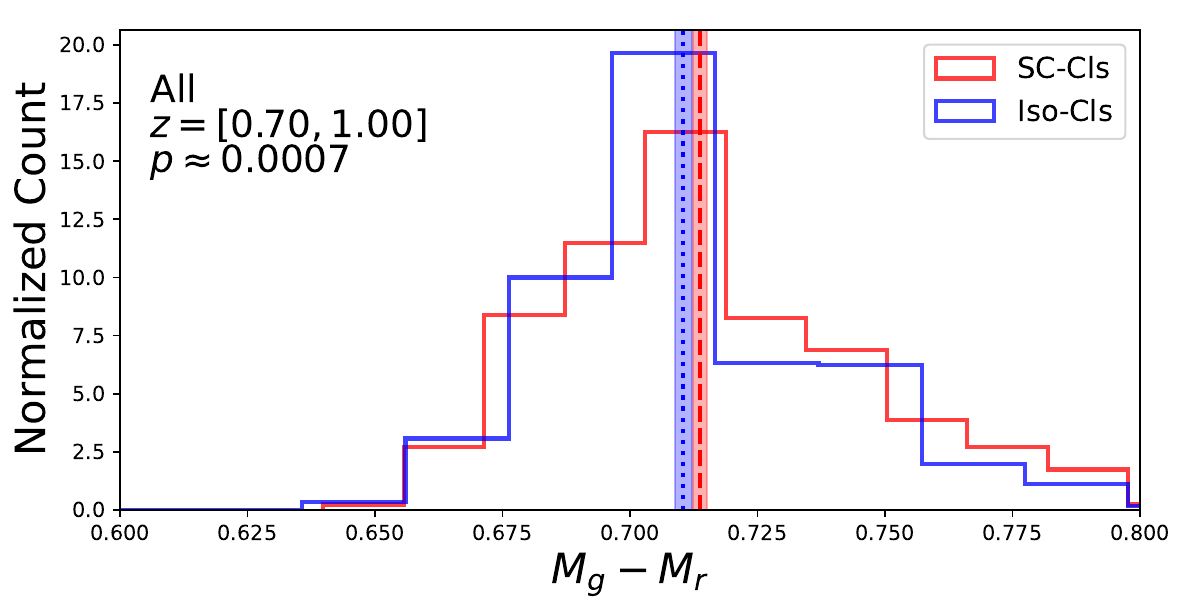}
            \includegraphics[width = 0.5 \hsize]{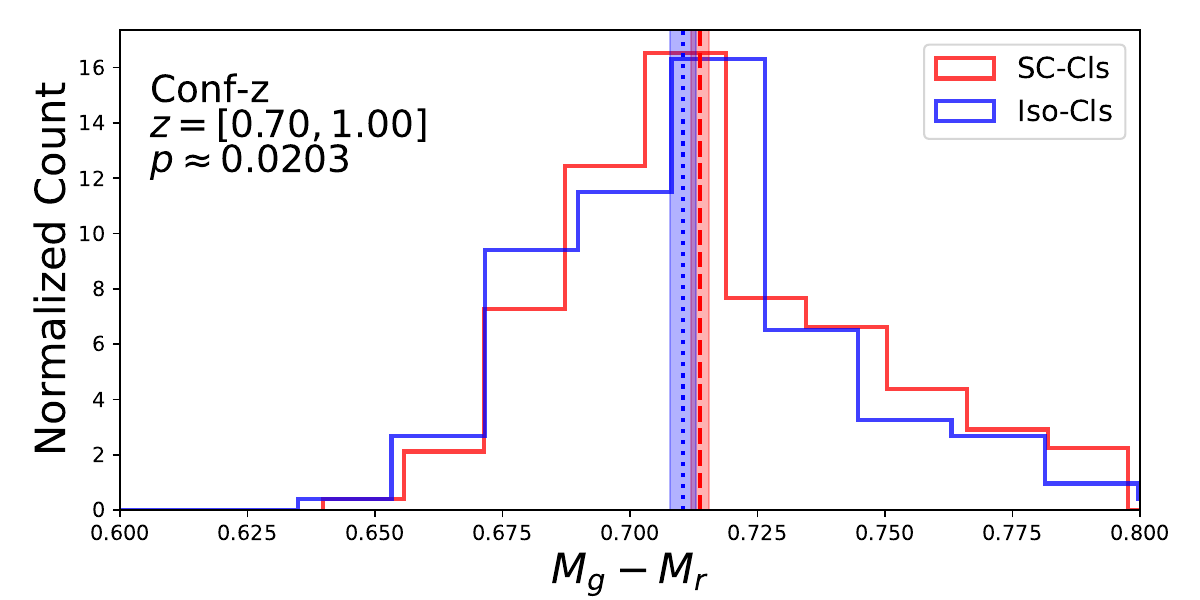}
        \end{minipage}
    \end{center}
    \caption{The normalized distributions of rest-frame color $M_g-M_r$ for supercluster clusters (SC-Cls; red) and isolated clusters (Iso-Cls; blue). The vertical dashed and the dotted lines show the median values of the distributions of SC-Cls and Iso-Cls, respectively. The width of the vertical band indicates the median uncertainty. The values of medians and median uncertainties are listed in Table \ref{tab:color_sfr_stats}. The corresponding redshift bin, sample set, and Kolmogorov-Smirnov test $p$ value of each panel are shown at the upper left corner. For demonstration purposes, 
    we only show the color range that encompasses 98\% of the sample.}
    \label{fig:color_hist}
\end{figure*}

\begin{figure*}[htb]
    \centering
    \begin{center}
        \begin{minipage}{0.95\textwidth}
            \includegraphics[width = 0.5 \hsize]{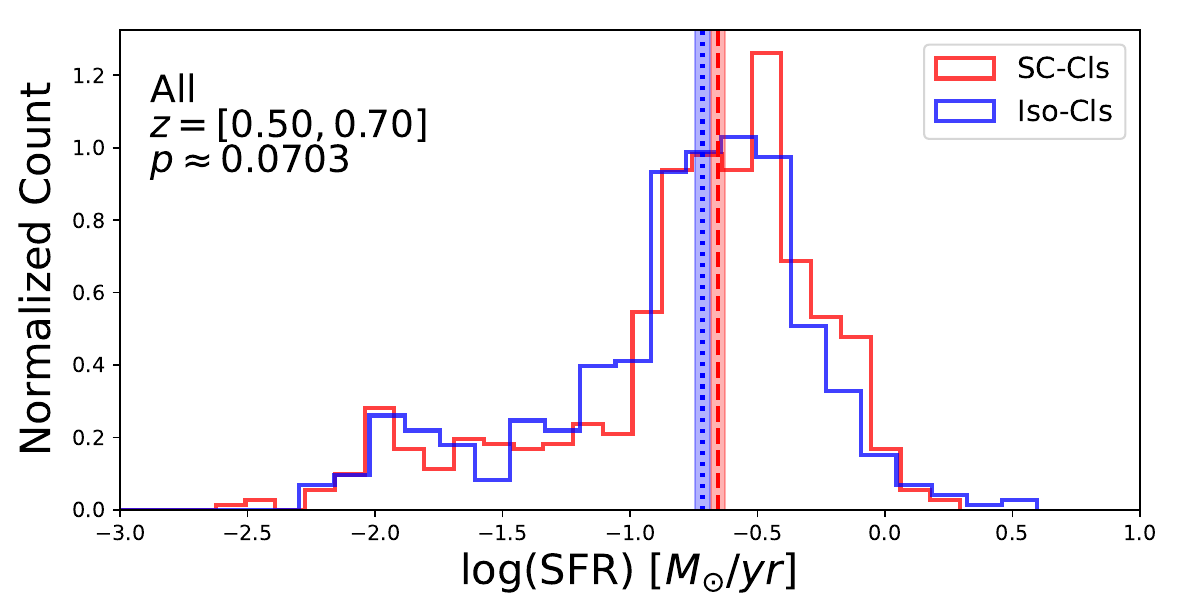}
            \includegraphics[width = 0.5 \hsize]{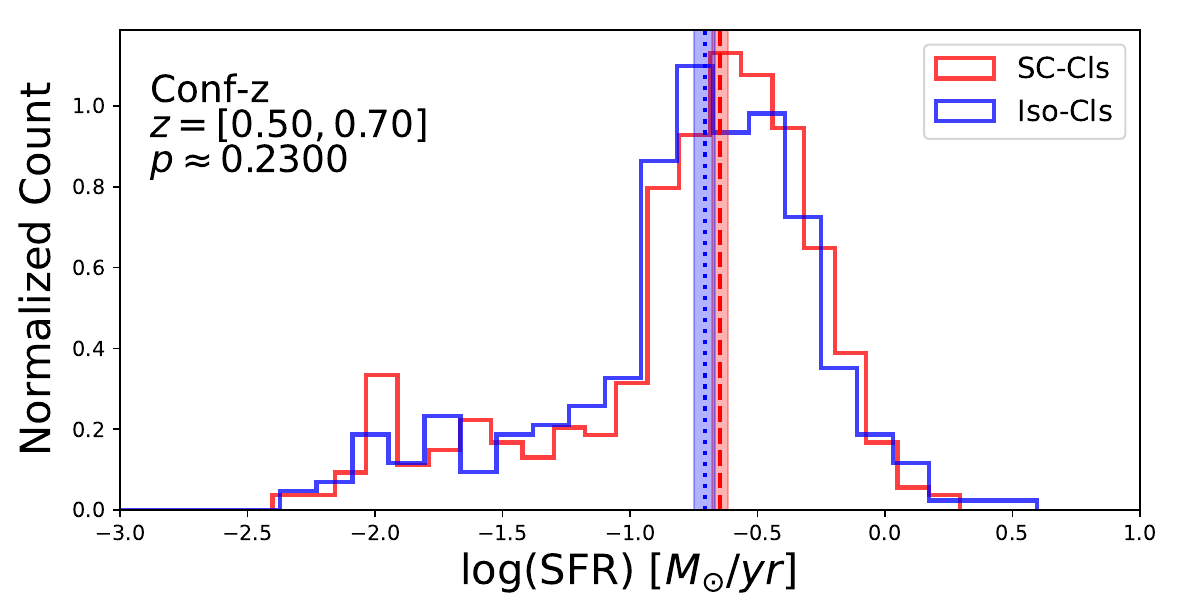}
            \includegraphics[width = 0.5 \hsize]{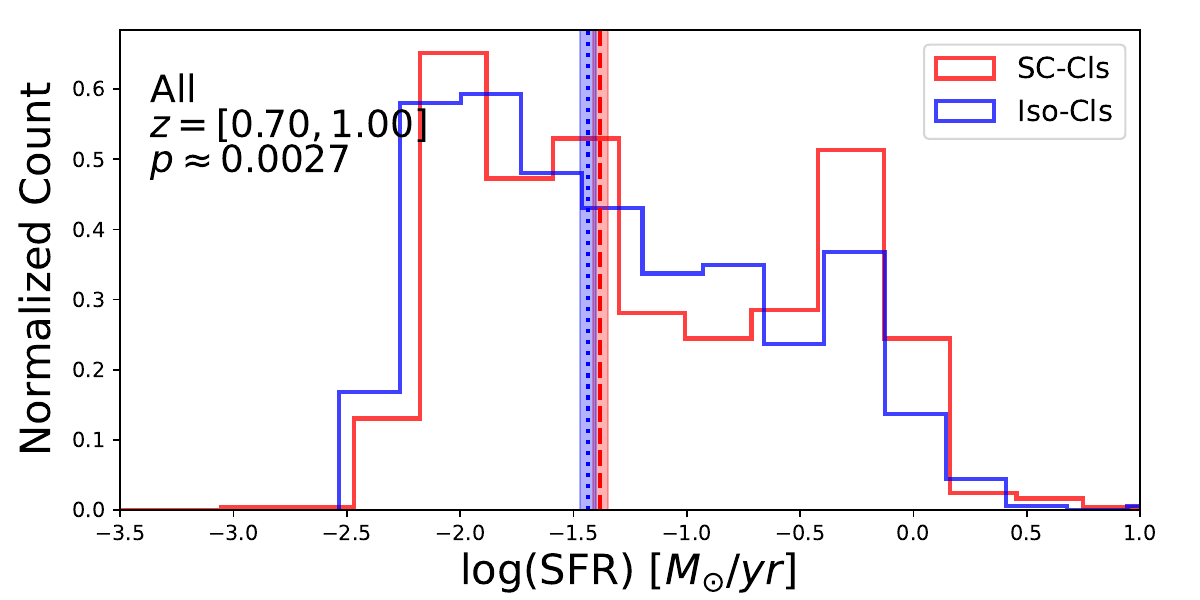}
            \includegraphics[width = 0.5 \hsize]{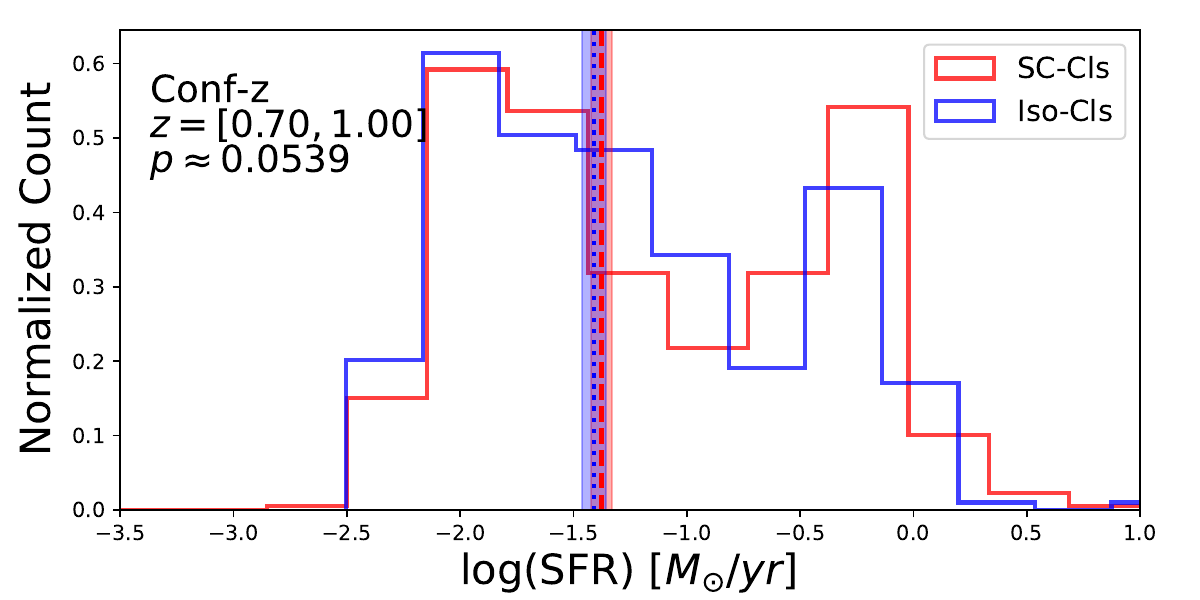}
        \end{minipage}
    \end{center}
    \caption{Same as Figure \ref{fig:color_hist}, but for SFR. For demonstration purposes, the range of SFR are chosen to encompasses $\gtrsim 99\%$ of the samples.}
    \label{fig:sfr_hist}
\end{figure*}
 
We further investigate the possible dependence of rest-frame color and SFR of BCGs on different global environments. For the rest-frame color, we adopt rest-frame magnitudes derived by the template fitting-code MIZUKI \citep{Tanaka15, Tanaka18}. In addition to the cluster selection criteria mentioned above, we further require the reduced chi-square of the best fit model \texttt{reduced chisq}\ $<5$ as recommended by \citet{Nishizawa20} when computing the rest-frame color. The SFR used here is derived by DEmP (described in section \ref{subsec:photo-z}). It is found that both color and SFR of BCGs do not seem to depend on richness.  We therefore do not perform any fitting to the distributions; rather, we simply compare the color and SFR distributions of SC-Cls and Iso-Cls using the KS test.
Figures~\ref{fig:color_hist} and \ref{fig:sfr_hist} show the distributions.
The  median values and their uncertainties of each redshift bin and sample set are presented in Table \ref{tab:color_sfr_stats}. In general, we find no significant difference in both rest-frame color and SFR between SC-Cls and Iso-Cls suggested by the  median and  $p$-values. 

In summary,  we have demonstrated that clusters living in superclusters are richer than isolated clusters, which leads to an apparent bias in stellar mass of the BCGs. After the correlation between BCG stellar mass and cluster richness is taken into account, there is only very marginal suggestion of BCGs associated with superclusters being slightly more massive than their cousins living in isolated clusters.  As for color and SFR, we do not find evidence for any differences either. Lastly, we remark that if blue BCGs are missed by the CAMIRA algorithm, our findings here could potentially be biased. Therefore, using our proposed methodology here, which could be adapted to cluster samples constructed by other means (e.g., X-rays, Sunyaev-Zel'dovich effect;  \citealt{2021ApJS..253....3H}; \citealt{2022A&A...661A...2L}; \citealt{2023MNRAS.526.3757K}), would be useful to confirm our findings in this section. 

\section{Discussion and Summary} \label{sec:summary}

In this work, we have adopted a physically-motivated definition of superclusters  proposed by \citet{Chon15} that superclusters should eventually collapse in an Universe dominated by dark energy. This definition is preferable as it relies on the hierarchical nature of structure formation with fewer subjective decisions (e.g., the choice of the linking length) in the process of identifying superclusters. Furthermore, the supercluster catalogs built following this definition facilitate a more straightforward comparison among different works. Our supercluster finding method is dedicated to fulfilling this definition with the aid of $N$-body simulations. In the following, we first discuss our optimized linking length, which reflects the limit of gravitational bound structures (Section \ref{subsec:discuss_limit}). We then show that the superclusters detected here are equivalent to the supercluster cores mentioned in the literature (Section \ref{subsec:discuss_multiplicity}). Finally, we interpret our analysis of BCGs in different global environments (Section \ref{subsec:discuss_BCGs}).

\subsection{On the Limit of Gravitational Bound Structures}\label{subsec:discuss_limit}

\citet{Dunner06} studied the limit of gravitationally bound structures, starting from the energy conservation equation of the spherical collapse model with a dark energy term. They ended up with the criterion that the mass density enclosed by the last bounded shell should be $\approx8$ times greater than the background matter density (see Eqn.~19 therein).  However, \citet{Dunner06} showed that this criterion could induce a $\approx30\%$ contamination rate, which they attributed mainly to the gravitational attraction from external structures. \citet{Luparello11} and \citet{Chon13} constructed supercluster catalog based on the criteria derived by \citet{Dunner06} and the spherical collapse model, respectively. In this sense, the superclusters they selected might not necessarily experience gravitational collapse due to external gravitational attraction. In contrast, our proposed method here empirically includes such effects during the optimization process for the linking length. Since we compare the future collapsed structures with those selected directly by the FoF algorithm, we take into account all different effects internally. This leads to a stricter selection and higher purity as shown in Figure \ref{fig:performanance}. 

The linking length derived in Figure \ref{fig:LD} reflects the length scale of superclusters, as it represents the most suitable length to connect  halos/clusters that will collapse in the (far) future. In general, the comoving size of the linking length increases as the redshift increases. The dependence of the linking length on redshift may imply that the sizes of superclusters decrease as time evolves. An intuitive description could be that it becomes more difficult for groups of clusters to collapse since the dark energy gradually dominates the dynamics of the Universe (at $z\approx 0.7$). \citet{Einasto19} investigated the evolution of superclusters in numerical simulation via a percolation analysis. They found a similar trend that the overall geometrical diameter of superclusters shrinks as time evolves (see Figure 4 in \citealt{Einasto19}). Interestingly, \citet{Einasto21} suggested that this trend is also found in models with different cosmological parameters (see their Figure 3). However, we note that the definition of supercluster and supercluster finding method of \citet{Einasto19} and \citet{Einasto21} are different from our approach. Understanding how cosmological parameters affect the evolution of supercluster is an interesting task, but that is beyond the scope of the present work.

\subsection{On the Multiplicity of Superclusters}\label{subsec:discuss_multiplicity}
The multiplicity function shown in Figure \ref{fig:multiplicity} reveals a notable result: The overall supercluster population is dominated by cluster pairs and cluster triplets. This result is also reflected in our comparison with the known massive structures in Figures \ref{fig:KG_SC} and \ref{fig:CL1604}. These structures (King Ghidorah Supercluster and CL1604 supercluster) are fragmented into several supercluster candidates of low multiplicity in our analysis. A similar physical picture that only the core regions of unbound over-dense structures will eventually collapse has already been revealed in several dynamical analyses of superclusters below  $z=0.5$ \citep{Chon15, Einasto16, Bagchi17, Einasto18}. These studies essentially analyzed the future collapsed regions by comparing the integrated mass profiles with the predictions from the spherical collapse model. Although the mass estimation approaches vary among the works, most of them reached the consistent conclusion that the radius of the future collapsed region is below $10\,h^{-1}$ cMpc. 

Recently, \citet[][hereafter E22]{Einasto22} identified High Density Cores (HDCs) in the BOSS Great Wall at $z\approx0.5$ \citep{Lietzen16} and estimated their radii and masses of future collapsed regions. The HDCs identified in E22 share several similarities with the superclusters identified in this work in terms of radius, multiplicity, and mass. In total, eight HDCs were found from massive unbounded structures of the BOSS Great Wall (see Table 2 in E22). The mean radius of future collapsed regions of these cores is $\approx 6.9\,h^{-1}\,$cMpc, while the calibrated linking length corresponding to the richness cut $N_{cut}=10.0$ at $z=0.5$ (Figure \ref{fig:LD}) in our work is approximately $7.2\,h^{-1}\,$cMpc. Furthermore, in E22, only $2-3$ galaxies more massive than $\log(M_*/M_{\odot})=11.3$,  corresponding to the completeness limit of their sample, were located in the turnaround regions for each HDCs. In E22, they assumed that galaxies with stellar mass  $\ge \log(M_*/M_{\odot})=11.3$ are the BGGs in  rich galaxy groups because of the survey depth. They further derived the corresponding halo masses using the stellar mass--halo mass relation \citep[e.g.,][]{2010ApJ...710..903M}, finding that the turnaround regions for each HDC contain 2-3 halos more massive than $\approx 2\times10^{13}\,h^{-1}\,M_{\odot}$, in agreement with our findings of the multiplicity function. Finally, the mean mass enclosed by the future collapsed region reported by E22 is  $\approx 1.5\times 10^{15}\,h^{-1}\,M_{\odot}$. Such a value was derived by first summing up all the halo masses inferred from galaxies above the completeness limit, then including a correction factor for incompleteness. Without such a correction factor, the sum of halo masses  above the completeness limit would be $\approx 1.3\times10^{14}\,h^{-1}\,M_{\odot}$, about the same order as our overall supercluster candidate mass (Table \ref{tab:SC_mass}). Given the similarity between these characteristics, we conclude that the supercluster candidates we have identified essentially correspond to future collapsed regions of HDCs. This is an independent confirmation that the supercluster candidates we have detected will eventually collapse because we do not rely on the spherical collapse model.

\subsection{On the BCGs in Superclusters}\label{subsec:discuss_BCGs}
As the first application of our new supercluster catalog, we study how superclusters affect their member clusters and BCGs. In Figure \ref{fig:mstar-Nmem}, we have shown that clusters residing in supercluster environments tend to be richer. This is not surprising, as many studies have found that richer and more luminous groups and clusters tend to be found in high-density environments \citep{Einasto03, Luparello13, Chon13}; the combination of great depth and wide sky coverage of the HSC survey enables us to confirm such a correlation up to $z=1.0$ in this study. 
We find only marginal differences in the stellar mass  of BCGs in SC-Cls and those in Iso-Cls (Figure \ref{fig:residual_richness_corr} and Figure \ref{fig:color_hist}), while the color and SFR appear to be quite similar. 

Our study here suggests that the evolution of BCGs is not significantly influenced by large-scale structures at the supercluster level below redshift $z=1$. An analogous study was conducted by \citet{Luparello15}, who investigated the dependence of the properties of BGGs on global environments. They found that late-type BGGs tend to be redder, more massive, and have lower SFR in supercluster environments, while early-type BGGs show no dependency on global environments. Given that BCGs of CAMIRA clusters presumably are red, early-type galaxies, our findings are consistent with their results (on BGGs). 

Recently, \citet{Einasto24} divide local galaxy groups and clusters from \citet{Tempel14} into four classes based on the group luminosity and study the dependence of the quiescent fraction of BGGs and BCGs  on the global environment quantified by the luminosity-density field \citep{Liivamagi12}. They find that once the group class is fixed, the relative ratio between quiescent galaxies, red star-forming galaxies, and blue star-forming galaxies does not vary with the global environment such as voids, supercluster outskirts, and supercluster cores (see their Table 6). Their results also suggest that the properties of BGGs are mainly shaped by the local environment, which is in agreement with our work at higher redshift. It will be interesting to investigate how supercluster environments affect other member galaxies. However, this is beyond the scope of the current study. 

Although the methodology proposed here could theoretically achieve a purity of $\approx 0.90$, we also demonstrate in Sections \ref{subsec:SC_n} and \ref{subsec:multiplicity} that the cluster photo-$z$ uncertainty hampers the power of our method. Both the ability to recover high-multiplicity superclusters and the overall abundance are reduced. Figure \ref{fig:contamination} further provides an estimate of contamination rate as a function of redshift. This suggests the importance of spectroscopic redshift in the search for superclusters at high redshift. We advise the users of the supercluster catalog to be aware of the redshift confidence of each supercluster candidate. Fortunately, several large spectroscopic surveys are either well on the way, or will take place in the near-future, including Dark Energy Spectroscopic Instrument \citep[DESI;][]{DESI16} and Subaru Prime Focus Spectrograph \citep[PFS;][]{Takada14, Greene22}. These surveys  will fully unleash the power of our method. 

\clearpage
\begin{acknowledgments}
We thank an anonymous referee for insightful comments that have improved the clarity of the manuscript. We acknowledge support from the National Science and Technology Council of Taiwan under grants MOST 111-2112-M-001-043, NSTC 111-2628-M-002-005-MY4,  NSTC 112-2112-M-001-061, and NSTC 113-2112-M-001-005.
This work was supported by JSPS KAKENHI Grant  JP22H01260. 
We are grateful to Shogo Ishikawa for providing mock catalogs for analysis.
TCC thanks Sut Ieng Tam and Guan-Ming Su for inspiring discussions, 
and YHC, YTC and TLC for constant support.
YTL thanks IH, LYL and ALL for constant encouragement and inspiration.
HYS acknowledges support from NTU Academic Research-Career Development Project under Grant No.~NTU-CDP-111L7779.

The Hyper Suprime-Cam (HSC) collaboration includes the astronomical communities of Japan and Taiwan, and Princeton University.  The HSC instrumentation and software were developed by the National Astronomical Observatory of Japan (NAOJ), the Kavli Institute for the Physics and Mathematics of the Universe (Kavli IPMU), the University of Tokyo, the High Energy Accelerator Research Organization (KEK), the Academia Sinica Institute for Astronomy and Astrophysics in Taiwan (ASIAA), and Princeton University.  Funding was contributed by the FIRST program from the Japanese Cabinet Office, the Ministry of Education, Culture, Sports, Science and Technology (MEXT), the Japan Society for the Promotion of Science (JSPS), Japan Science and Technology Agency  (JST), the Toray Science  Foundation, NAOJ, Kavli IPMU, KEK, ASIAA, and Princeton University.

This paper is based  on data collected at the Subaru Telescope and retrieved from the HSC data archive system, which is operated by Subaru Telescope and Astronomy Data Center (ADC) at NAOJ. Data analysis was in part carried out with the cooperation of Center for Computational Astrophysics (CfCA) at NAOJ.  We are honored and grateful for the opportunity of observing the Universe from Maunakea, which has the cultural, historical and natural significance in Hawaii.

This paper makes use of software developed for Vera C.~Rubin Observatory. We thank the Rubin Observatory for making their code available as free software at http://pipelines.lsst.io/. 
\end{acknowledgments}

\software{astropy \citep{2022ApJ...935..167A},
          colossus  \citep{2018ApJS..239...35D},
          matplotlib \citep{Hunter:2007},
          numpy \citep{harris2020array}
          }
          
\appendix
\section{Convergence Test} \label{sec:convergence}
\begin{table}[htb]
    \centering
    \begin{tabular}{|c|c|}
        \hline 
          & Configurations: ($m_{DM}\,[h^{-1}M_{\odot}]$, $\epsilon\,[h^{-1}\,$ckpc$]$)\\ \hline \hline
        Mass resolution & $(2.48\times10^{10}, 15)$ v.s. $(3.10\times10^{9}, 7.5)$ \\ \hline
        Softening length & $(2.48\times10^{10}, 15)$ v.s. $(2.48\times10^{10}, 7.5)$\\ \hline
    \end{tabular}
    \caption{The configurations of $N$-body simulations that are used in the convergence test. All the simulations are fixed to a box size $L=350\,h^{-1}$cMpc. $m_{DM}$ is the dark matter particle mass in unit of $h^{-1}\,M_{\odot}$. $\epsilon$ denotes the gravitational softening length of dark matter particles in unit of $h^{-1}$ckpc. The first row lists the simulation configurations for the mass resolution test. The second row lists the simulation configurations for the softening length test.}
    \label{tab:sim_config}
\end{table}

\begin{figure*}[htb]
    \centering
    \begin{center}
        \begin{minipage}{0.95\textwidth}
            \includegraphics[width = 0.5\hsize]{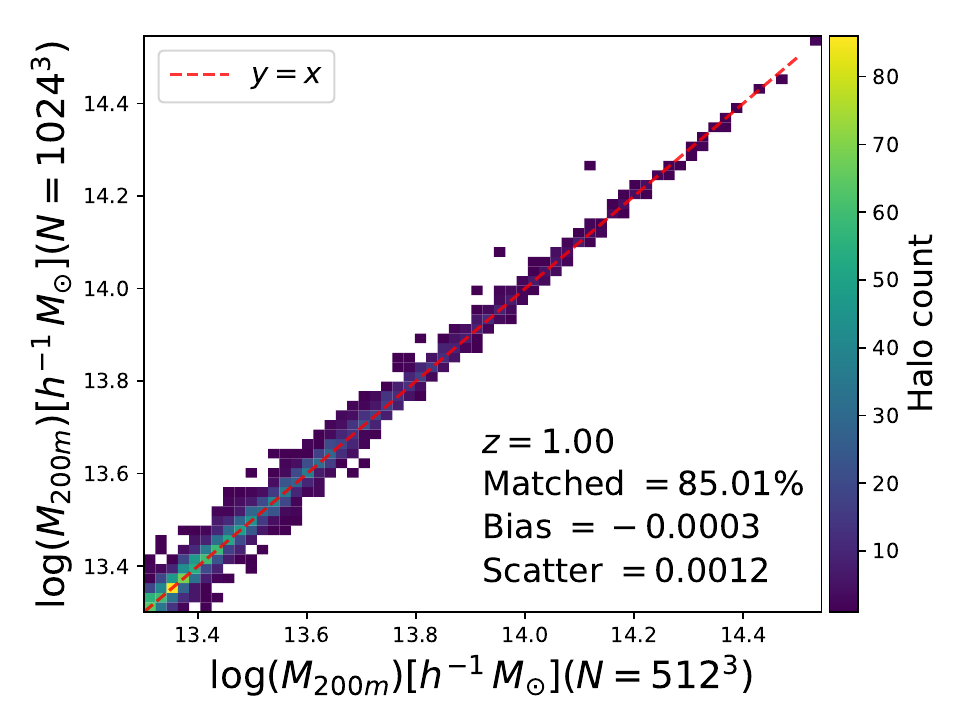}
            \includegraphics[width = 0.5\hsize]{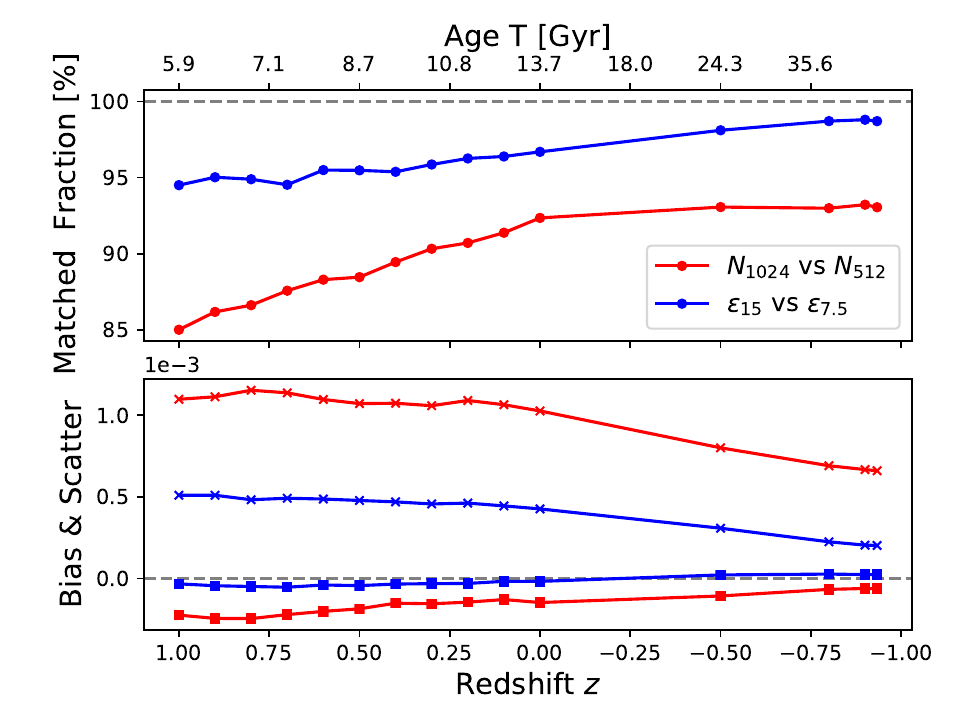}
        \end{minipage}
    \end{center}
    \caption{Left: Halo mass $M_{200m}$ cross-match between \texttt{ROCKSTAR} halo catalogs of low-resolution $(N=512^3)$ and high-resolution $(N=1024^3)$ runs using a maximum separation $0.2\,h^{-1}\,$cMpc at redshift $z=1.0$. The matched halo fraction, bias, and scatter are shown in the lower right corner. The color map indicates the halo count in each 2D histogram bin. Right: In both panels, the cross-match between halo catalogs of different softening lengths are indicated by the red curves. The cross-match between halo catalogs of different mass resolutions are indicated by the blue curves. Upper panel: The matched fraction as a function of redshift. Lower panel: Bias (squares) and scatter (circles) as a function of redshift.}
    \label{fig:convergent_1}
\end{figure*}

Our convergence tests examine the convergence of mass resolution and softening length. We set a halo mass lower limit $M_{200m, min}=2.0\times10^{13}\,h^{-1}\,M_{\odot}$ (see Section \ref{subsec:mock}) and conducted the tests by cross-matching halos with $M_{200m}\geq M_{200m, min}$ between low resolution and high resolution runs by a maximum separation $\Delta d=0.2\,h^{-1}$cMpc. The simulation configurations are listed in Table \ref{tab:sim_config}. The first and second rows show the configurations for mass resolution test and the softening length test, respectively. The random seed in these simulations is fixed and the box size is $L=350\,h^{-1}$cMpc.

The  plot on the left of Figure \ref{fig:convergent_1} shows the comparison of the halo mass of the matched halos between the low mass resolution run and the high mass resolution run at redshift $z=1.00$. We indicate the matched fraction, bias, and scatter in the bottom right corner. We define bias $\delta$ and scatter $\sigma$ of residual $\Delta$ as following:
\begin{equation}
    \delta = \mbox{median}(\Delta)
\end{equation}
\begin{equation}
    \sigma = 1.48\times \mbox{MAD}(\Delta)
\end{equation}
where $\Delta=[\log(M_{200m, low})-\log(M_{200m, high})]/(1+\log(M_{200m, high}))$. The cross-match result shows that 85\% of halos in the higher mass resolution run can have counterparts in the lower mass resolution run. Furthermore, the halo masses between two simulations agree well with scatter $0.0012(1+\log(M_{200m, high}))$. The panels on the right show the matched fraction, bias, and scatter as a function of redshift for both the mass resolution test (red curves) and the softening length test (blue curves). Throughout all snapshots, the upper panel shows that the matched fractions are greater than 85\%, and the lower panel further shows that the masses of the matched halos are in good agreement in both the mass resolution test and the softening test. This suggests that the positions and masses of halos, on which our analysis is based, have reached sufficient convergence. Hence, we set our fiducial configuration to dark matter particle mass of $m_{DM}=2.48\times10^{10}\,h^{-1}\,M_{\odot}$ and softening length of $\epsilon=15\,h^{-1}$ckpc.

\section{Unweighted purity and completeness} \label{sec:appendix_uw_metric}
\begin{figure*}[htb]
    \centering
    \begin{center}
        \begin{minipage}{0.95\textwidth}
            \includegraphics[width = 0.5 \hsize]{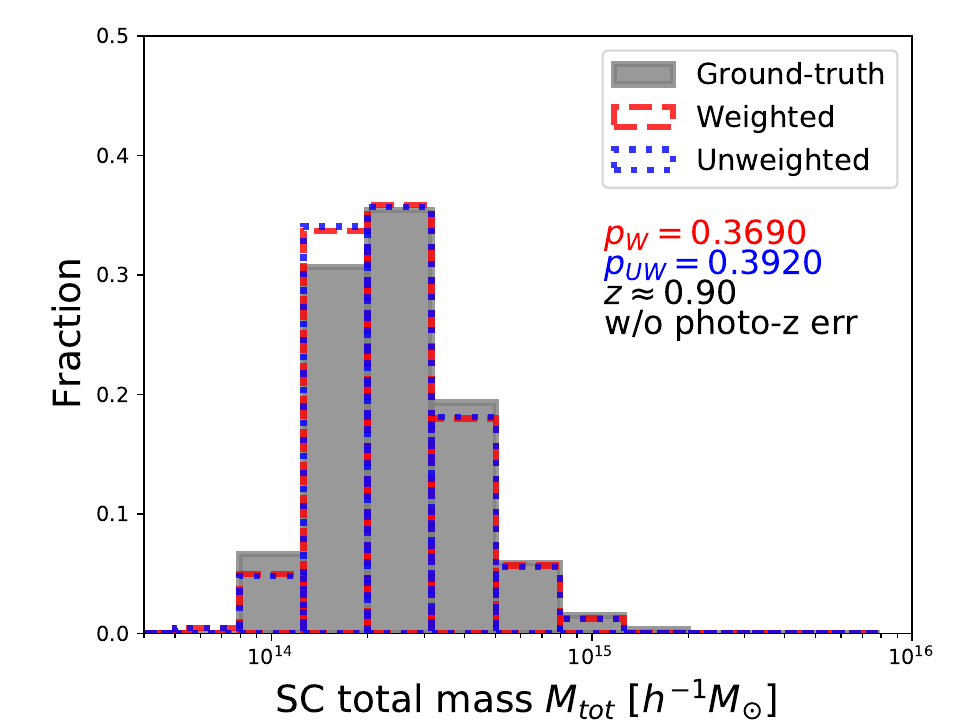}
            \includegraphics[width = 0.5 \hsize]{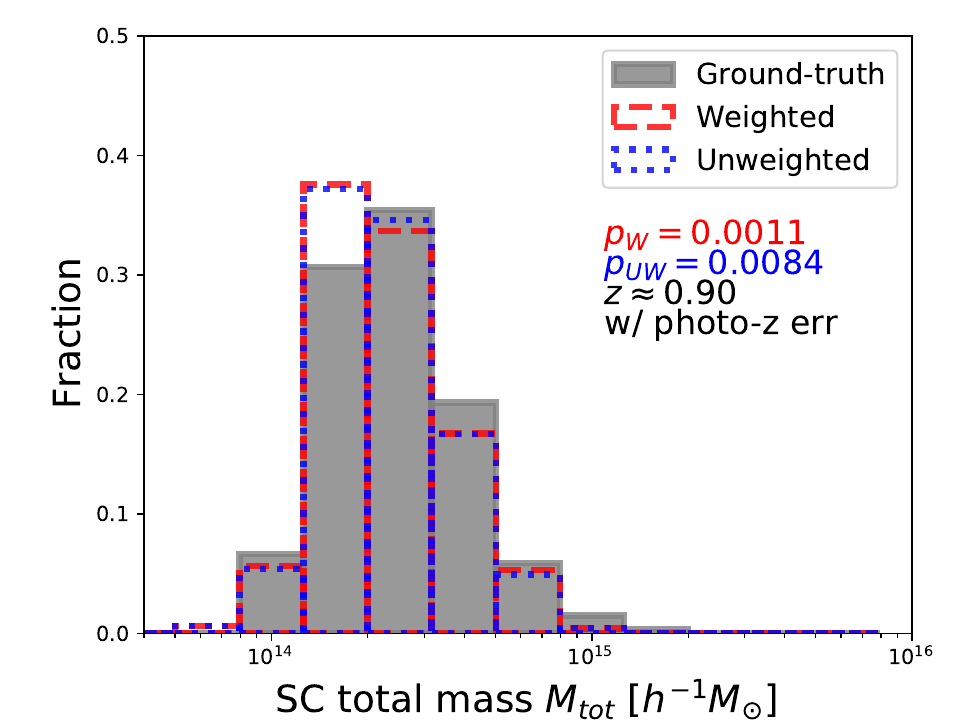}
            \includegraphics[width = 0.5 \hsize]{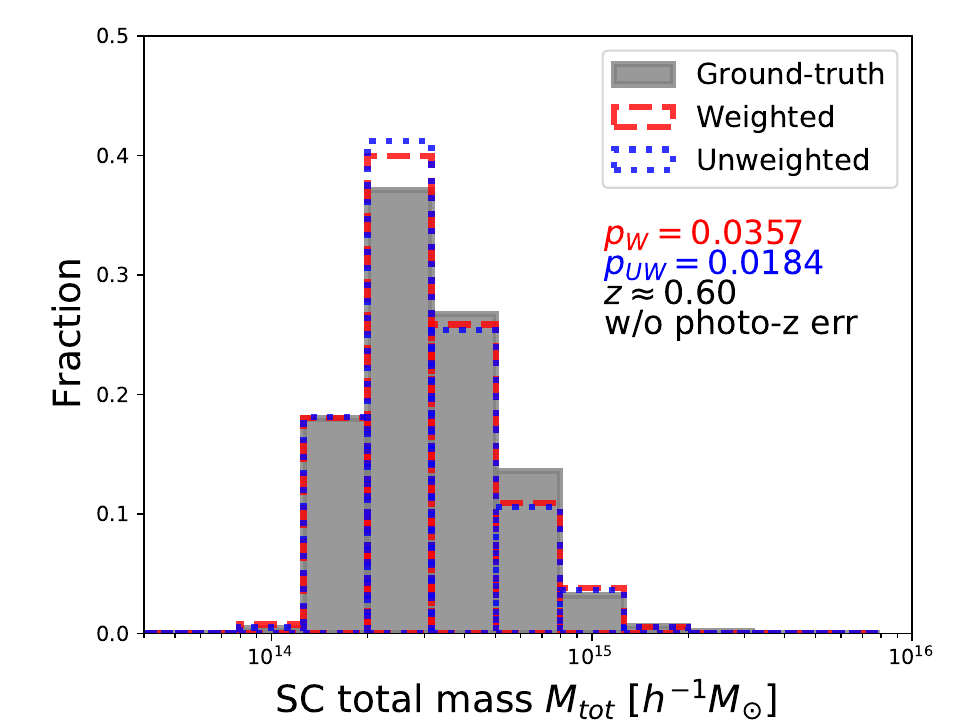}
            \includegraphics[width = 0.5 \hsize]{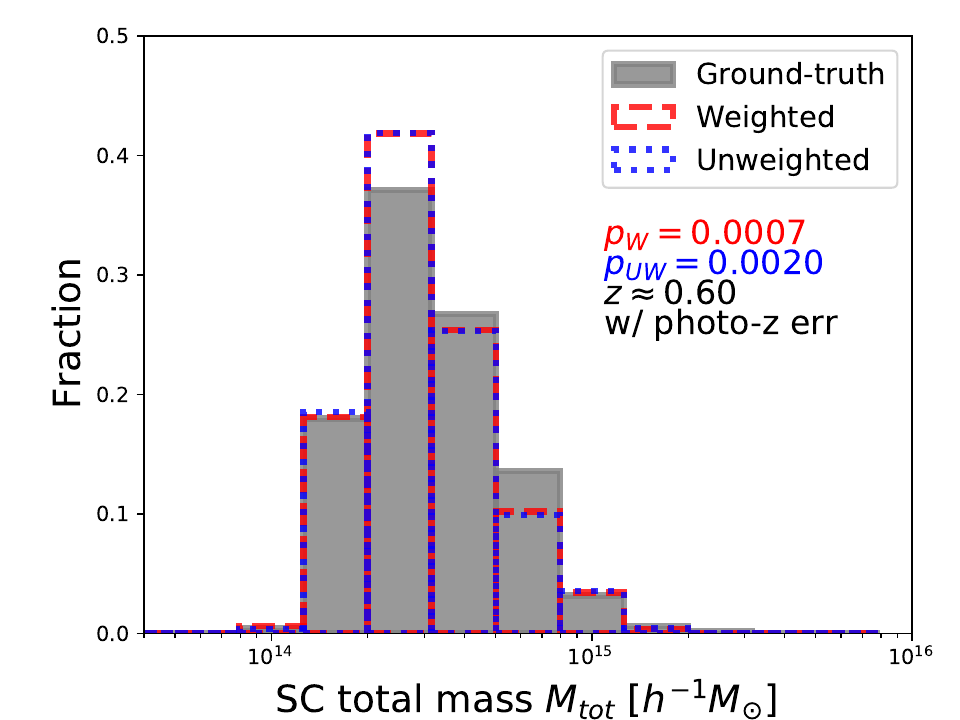}
        \end{minipage}
    \end{center}
    \caption{Supercluster total mass $M_{tot}$ distribution in the unperturbed mock catalog (left column) and perturbed mock catalog (right column). The top  and bottom rows show distributions at redshift $z=0.9$ and $z=0.6$, respectively. In each panel, the gray histogram shows the $M_{tot}$ distribution of the ground-truth superclusters. The red dashed histogram and blue dotted histogram are $M_{tot}$ distributions  obtained by the mass-weighted linking length and the unweighted linking length, respectively.}
    \label{fig:SC_mass_uw_metric}
\end{figure*}

As the purity and completeness used for the linking length optimization are weighted by the halo mass (equations \ref{purity} and \ref{completeness}), one might be concerned about any potential biases resulted from these metrics. Below we show that the mass distribution of our supercluster sample is unbiased.

We first rerun the optimization process using unweighted purity and completeness. The resulting optimized linking length is systematically shorter than the optimized linking length as shown in Figure \ref{fig:LD}, but the different is $\lessapprox 0.7\,h^{-1}\,$cMpc. We then further apply the unweighted linking length to the light cones described in Section \ref{subsec:mock}. Figure \ref{fig:SC_mass_uw_metric} shows the total mass distribution of superclusters at different redshifts in the unperturbed and perturbed light cones. In each panel, the gray histogram is the mass distribution of the ground-truth superclusters. The red dashed and blue dotted histograms are mass distributions of supercluster candidates obtained by mass-weighted linking length and unweighted linking length, respectively. Although using the unweighted metric slightly reduces the optimized linking length, the resulting mass distributions are nearly identical to the mass distributions constructed by the mass-weighted linking length. Furthermore, both supercluster candidate mass distributions appear to be consistent with the ground-truth supercluster mass distribution.

To be more quantitative, we  perform a KS test comparing the mass distributions of ground-truth superclusters with supercluster candidates obtained with unweighted and mass-weighted linking lengths. The resulting p-values ($p_{uw}$ for the unweighted metric and $p_w$ for the mass-weighted metric) are shown in the texts of each panel, and they suggest these three distributions are consistent with each other when the light cones are not perturbed (left column of Figure \ref{fig:SC_mass_uw_metric}). We therefore conclude that the supercluster candidates we capture are unbiased regardless of the metric adopted for the optimization process. However, introducing photometric redshift uncertainties might marginally reduce the consistency as shown in the right column of Figure \ref{fig:SC_mass_uw_metric}.

\section{Photometric Redshift Confirmation} \label{sec:appendix_photoz_confirmation}

Here we present more details on our photo-$z$ examination process described in Section \ref{sec:data}. We begin by justifying the hyperparameter choice for the DESI examination in Section \ref{subsec:appendix_DESI}. Then, using the cluster redshift derived from DESI EDR, we show that the spectroscopic redshift of a BCG can be regarded as a unbiased proxy of the  redshift of its host cluster  (Section \ref{subsec:appendix_BCG}). Lastly, in Section \ref{subsec:appendix_DEmP}, we validate the hyperparameter selection for the DEmP photo-$z$ examination.

\subsection{DESI} \label{subsec:appendix_DESI}

\begin{figure*}[htb]
    \centering
    \begin{minipage}{1.0\textwidth}
        \begin{center}
            \includegraphics[width = 0.49\hsize]{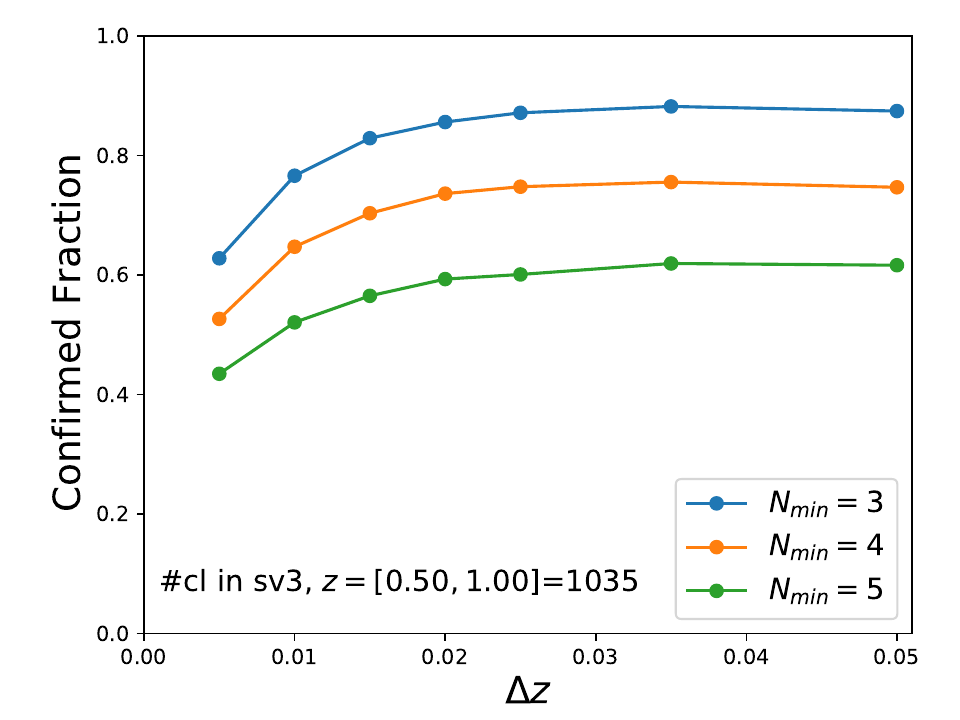}
            \includegraphics[width = 0.49\hsize]{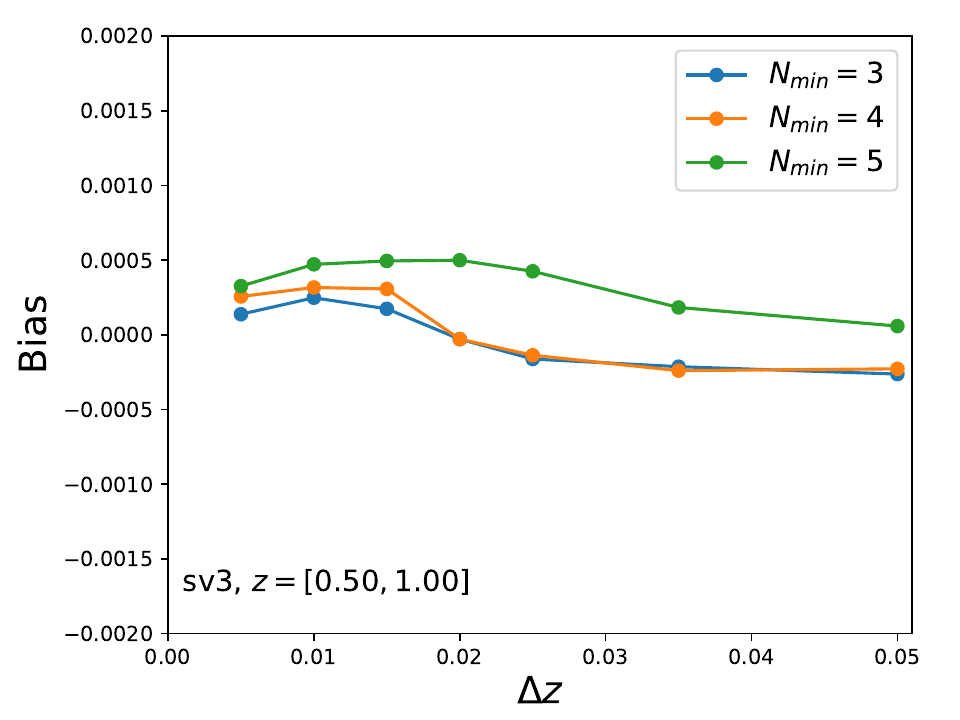}
            \includegraphics[width = 0.49\hsize]{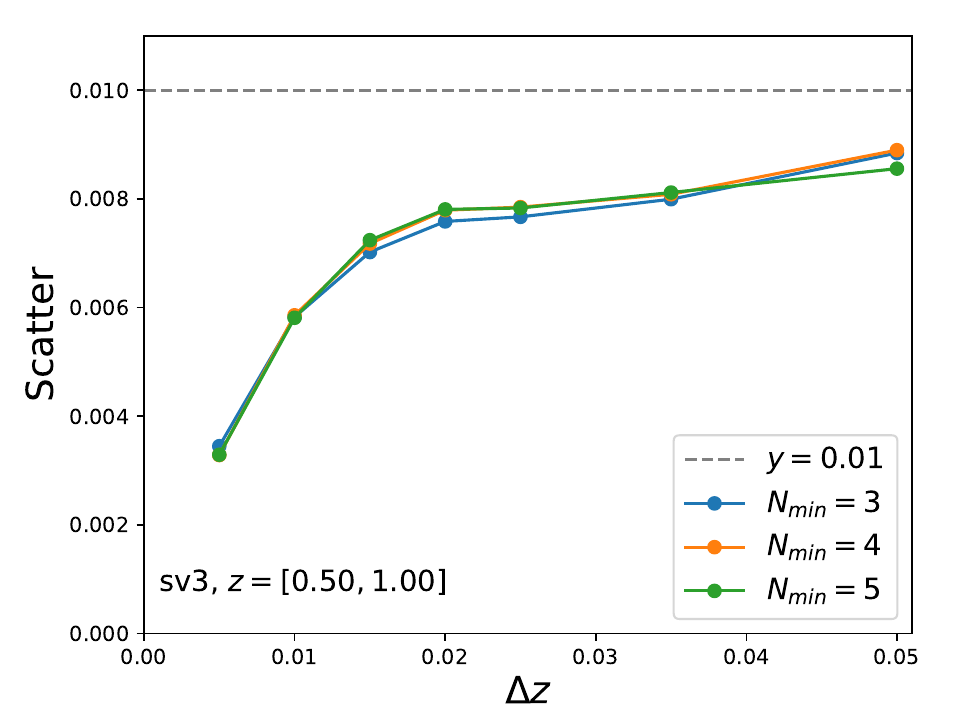}
            \includegraphics[width = 0.49\hsize]{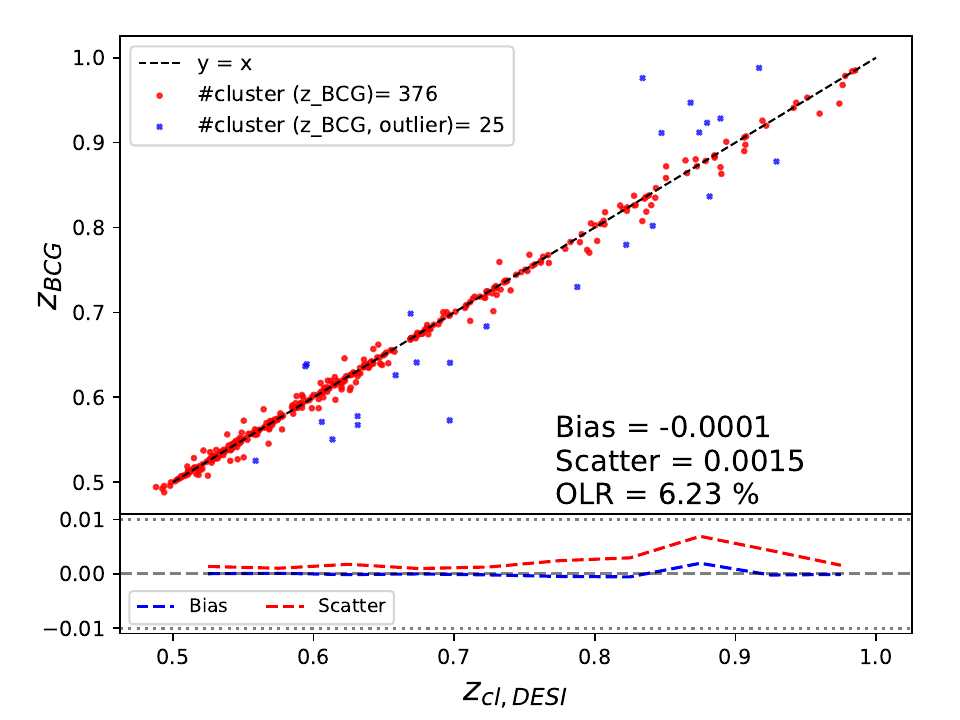}
        \end{center}
    \end{minipage}
    \caption{Upper left panel: The confirm fraction of CAMIRA clusters in the DESI SV3 footprint validated by DESI EDR as a function of redshift interval size $\Delta z$ (see Section \ref{subsec:DESI} for detail). The curves of different colors represent different minimum DESI galaxies required in our DESI examination process. Upper right panel: similar to the upper left panel, but for the bias between CAMIRA photo-$z$ $z_{CAM}$ and cluster redshift derived from DESI EDR $z_{cl,DESI}$. Lower left panel: similar to the upper left panel, but for the scatter between CAMIRA photo-$z$ $z_{CAM}$ and cluster redshift derived from DESI EDR $z_{cl,DESI}$. Lower right panel: a comparison between spectroscopic redshift of BCGs $z_{BCG}$ and $z_{cl,DESI}$. The bias and scatter are calculated by equations \ref{eqn:photoz_bias} and \ref{eqn:photoz_scatter}, respectively. Outlier rate (OLR) is the fraction of redshift $4\sigma-$clipped when calculating the scatter as defined in \citet{Oguri18}.}
    \label{fig:photoz_DESI}
\end{figure*}

For Section \ref{subsec:DESI}, we need to determine the appropriate $\Delta z$ and $N_{min}$ values by conducting a parameter search. We first collect $1035$ high-$z$ ($z=0.5-1.0$) CAMIRA clusters located in the one-percent survey footprint of DESI \citep[hereafter SV3]{DESI_EDR23}, which has the highest completeness, as our test sample. We then repeat the procedure described in Section \ref{subsec:DESI} for different $\Delta z$ and $N_{min}=3,4,5$. The upper left panel of Figure \ref{fig:photoz_DESI} shows the confirmed cluster fraction as a function of $\Delta z$ and $N_{min}$. The upper right  and lower left panels show the bias and scatter as a function of $\Delta z$ and $N_{min}$. The bias and scatter are calculated using the equations \ref{eqn:photoz_bias} and \ref{eqn:photoz_scatter} where $z_{BCG}$ is replaced by $z_{cl, DESI}$.

We see that both the confirmed fraction and the scatter increase initially with $\Delta z$ but become saturated when $\Delta z\ge 0.02$. The saturation suggests that statistically no new clusters could be confirmed, even if we search for structures within a larger initial volume. The saturated scatter is around $0.008$, which is close to the estimate of \citet{Oguri18}. Therefore, in this work, we set $\Delta z$ to $0.02$.

As for $N_{min}$, the scatter seems to be independent of $N_{min}$ as shown in lower left panel of Figure \ref{fig:photoz_DESI}. We estimate the background value by randomly placing $5000$ cylinders with a radius $1h^{-1}\,$pMpc and line-of-sight distance of $\Delta v = \pm 3000\,$km/s in the SV3 footprint and find that about $1.3$ galaxies are enclosed within such a cylinder on average. Therefore, we set $N_{min}=4$, which is approximately three times the background value.

\subsection{BCG} \label{subsec:appendix_BCG}

BCGs are of a special galaxy population residing in galaxy clusters. Although a significant number of BCGs are found in motion relative to their host clusters, the typical peculiar velocity is around $100$ to $200$ km s$^{-1}$ \citep{Coziol09, DePropris21}, which is much smaller than the cluster photo-$z$ scatter. In Section \ref{subsec:camira}, we use the spectroscopic redshift of BCG $z_{BCG}$ as the cluster redshift if it is available {\it and} if $|z_{BCG}-z_{CAM}|<0.01(1+z_{CAM})$ where $z_{CAM}$ is the cluster photo-$z$ provided in the CAMIRA catalog.

Here, we investigate whether the redshift of BCGs $z_{BCG}$ is representative of cluster redshift by comparing the available BCG spectroscopic redshift $z_{BCG}$ with clusters validated by DESI EDR (see Section \ref{subsec:DESI}). The lower right panel of Figure \ref{fig:photoz_DESI}  shows the redshift comparison for clusters with available $z_{BCG}$  and are confirmed by our DESI photo-$z$ examination process described above (i.e., $z_{cl, DESI}$ is available). The bias and scatter are calculated by equations \ref{eqn:photoz_bias} and \ref{eqn:photoz_scatter} with $z_{CAM}$ replaced by $z_{BCG}$ and $z_{BCG}$ replaced by $z_{cl, DESI}$. The small bias and scatter are $\approx -0.0001(1+z_{cl, DESI})$ and $\approx 0.0015(1+z_{cl, DESI})$, suggesting using $z_{BCG}$ to represent cluster's redshift is robust.

\subsection{DEmP} \label{subsec:appendix_DEmP}

\begin{table*}
    \centering
    \begin{tabular}{| l | c | c |}
        \hline
           & Denominator & Numerator \\
        \hline
        Purity & 
        \makecell[l]{
            Richness $N_{mem}\geq N_{min}$\\
            In SV3 footprint\\
            Validated by DEmP examination }
        & 
        \makecell[l]{
        All conditions in the purity denominator\\
        Validated by DESI examination
        } \\
        \hline
        Completeness & 
        \makecell[l]{
            Richness $N_{mem}\geq N_{min}$\\
            Validated by DESI examination}
        & 
        \makecell[l]{
        All conditions in the completeness denominator\\
        Validated by DEmP examination
        } \\
        \hline
    \end{tabular}
    \caption{Definition of purity and completeness in Figure \ref{fig:photoz_DEmP}}
    \label{tab:purity_completeness_DEmP}
\end{table*}

\begin{figure*}[htb]
    \centering
    \begin{minipage}{1.0\textwidth}
        \begin{center}
            \includegraphics[width = 0.49\hsize]{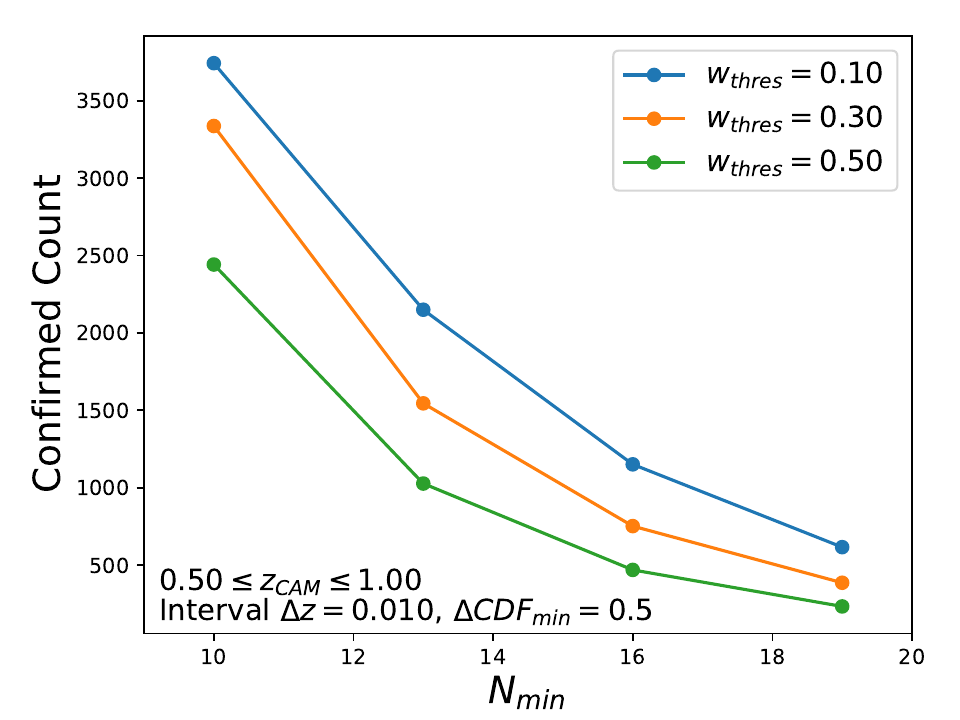}
            \includegraphics[width = 0.49\hsize]{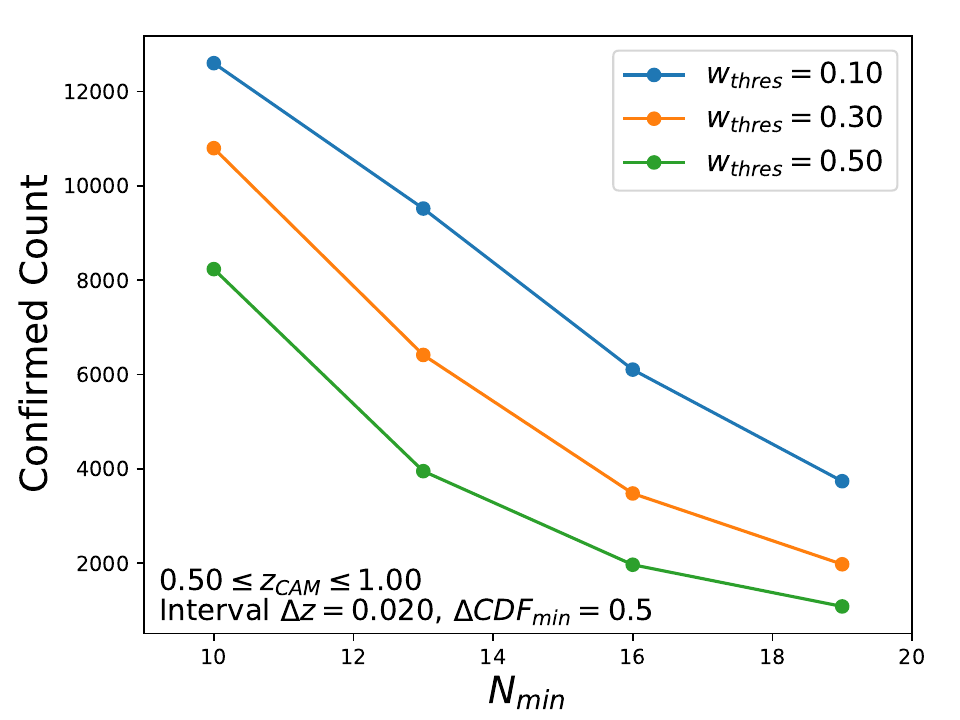}
            \includegraphics[width = 0.49\hsize]{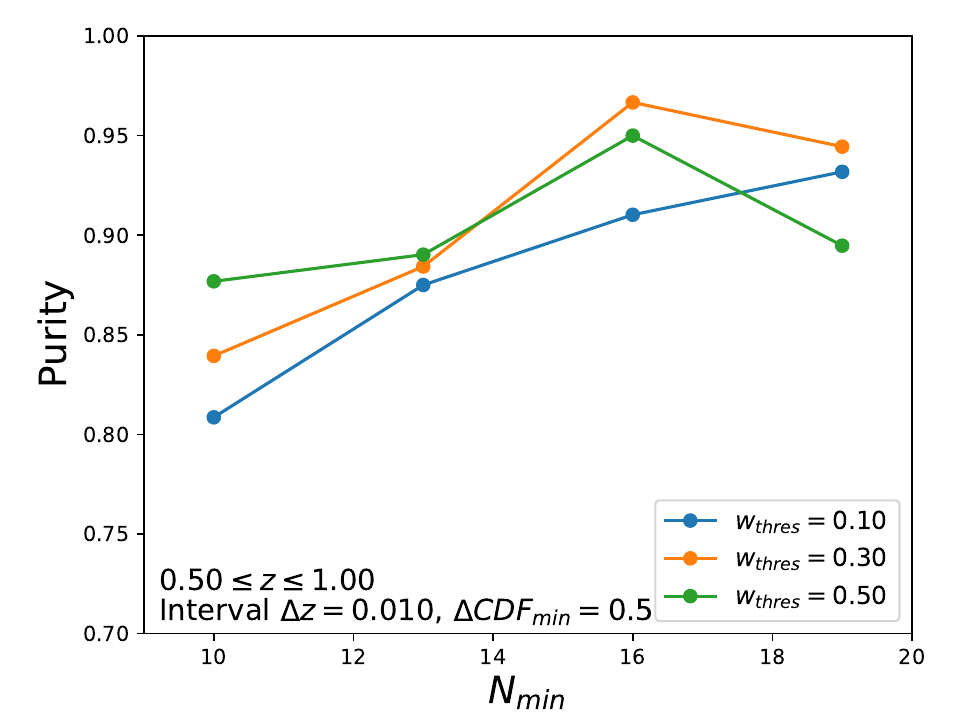}
            \includegraphics[width = 0.49\hsize]{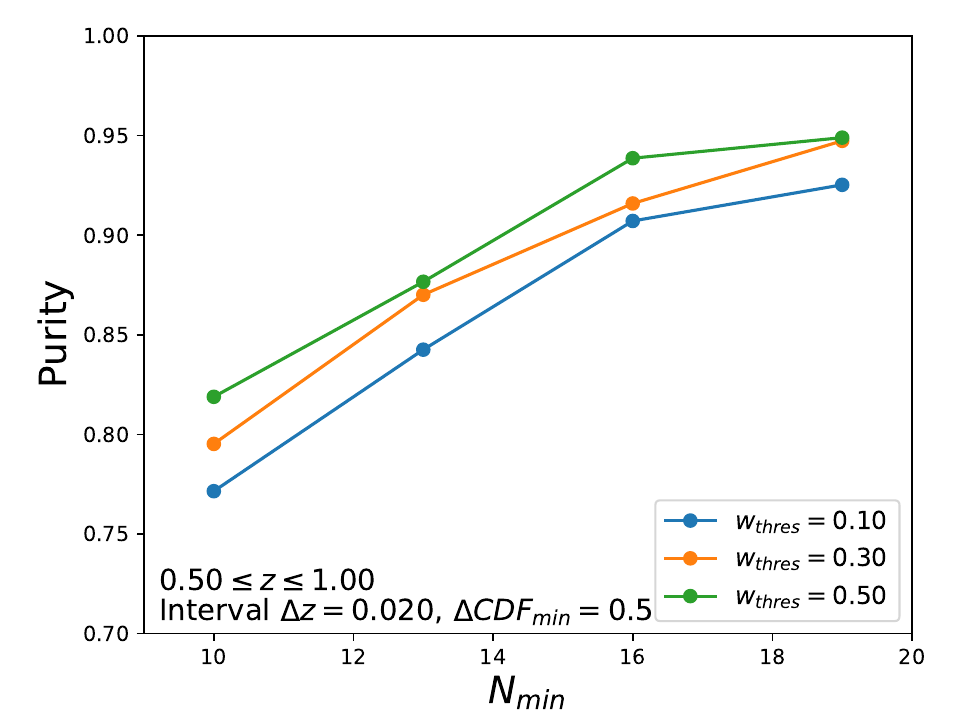}
            \includegraphics[width = 0.49\hsize]{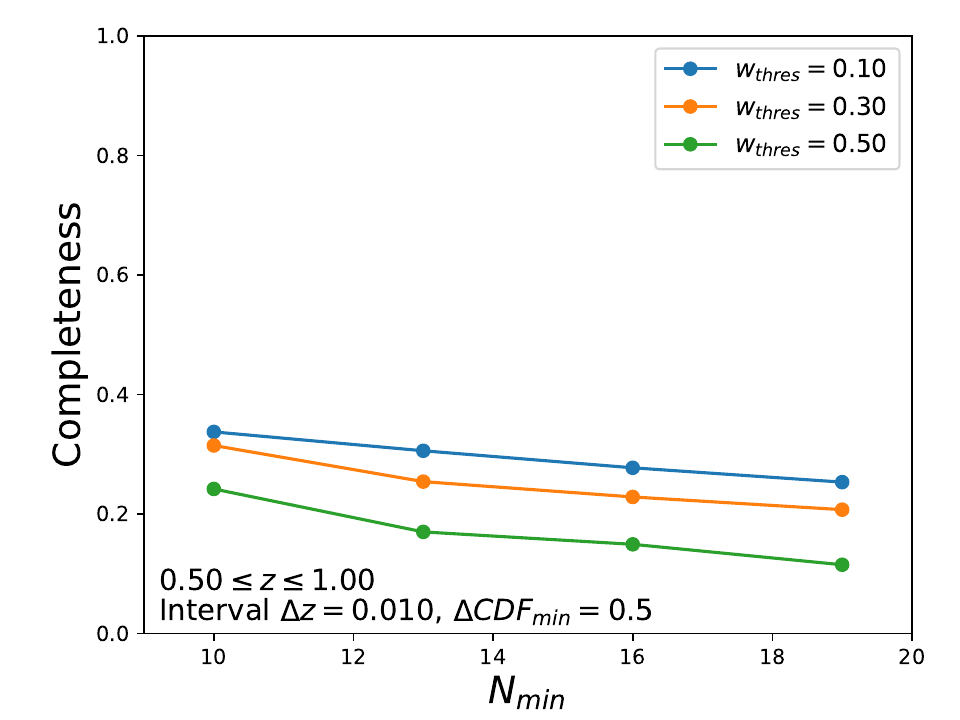}
            \includegraphics[width = 0.49\hsize]{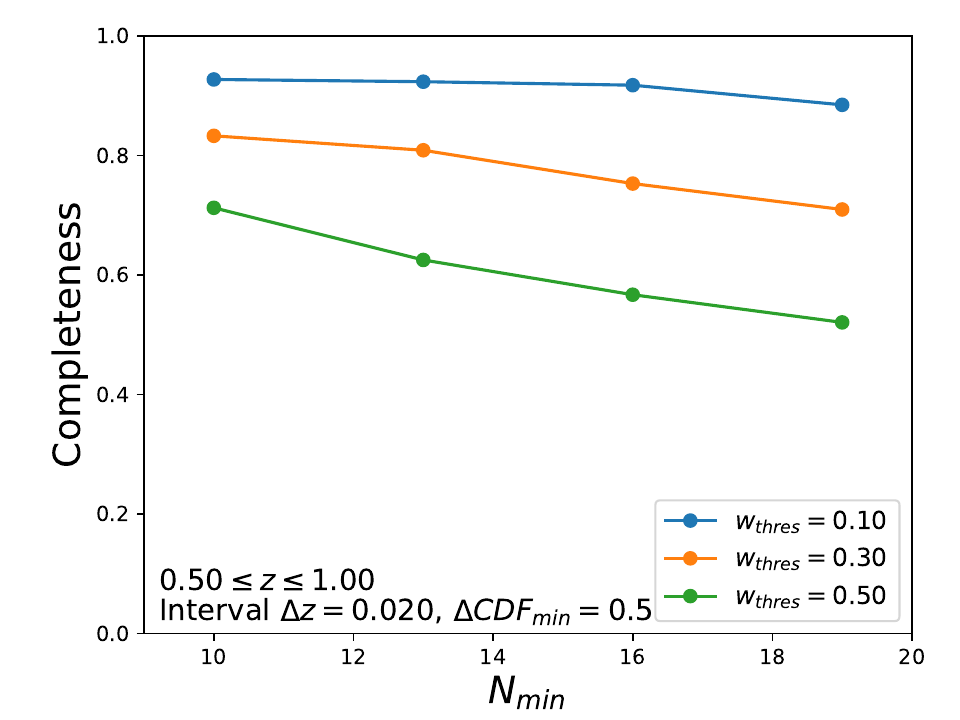}
        \end{center}
    \end{minipage}
    \caption{The confirmed count, purity, and completeness as a function of hyperparameters $(\Delta z, w_{thres}, N_{min})$ for DEmP examination described in Section \ref{subsec:photo-z}. Top row: the confirmed count of CAMIRA clusters as a function of the minimum member $N_{min}$. Curves of different colors represent different membership probability thresholds $w_{thres}$ adopted. Middle row: same as the top row, but for the purity defined in Table \ref{tab:purity_completeness_DEmP}. Bottom row: same as the top row, but for the completeness defined in Table \ref{tab:purity_completeness_DEmP}. The size of redshift interval $\Delta z=0.01$ and $\Delta z=0.02$ are used for the left and right column, respectively.}
    \label{fig:photoz_DEmP}
\end{figure*}

Here we conduct a parameter search in order to find the most suitable hyperparameters $(\Delta z, w_{thres}, N_{min})$ for the DEmP examination described in Section \ref{subsec:photo-z}. Figure \ref{fig:photoz_DEmP} shows the confirmed cluster count, purity, and completeness as a function of $(\Delta z, w_{thres}, N_{min})$. The confirmed count is the number of clusters in our whole sample validated by the DEmP examination for different hyperparameter choices. As for the purity and completeness, in table \ref{tab:purity_completeness_DEmP}, we list  the conditions that clusters have to satisfy {\it simultaneously} when considering the clusters for denominator and numerator of purity and completeness. In this way, the purity quantifies the fraction of clusters confirmed by the DESI examination among the clusters confirmed by the DEmP examination. The completeness quantifies the fraction of clusters confirmed by the DEmP examination among the clusters confirmed by the DESI examination.

In Figure \ref{fig:photoz_DEmP}, we first find that when we adopt $\Delta z=0.02$ instead of $\Delta z=0.01$, the purity is slightly reduced, but the completeness is  increased dramatically. Then, considering that both purity and completeness are not a strong function of $N_{min}$ (the variation is $\lesssim 10\%$), we adopt $N_{min}=10$ since a higher $N_{min}$ would strongly reduce the confirmed count. Lastly, as we prioritize purity over completeness, we settle the membership probability threshold to $w_{thres}=0.5$, i.e., we only use reliable member galaxies for this analysis. In summary, we set our hyperparameter choice to $(\Delta z, w_{thres}, N_{min})=(0.02, 0.5, 10)$, leading to purity and completeness of $\approx 82\%$ and $\approx 71\%$, respectively.

\section{Fragmentation} \label{sec:appendix_fragmentation}

\begin{figure*}[htb]
    \centering
    \begin{minipage}{1.0\textwidth}
        \begin{center}
            \includegraphics[width = 0.49\hsize]{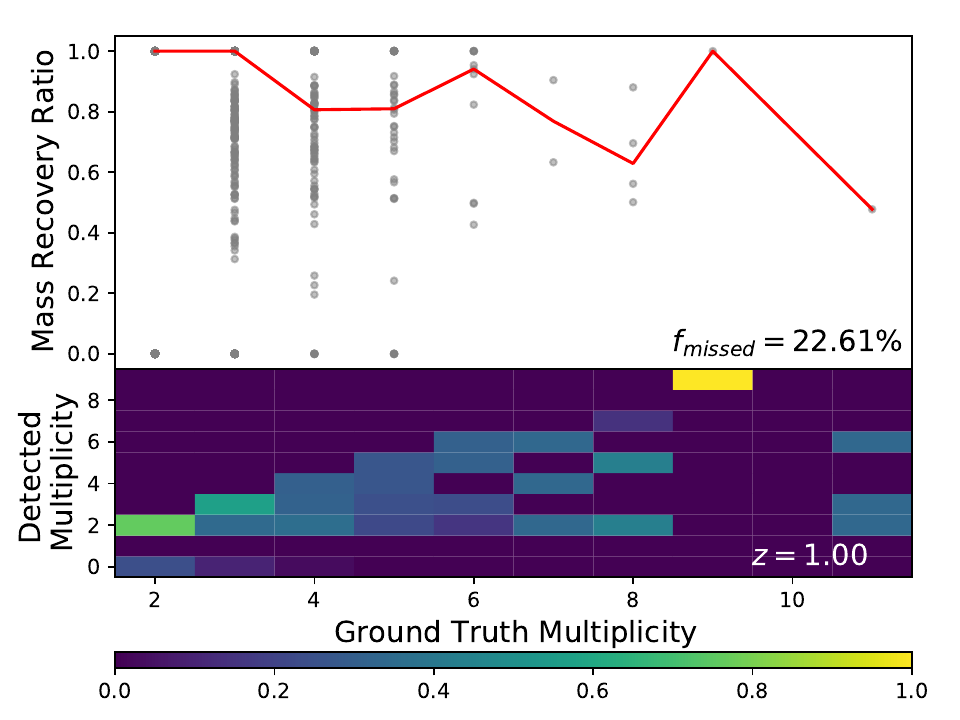}
            \includegraphics[width = 0.49\hsize]{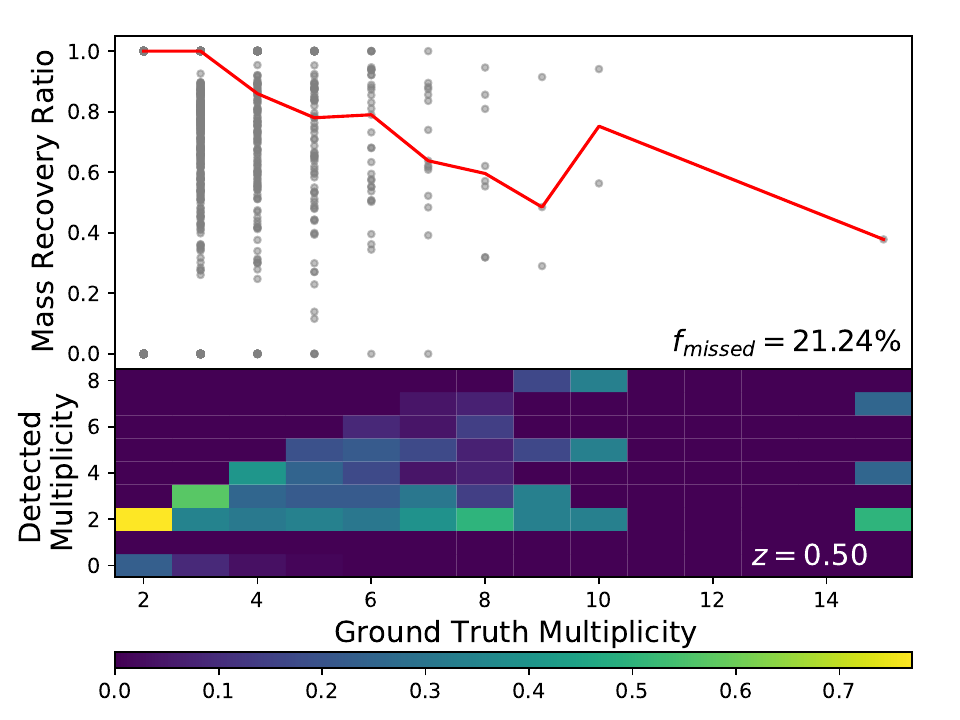}
        \end{center}
    \end{minipage}
    \caption{Left panel: The upper subplot shows the mass recovery ratio  as a function of the multiplicity $N_{\rm SC}$ of the ground-truth superclusters. Each gray point represents a ground-truth supercluster in our six simulation boxes. The red curve shows the median as a function of $N_{\rm SC}$. The fraction of ground-truth superclusters that are not detected $f_{missed}$ is shown in the lower right corner. The lower subplot shows a 2D histogram between the detected multiplicity and the ground-truth multiplicity. Normalization is set such that each column sums up to unity. Right panel: same as left panel, but at redshift $z=0.50$.}
    \label{fig:fragmentation_box}
\end{figure*}

It is possible that a high-multiplicity supercluster is split or identified as several low-multiplicity superclusters. We investigate this ``fragmentation'' issue here. We begin our analysis by comparing the ground-truth supercluster catalog with the FoF groups (supercluster candidates) catalog in the simulation boxes. The upper panel of Figure~\ref{fig:fragmentation_box} shows the mass recovery ratio as a function of the multiplicity $N_{\rm SC}$ of ground-truth supercluster at redshift $z=1.0$ (left panel) and $z=0.5$ (right panel).  For each ground-truth supercluster (represented by a gray point in the Figure), we calculate the mass recovery ratio by picking up their most massive FoF group counterpart, which is a subset of the ground-truth supercluster cluster members. As suggested by the median shown as the red curve, for most ground-truth supercluster of  $N_{\rm SC}=2-3$, which dominate the supercluster population (Figure \ref{fig:multiplicity}), our method can recover their cluster members without splitting them. However, for ground-truth superclusters of   $N_{\rm SC}>3$, our method generally recovers $\approx 80\%$ of their mass, reflecting the fragmentation problem. The text in the lower right corner shows the fraction $f_{missed}$ of ground-truth superclusters not detected at all by our method, i.e., the mass recovery ratio equals zero. The value $f_{missed}\approx 20\%$ is consistent with our optimized completeness shown in Figure \ref{fig:performanance}.

The lower panel of Figure \ref{fig:fragmentation_box} shows 2D histograms between the detected multiplicity and the multiplicity of the ground-truth superclusters. Normalization is set such that the sum of value in each column is unity. For example, at redshift $z=1.0$, there is one ground-truth supercluster with $N_{\rm SC}=9$ completely captured. As another example, at redshift $z=0.5$, there is one ground-truth supercluster with $N_{\rm SC}=15$ (the right most column) identified as a $N_{\rm SC}=7$ FoF group, one $N_{\rm SC}=4$ FoF group, and two $N_{\rm SC}=2$ FoF groups. We note that the FoF groups with multiplicity higher than their ground-truth supercluster counterparts are excluded in the analysis. This is because such FoF groups are regarded as contamination and only account for less than $5\%$ of all FoF groups by number. In summary, from the 2D histograms, we see that while the ground-truth superclusters of  $N_{\rm SC} \leq 4$ are generally not split, the fragmentation issue becomes worse for ground-truth supercluster of higher multiplicity, as can be expected. Although the high-multiplicity superclusters are extremely rare, this analysis  illustrates the limitation of using distance solely to connect  structures that will collapse in the future.

In Figure \ref{fig:fragmentation_box}, the effect of photo-$z$ uncertainties is not accounted for, but it is expected to worsen the issue. This also explains the absence of  high-multiplicity superclusters in Figure \ref{fig:multiplicity}. This again reveals the importance of using spectroscopic data to precisely locate large-scale structures at higher redshift.

\bibliography{sample631}{}
\bibliographystyle{aasjournal}

\end{document}